\documentclass[fleqn,usenatbib]{mnras}

\usepackage[T1]{fontenc}

\usepackage{graphicx}	
\usepackage{amsmath}	
\usepackage{amssymb}	

\usepackage{float}

\title[A new nook at local ULIRGs]
{A new look at local ultraluminous infrared galaxies: the atlas and radiative transfer models of their complex physics}

\author[A. Efstathiou et al.]{\parbox{\linewidth}{A. Efstathiou$^{1}$\thanks{E-mail: a.efstathiou@euc.ac.cy}, D. Farrah$^{2,3}$, J. Afonso$^{4,5}$, D. L. Clements$^{6}$, E. Gonz\'{a}lez-Alfonso$^{7}$, M. Lacy$^{8}$, S. Oliver$^{9}$, V. Papadopoulou Lesta$^{1}$, C. Pearson$^{10,11,12}$, D. Rigopoulou$^{12}$, M. Rowan-Robinson$^{6}$,  H.W.W. Spoon$^{13}$, A. Verma$^{12}$, L. Wang$^{14,15}$}
\\
\\
$^{1}$School of Sciences, European University Cyprus, Diogenes street, Engomi, 1516 Nicosia, Cyprus\\
$^{2}$Department of Physics and Astronomy, University of Hawaii, 2505 Correa Road, Honolulu, HI 96822, USA\\
$^{3}$Institute for Astronomy, 2680 Woodlawn Drive, University of Hawaii, Honolulu, HI 96822, USA\\
$^{4}$Instituto de Astrof\'{ı}sica e Ci\^{e}ncias do Espaco, Universidade de Lisboa, OAL, Tapada da Ajuda, P-1349-018 Lisboa, Portugal\\
$^{5}$Departamento de F\'{ı}sica, Faculdade de Ci\^{e}ncias, Universidade de Lisboa, Edif\'{ı}cio C8, Campo Grande, P-1749-016 Lisbon, Portugal\\
$^{6}$Astrophysics Group, Imperial College London, Blackett Laboratory, Prince Consort Road, London SW7 2AZ, UK\\
$^{7}$Departamento de F\'{i}sica y Matem\'{a}ticas, Universidad de Alcal\'{a}, Campus Universitario, E-28871 Alcal\'{a} de Henares, Madrid, Spain\\
$^{8}$National Radio Astronomy Observatory, 520 Edgemont Road, Charlottesville, VA 22903, USA\\
$^{9}$Astronomy Centre, Department of Physics and Astronomy, University of Sussex, Falmer, Brighton BN1 9QH, UK\\
$^{10}$RAL Space, CCLRC, Rutherford Appleton Laboratory, Chilton, Didcot, Oxfordshire OX11 0QX, UK\\
$^{11}$School of Physical Sciences, The Open University, Milton Keynes MK7 6AA, UK\\
$^{12}$Oxford Astrophysics, Denys Wilkinson Building, University of Oxford, Keble Road, Oxford OX1 3RH, UK\\
$^{13}$Cornell Center for Astrophysics and Planetary Science, Ithaca, NY 14853, USA
\\
$^{14}$SRON Netherlands Institute for Space Research, Landleven 12, NL-9747AD, Groningen, the Netherlands\\
$^{15}$Kapteyn Astronomical Institute, University of Groningen, Postbus 800, NL-9700 AV, Groningen, the Netherlands\\
}

\vspace{-40pt}

\date{Accepted 2021 November 29; Received 2021 November 29; in original form 2021 April 6}

\pubyear{2022}

\begin{document}

\pagerange{\pageref{firstpage}--\pageref{lastpage}} \pubyear{2022}

\maketitle

\label{firstpage}

\begin{abstract}
We present the ultraviolet to submillimetre spectral energy distributions (SEDs) of the HERschel Ultra Luminous Infrared Galaxy Survey (HERUS) sample of 42 local Ultraluminous Infrared Galaxies (ULIRGs) and fit them with a Markov chain Monte Carlo (MCMC) code using the CYprus models for Galaxies and their NUclear Spectra (CYGNUS) radiative transfer models for starbursts, active galactic nucleus (AGN) tori and host galaxy. The Spitzer IRS spectroscopy data are included in the fitting. Our bayesian SED fitting method takes comparable time to popular energy balance methods but it is more physically motivated and versatile.
All HERUS galaxies harbor high rates of star formation but we also find bolometrically significant AGN in all of the galaxies of the sample. We estimate the correction of the luminosities of the AGN in the ULIRGs due to the anisotropic emission of the torus and find that it could be up to a factor of $\sim10$ for nearly edge-on tori.
We present a comparison of our results with the smooth torus model of Fritz et al. and the two-phase models of Siebenmorgen et al. and SKIRTOR. We find that the CYGNUS AGN torus models fit significantly better the SEDs of our sample compared to all other models.
We find no evidence that strong AGN appear either at the beginning or end of a starburst episode or that starbursts and AGN affect each other.  IRAS~01003-2238 and Mrk~1014 show evidence for dual AGN in their SED fits suggesting a minimum dual AGN fraction in the sample of 5\%.
\end{abstract}


\vspace{-50pt}

\section{Introduction}\label{intro}
Local Ultraluminous Infrared Galaxies (ULIRGs), with infrared luminosities exceeding $10^{12} L_\odot$ are an important population in their own right as they are the most luminous galaxies in the local Universe. ULIRGs also allow us to study in detail the role of mergers in triggering extreme star formation and active galactic nucleus (AGN) activity which is deeply obscured by dust. They are therefore important laboratories for understanding these processes in order to aid the interpretation of observations of similar events which took place in the most luminous galaxies in the history of the Universe such as hyperluminous infrared galaxies 
\citep{mrr93,mrr00,ver02,far02,efstathiou06}, submillimetre galaxies \citep{hug98,bar98,efs03,cas14,mrr18}, quasars, and Hot Dust Obscured Galaxies (Hot DOGs; \citealt{eis12,bri13,far17,efs21}). For these and other reasons ULIRGs have received a lot of attention since their discovery by the Infrared Astronomical Satellite (IRAS) in the 1980s \citep{hou85,soi85}. Excellent reviews of observational and theoretical work are given in \citet{san96}, \citet{lon06}, and more recently \citet{perez21}.

\begin{table*}
\label{tab:basic}
\begin{center}
\begin{tabular}{ccccccccccc}
\hline 
ID  & IRAS      & Other       & RA            & Dec         & z        & Opt.    & \multicolumn{4}{c}{Maximum Likelihoods} \\  
    & Name      & Names       &              &              &          & Class   & CYGNUS  &  F06   & S15  &  SKIRTOR  \\  
\hline 
1  & 00188-0856 & ---         &  5.360507    &  -8.657217   &   0.128  & LINER &  -459.7  &  -467.7  &  -519.9  &  -567.3 \\ 
2  & 00397-1312 & ---         &  10.564704   &  -12.934122  &   0.262  & HII   &  -485.5  &  -342.5  &  -1093.6  &  -848.9 \\ 
3  & 01003-2238 & ---         &  15.708365   &  -22.365895  &   0.118  & HII   &  -137.6  &  -1227.6  &  -125.7  &  -1074.0 \\ 
4  & 03158+4227 & ---         &  49.801667   &   42.641111  &   0.134  & Sy2   &  -262.8  &  -702.0  &  -291.9  &  -973.9 \\ 
5  & 03521+0028 & ---         &  58.675800   &   0.617611   &   0.152  & LINER &  -196.5  &  -373.3  &  -199.8  &  -402.8 \\ 
6  & 05189-2524 & ---         &  80.255800   &  -25.362600  &   0.043  & Sy2   &  -97.2  &  -260.8  &  -63.3  &  -339.8 \\ 
7  & 06035-7102 & ---         &  90.725042   &  -71.052833  &   0.079  & HII   &  -126.9  &  -158.4  &  -223.1  &  -188.2 \\ 
8  & 06206-6315 & ---         &  95.255042   &  -63.289861  &   0.092  & Sy2   &  -213.6  &  -302.4  &  -229.8  &  -379.7 \\ 
9  & 07598+6508 & ---         &  121.137833  &   64.996833  &   0.148  & Sy1   &  -33.9  &  -26.5  &  -181.4  &  -33.4 \\ 
10 & 08311-2459 & ---         &  128.335833  &  -25.159361  &   0.100  & Sy1   &  -111.8  &  -275.3  &  -82.7  &  -300.5 \\ 
11 & 08572+3915 & ---         &  135.105792  &   39.065111  &   0.058  & Sy2   &  -218.6  &  -618.5  &  -1365.0  &  -1353.1 \\ 
12 & 09022-3615 & ---         &  136.052961  &  -36.450274  &   0.060  & HII   &  -138.5  &  -189.8  &  -145.6  &  -164.4 \\ 
13 & 10378+1109 & ---         &  160.121539  &   10.888415  &   0.136  & LINER &  -182.8  &  -237.4  &  -164.2  &  -370.6 \\ 
14 & 10565+2448 & ---         &  164.825493  &   24.542905  &   0.043  & HII   &  -322.9  &  -512.7  &  -363.5  &  -493.1 \\ 
15 & 11095-0238 & ---         &  168.014071  &  -2.906219   &   0.107  & LINER &  -187.1  &  -773.2  &  -547.7  &  -983.2 \\ 
16 & 12071-0444 & ---         &  182.438049  &  -5.020490   &   0.128  & Sy2   &  -177.8  &  -643.3  &  -168.6  &  -662.0 \\ 
17 & 13120-5453 & ---         &  198.776494  &  -55.156452  &   0.031  & Sy2   &  -349.0  &  -503.4  &  -343.4  &  -508.9 \\ 
18 & 13451+1232 & 4C 12.50    &  206.889007  &   12.290067  &   0.122  & Sy2   &  -107.3  &  -524.5  &  -87.6  &  -671.4 \\ 
19 & 14348-1447 & ---         &  219.410000  &  -15.005556  &   0.083  & LINER &  -119.7  &  -233.0  &  -153.6  &  -299.9 \\ 
20 & 14378-3651 & ---         &  220.245867  &  -37.075538  &   0.068  & Sy2   &  -279.1  &  -581.0  &  -255.2  &  -602.9 \\ 
21 & 15250+3609 & ---         &  231.747517  &   35.977092  &   0.055  & HII   &  -370.5  &  -854.7  &  -595.7  &  -1198.3 \\ 
22 & 15462-0450 & ---         &  237.236721  &  -4.992669   &   0.100  & Sy1   &  -174.7  &  -104.4  &  -54.6  &  -193.5 \\ 
23 & 16090-0139 & ---         &  242.918469  &  -1.785156   &   0.134  & LINER &  -185.5  &  -284.5  &  -230.7  &  -290.2 \\ 
24 & 17208-0014 & ---         &  260.841481  &  -0.283594   &   0.043  & HII   &  -452.6  &  -620.9  &  -461.0  &  -601.8 \\ 
25 & 19254-7245 & SuperAntena &  292.839167  &  -72.655000  &   0.062  & Sy2   &  -165.3  &  -134.5  &  -125.8  &  -195.2 \\ 
26 & 19297-0406 & ---         &  293.088544  &  -3.998962   &   0.086  & HII   &  -179.3  &  -265.9  &  -186.9  &  -340.0 \\ 
27 & 20087-0308 & ---         &  302.849458  &  -2.997417   &   0.106  & LINER &  -333.2  &  -343.5  &  -386.3  &  -352.2 \\ 
28 & 20100-4156 & ---         &  303.373083  &  -41.793028  &   0.130  & HII   &  -511.7  &  -542.4  &  -605.0  &  -721.8 \\ 
29 & 20414-1651 & ---         &  311.075888  &  -16.671172  &   0.087  & HII   &  -199.7  &  -367.5  &  -195.0  &  -292.7 \\ 
30 & 20551-4250 & ESO 286-19  &  314.611589  &  -42.650056  &   0.043  & HII   &  -198.3  &  -581.2  &  -433.6  &  -922.9 \\ 
31 & 22491-1808 & ---         &  342.955267  &  -17.873183  &   0.078  & HII   &  -188.8  &  -534.9  &  -208.1  &  -607.2 \\ 
32 & 23128-5919 & ESO 148-2   &  348.944790  &  -59.054320  &   0.045  & HII   &  -208.2  &  -573.4  &  -205.3  &  -517.3 \\ 
33 & 23230-6926 & ---         &  351.515083  &  -69.171889  &   0.107  & LINER &  -298.4  &  -662.7  &  -389.8  &  -836.7 \\ 
34 & 23253-5415 & AM 2325-541 &  352.025417  &  -53.975278  &   0.130  & LINER &  -190.9  &  -612.1  &  -141.5  &  -688.8 \\ 
35 & 23365+3604 & ---         &  354.755305  &   36.352308  &   0.064  & LINER &  -274.9  &  -898.9  &  -276.3  &  -936.6 \\ 
36 & 09320+6134 & UGC 5101    &  143.964981  &   61.353182  &   0.039  & LINER &  -475.7  &  -548.2  &  -520.2  &  -558.4 \\ 
37 & 12540+5708 & Mrk 231     &  194.059308  &   56.873677  &   0.042  & Sy1   &  -49.7  &  -67.8  &  -46.8  &  -186.7 \\ 
38 & 13428+5608 & Mrk 273     &  206.175463  &   55.886847  &   0.037  & Sy2   &  -218.0  &  -288.8  &  -262.2  &  -411.6 \\ 
39 & 13536+1836 & Mrk 463     &  209.011963  &   18.372078  &   0.049  & Sy2   &  -141.2  &  -150.0  &  -161.1  &  -122.4 \\ 
40 & 15327+2340 & Arp 220     &  233.738563  &   23.503139  &   0.018  & Sy2   &  -1099.8  &  -1938.0  &  -868.3  &  -2143.5 \\ 
41 & 16504+0228 & NGC 6240    &  253.245295  &   2.400926   &   0.024  & LINER &  -257.1  &  -380.8  &  -280.0  &  -484.1 \\ 
42 & 01572+0009 & Mrk 1014    &  29.959214   &   0.394615   &   0.163  & Sy1   &  -483.5  &  -425.6  &  -96.3  &  -536.9 \\ 
\hline	
\end{tabular}
\caption{The sample, their basic data, and the log-likelihood (ML) of the best fit resulting from fitting the CYGNUS, FR06, S15 and SKIRTOR combinations of models. The sample selection criteria are given in section 2.
}
\end{center}
\end{table*}

The HERschel Ultra Luminous Infrared Galaxy Survey (HERUS) consisted of observations with the Herschel Space Observatory \citep{pilb10} of the 42 most luminous ULIRGs in the local Universe in spectroscopy and photometry mode. \citet{far13} and \citet{spo13} discussed the diagnostics provided by PACS spectroscopy and \citet{pea16} discussed the corresponding diagnostics from SPIRE spectroscopy. \citet{efs14} presented a detailed study of IRAS~08572+3915 which exhibits one of the deepest silicate absorption features observed in a galaxy. \citet{cle18} presented the SPIRE photometry of the HERUS galaxies. 

As ULIRGs are deeply obscured by dust, radiative transfer models for both the starburst and AGN activity are needed for interperting their SEDs. \citet{mrr93b} presented the first such models for the local ULIRG population and fitted the SED of the prototypical ULIRG Arp~220. They showed that the SED of Arp~220 could be explained by a starburst model which was a factor of four more optically thick than the model for the starbursts in M~82 and NGC~1068 presented in the same paper. They also argued on the basis of fitting the IRAS colours that all ULIRGs in the IRAS Bright Galaxy sample \citep{san03} required a similar higher optical depth starburst model to explain the far-infrared emission. \citet{rig96} fitted the SEDs of the first ULIRGs to be detected in the submillimetre with the \citet{mrr93b} models and arrived at similar conclusions. \citet{far03} carried out for the first time starburst/AGN de-composition of the near to far-infared SEDs of a sample of 41 local ULIRGs using the starburst library of \citet{efstathiou00} and the AGN torus models of \citet{efstathiour95}. They concluded from this study that whereas most of the luminosity in local ULIRGs is due to starburst activity in about half of the sample there was significant emission from an AGN. \citet{vega08} fitted a sample of ULIRGs and Luminous Infrared Galaxies (LIRGs; $10^{12}L_\odot  > L_{IR} > 10^{11} L_\odot$) with the GRASIL models \citep{sil98} reaching a similar conclusion. \citet{efs14} de-composed the SED of the deep-silicate ULIRG IRAS~08572+3915 concluding that its emission is dominated by an obscured AGN with its torus viewed almost edge-on.

\begin{table*}
\centering
\caption{Parameters of the models used in this paper, symbols used, their assumed ranges and summary of other information about the models. The Fritz et al. model assumes two additional parameters that define the density distribution in the radial direction ($\beta$) and azimuthal direction ($\gamma$). In this paper we assume $\beta=0$ and $\gamma=4$. The SKIRTOR model assumes two additional parameters that define the density distribution in the radial direction ($p$) and azimuthal direction ($q$). In this paper we assume $p=1$ and $q=1$. In addition the SKIRTOR library we used assumes the fraction of mass inside clumps is 97\%. There are 4 additional scaling parameters for the starburst, spheroidal, AGN and polar dust models, $f_{SB}$,  $f_s$, $f_{AGN}$ and $f_{p}$ respectively.}
	\label{tab:example_table}
	\begin{tabular}{llll} 
		\hline
		Parameter &  Symbol & Range &  Comments\\
		\hline
                 &  &  & \\
{\bf CYGNUS Starburst}  &  &  & \\
                 &  &  \\
Initial optical depth of giant molecular clouds & $\tau_V$  &  50-250  &  \cite{efstathiou00}, \cite{efstathiou09} \\
Starburst star formation rate e-folding time       & $\tau_{*}$  & 10-30Myr  & Incorporates \citet{bru93,bru03}  \\
Starburst age      & $t_{*}$   &  5-35Myr &  metallicity=solar, Salpeter Initial Mass Function (IMF) \\
                  &            &  & Standard galactic dust mixture with PAHs \\
                   &           &         &    \\
{\bf CYGNUS Spheroidal Host}  &  &  &  \\
                 &  &  \\
Spheroidal star formation rate e-folding time      & $\tau^s$  &  0.125-8Gyr  & \cite{efs03}, \cite{efs21}  \\
Starlight intensity      & $\psi^s$ &  1-17 &  Incorporates \citet{bru93,bru03} \\ 
Optical depth     & $\tau_{v}^s$ & 0.1-15 &  metallicity=40\% of solar, Salpeter IMF\\ 
                  &            &  & Standard galactic dust mixture with PAHs \\
                  &            &  &  \\
{\bf CYGNUS AGN torus}  &  &    &  \\
                 &  &  &  \\
Torus equatorial UV optical depth   & $\tau_{uv}$  &  250-1450 &  Smooth tapered discs\\  
Torus ratio of outer to inner radius & $r_2/r_1$ &  20-100 & \cite{efstathiour95}, \cite{efstathiou13} \\   
Torus half-opening angle  & $\theta_o$  &  30-75\degr & Standard galactic dust mixture without PAHs\\ 
Torus inclination     & $\theta_i$  &  0-90\degr &  \\ 
                 &            & \\
{\bf Fritz AGN torus}  &  &  &   \\
                 &  &  & \\
Torus equatorial optical depth at 9.7$\mu m$  &  &  0.1-10 & Smooth flared discs \\  
Torus ratio of outer to inner radius &  &  10-150 &  \cite{fritz06}\\   
Torus half-opening angle  &   &  20-70\degr & Standard galactic dust mixture without PAHs\\ 
Torus inclination     &   &  0-90\degr &  \\   
                 &            & &  \\
{\bf SKIRTOR AGN torus}  &  &   &  \\
                 &  &  &  \\
Torus equatorial optical depth at 9.7$\mu m$  &   &  3-11 & Two-phase flared discs \\  
Torus ratio of outer to inner radius &  &  10-30 &  \cite{sta12}, \cite{stal16} \\   
Torus half-opening angle  &  &  20-70\degr &  Standard galactic dust mixture without PAHs\\ 
Torus inclination     &   &  0-90\degr &  \\ 
                 &            & &  \\
{\bf Siebenmorgen15 AGN torus}  &  &   &  \\
                 &  &  &  \\
Cloud volume filling factor (\%)   &  &  1.5-77.7  & Two-phase anisotropic spheres \\  
Optical depth of the individual clouds &   &  0-45 & \cite{sie15}\\
Optical depth of the disk mid-plane &   &  0-1000 &  Fluffy dust mixture without PAHs\\ 
     Inclination     &   &   0-90\degr & \\   
		\hline
	\end{tabular}
\end{table*}

\begin{table}
	\centering
\caption{Derived physical quantities and the symbol used in both papers A and B. The luminosities are integrated over 1-1000$\mu m$ except where indicated in the symbol that they are bolometric.}
	\label{tab:example_table}
	\begin{tabular}{ll} 
		\hline
		Physical Quantity &  Symbol \\
		\hline
                                       &                \\     
Observed AGN torus Luminosity          & $L_{AGN}^{o}$  \\  
Corrected AGN torus Luminosity         & $L_{AGN}^{c}$  \\
Polar dust AGN Luminosity              & $L_{p}$  \\  
Starburst Luminosity                   & $L_{Sb}$ \\   
Spheroidal host Luminosity                  & $L_{host}$   \\ 
Total observed Luminosity              & $L_{Tot}^{o}$    \\  
Total corrected Luminosity             & $L_{Tot}^{c}$    \\   
Starburst SFR (averaged over 50Myr)    & $\dot{M}_*^{50}$   \\
Starburst SFR (averaged over SB age)   & $\dot{M}_*^{age}$   \\
Spheroidal SFR                         & $\dot{M}_*^{Sph}$    \\
Total SFR                              & $\dot{M}_{tot}$    \\ 
Starburst Stellar Mass                 & $M^{*}_{sb}$  \\
Spheroidal Stellar Mass                & $M^{*}_{sph}$  \\ 
Total Stellar Mass                     & $M^{*}_{tot}$ \\   
AGN fraction                           & $F_{AGN}$  \\ 
Anisotropy correction factor           & $A$  \\
		\hline
	\end{tabular}
\end{table}

\begin{table*}
\label{tab:cyg:irlums}
\begin{center}
\begin{tabular}{cccccccc}
\hline
ID & L$_{AGN}^{o}$        & L$_{AGN}^{c}$        & L$_{Sb}$             & L$_{host}$           & L$_{Tot}^{o}$        & L$_{Tot}^{c}$        & A($\theta_{i}$) \\ 
   & \multicolumn{3}{c}{$10^{12}$L$_{\odot}$}                           & $10^{11}$L$_{\odot}$ & \multicolumn{2}{c}{$10^{12}$L$_{\odot}$}    & \\  
\hline	
1 & $ 0.28^{+0.03}_{-0.05}$  &  $ 0.83^{+0.14}_{-0.18}$  &  $ 1.86^{+0.03}_{-0.02}$  &  $ 1.06^{+0.12}_{-0.23}$  &  $ 2.25^{+0.04}_{-0.05}$  &  $ 2.80^{+0.14}_{-0.18}$  &  $ 2.95^{+0.17}_{-0.27 }$ \\ 
2 & $ 2.10^{+0.55}_{-0.31}$  &  $ 15.62^{+7.27}_{-3.51}$  &  $ 7.14^{+0.18}_{-0.35}$  &  $ 1.55^{+0.49}_{-0.49}$  &  $ 9.40^{+0.34}_{-0.24}$  &  $ 22.91^{+6.95}_{-3.37}$  &  $ 7.43^{+1.41}_{-0.95 }$ \\ 
3 & $ 0.81^{+0.22}_{-0.09}$  &  $ 1.47^{+0.17}_{-0.21}$  &  $ 0.77^{+0.02}_{-0.12}$  &  $ 0.23^{+0.05}_{-0.03}$  &  $ 1.60^{+0.12}_{-0.08}$  &  $ 2.26^{+0.12}_{-0.22}$  &  $ 1.81^{+0.03}_{-0.27 }$ \\ 
4 & $ 1.39^{+0.18}_{-0.23}$  &  $ 6.07^{+0.81}_{-1.27}$  &  $ 2.63^{+0.10}_{-0.06}$  &  $ 0.78^{+0.09}_{-0.10}$  &  $ 4.09^{+0.13}_{-0.15}$  &  $ 8.78^{+0.75}_{-1.17}$  &  $ 4.38^{+0.15}_{-0.27 }$ \\ 
5 & $ 0.26^{+0.04}_{-0.03}$  &  $ 0.95^{+0.20}_{-0.19}$  &  $ 2.16^{+0.03}_{-0.02}$  &  $ 0.11^{+0.02}_{-0.01}$  &  $ 2.42^{+0.03}_{-0.03}$  &  $ 3.12^{+0.18}_{-0.18}$  &  $ 3.70^{+0.52}_{-0.69 }$ \\ 
6 & $ 0.55^{+0.08}_{-0.06}$  &  $ 2.49^{+0.32}_{-0.30}$  &  $ 0.77^{+0.07}_{-0.08}$  &  $ 0.68^{+0.32}_{-0.20}$  &  $ 1.39^{+0.07}_{-0.05}$  &  $ 3.33^{+0.29}_{-0.25}$  &  $ 4.52^{+0.03}_{-0.27 }$ \\ 
7 & $ 0.21^{+0.09}_{-0.03}$  &  $ 0.38^{+0.86}_{-0.07}$  &  $ 1.21^{+0.03}_{-0.09}$  &  $ 1.82^{+0.16}_{-1.16}$  &  $ 1.60^{+0.03}_{-0.11}$  &  $ 1.78^{+0.71}_{-0.08}$  &  $ 1.82^{+2.40}_{-0.11 }$ \\ 
8 & $ 0.17^{+0.02}_{-0.02}$  &  $ 0.41^{+0.09}_{-0.06}$  &  $ 1.18^{+0.01}_{-0.03}$  &  $ 1.19^{+0.28}_{-0.12}$  &  $ 1.47^{+0.02}_{-0.03}$  &  $ 1.71^{+0.08}_{-0.07}$  &  $ 2.38^{+0.24}_{-0.15 }$ \\ 
9 & $ 1.98^{+0.17}_{-0.19}$  &  $ 1.17^{+0.10}_{-0.12}$  &  $ 2.18^{+0.01}_{-0.05}$  &  $ 0.04^{+0.03}_{-0.01}$  &  $ 4.16^{+0.16}_{-0.19}$  &  $ 3.35^{+0.09}_{-0.13}$  &  $ 0.59^{+0.00}_{-0.01 }$ \\ 
10 & $ 0.56^{+0.07}_{-0.08}$  &  $ 1.61^{+0.20}_{-0.21}$  &  $ 2.26^{+0.03}_{-0.05}$  &  $ 1.06^{+0.17}_{-0.09}$  &  $ 2.93^{+0.05}_{-0.08}$  &  $ 3.97^{+0.18}_{-0.19}$  &  $ 2.86^{+0.08}_{-0.02 }$ \\ 
11 & $ 0.78^{+0.15}_{-0.11}$  &  $ 8.77^{+3.02}_{-1.36}$  &  $ 0.66^{+0.05}_{-0.09}$  &  $ 0.19^{+0.09}_{-0.02}$  &  $ 1.47^{+0.08}_{-0.05}$  &  $ 9.45^{+2.95}_{-1.30}$  &  $ 11.19^{+2.57}_{-1.07 }$ \\ 
12 & $ 0.26^{+0.03}_{-0.02}$  &  $ 1.43^{+0.29}_{-0.19}$  &  $ 1.41^{+0.01}_{-0.02}$  &  $ 0.36^{+0.06}_{-0.05}$  &  $ 1.71^{+0.02}_{-0.02}$  &  $ 2.88^{+0.28}_{-0.19}$  &  $ 5.57^{+0.53}_{-0.26 }$ \\ 
13 & $ 0.83^{+0.12}_{-0.14}$  &  $ 10.72^{+5.02}_{-3.46}$  &  $ 0.81^{+0.12}_{-0.13}$  &  $ 0.90^{+0.43}_{-0.35}$  &  $ 1.73^{+0.04}_{-0.06}$  &  $ 11.63^{+4.93}_{-3.37}$  &  $ 12.91^{+4.01}_{-3.04 }$ \\ 
14 & $ 0.09^{+0.02}_{-0.01}$  &  $ 0.17^{+0.03}_{-0.01}$  &  $ 0.93^{+0.02}_{-0.01}$  &  $ 0.93^{+0.12}_{-0.13}$  &  $ 1.11^{+0.02}_{-0.00}$  &  $ 1.19^{+0.04}_{-0.01}$  &  $ 1.86^{+0.15}_{-0.06 }$ \\ 
15 & $ 0.94^{+0.12}_{-0.10}$  &  $ 12.66^{+4.71}_{-2.44}$  &  $ 0.59^{+0.08}_{-0.10}$  &  $ 0.19^{+0.04}_{-0.02}$  &  $ 1.55^{+0.03}_{-0.04}$  &  $ 13.27^{+4.62}_{-2.36}$  &  $ 13.41^{+3.52}_{-1.52 }$ \\ 
16 & $ 0.84^{+0.18}_{-0.10}$  &  $ 2.59^{+1.08}_{-0.28}$  &  $ 1.19^{+0.06}_{-0.17}$  &  $ 1.61^{+0.47}_{-0.35}$  &  $ 2.20^{+0.06}_{-0.09}$  &  $ 3.94^{+0.92}_{-0.25}$  &  $ 3.07^{+0.78}_{-0.21 }$ \\ 
17 & $ 0.10^{+0.02}_{-0.01}$  &  $ 0.22^{+0.12}_{-0.04}$  &  $ 1.52^{+0.02}_{-0.02}$  &  $ 1.39^{+0.18}_{-0.27}$  &  $ 1.76^{+0.01}_{-0.02}$  &  $ 1.88^{+0.11}_{-0.04}$  &  $ 2.23^{+0.68}_{-0.15 }$ \\ 
18 & $ 1.33^{+0.25}_{-0.22}$  &  $ 8.39^{+1.47}_{-1.42}$  &  $ 0.35^{+0.10}_{-0.12}$  &  $ 2.39^{+0.60}_{-0.56}$  &  $ 1.92^{+0.19}_{-0.17}$  &  $ 8.98^{+1.41}_{-1.36}$  &  $ 6.30^{+0.33}_{-0.15 }$ \\ 
19 & $ 0.39^{+0.04}_{-0.14}$  &  $ 2.59^{+0.25}_{-1.21}$  &  $ 1.33^{+0.12}_{-0.01}$  &  $ 2.68^{+0.11}_{-0.67}$  &  $ 1.99^{+0.02}_{-0.09}$  &  $ 4.19^{+0.23}_{-1.16}$  &  $ 6.61^{+0.02}_{-1.17 }$ \\ 
20 & $ 0.14^{+0.01}_{-0.01}$  &  $ 0.53^{+0.08}_{-0.07}$  &  $ 0.95^{+0.01}_{-0.01}$  &  $ 0.43^{+0.13}_{-0.11}$  &  $ 1.13^{+0.01}_{-0.01}$  &  $ 1.52^{+0.08}_{-0.07}$  &  $ 3.73^{+0.27}_{-0.30 }$ \\ 
21 & $ 0.37^{+0.04}_{-0.05}$  &  $ 2.59^{+0.46}_{-0.42}$  &  $ 0.66^{+0.03}_{-0.02}$  &  $ 0.25^{+0.02}_{-0.03}$  &  $ 1.06^{+0.03}_{-0.03}$  &  $ 3.28^{+0.45}_{-0.40}$  &  $ 6.98^{+0.62}_{-0.39 }$ \\ 
22 & $ 0.25^{+0.05}_{-0.04}$  &  $ 0.22^{+0.04}_{-0.03}$  &  $ 0.96^{+0.09}_{-0.04}$  &  $ 0.41^{+0.73}_{-0.14}$  &  $ 1.26^{+0.13}_{-0.06}$  &  $ 1.22^{+0.12}_{-0.06}$  &  $ 0.88^{+0.05}_{-0.03 }$ \\ 
23 & $ 0.26^{+0.07}_{-0.05}$  &  $ 0.68^{+0.24}_{-0.13}$  &  $ 2.69^{+0.09}_{-0.11}$  &  $ 0.64^{+0.24}_{-0.15}$  &  $ 3.02^{+0.06}_{-0.07}$  &  $ 3.44^{+0.16}_{-0.10}$  &  $ 2.58^{+0.25}_{-0.16 }$ \\ 
24 & $ 0.06^{+0.02}_{-0.01}$  &  $ 0.13^{+0.07}_{-0.02}$  &  $ 2.10^{+0.01}_{-0.02}$  &  $ 0.67^{+0.31}_{-0.11}$  &  $ 2.22^{+0.03}_{-0.01}$  &  $ 2.29^{+0.07}_{-0.03}$  &  $ 2.15^{+0.41}_{-0.06 }$ \\ 
25 & $ 0.28^{+0.20}_{-0.04}$  &  $ 1.00^{+0.50}_{-0.15}$  &  $ 0.71^{+0.04}_{-0.14}$  &  $ 3.24^{+0.83}_{-0.56}$  &  $ 1.32^{+0.14}_{-0.04}$  &  $ 2.04^{+0.43}_{-0.15}$  &  $ 3.61^{+0.06}_{-0.62 }$ \\ 
26 & $ 0.34^{+0.17}_{-0.05}$  &  $ 2.31^{+5.11}_{-0.53}$  &  $ 1.96^{+0.07}_{-0.13}$  &  $ 1.96^{+0.74}_{-0.51}$  &  $ 2.50^{+0.08}_{-0.04}$  &  $ 4.47^{+4.98}_{-0.51}$  &  $ 6.78^{+8.54}_{-1.48 }$ \\ 
27 & $ 0.10^{+0.03}_{-0.03}$  &  $ 0.25^{+0.13}_{-0.10}$  &  $ 2.52^{+0.07}_{-0.02}$  &  $ 0.97^{+0.25}_{-0.68}$  &  $ 2.71^{+0.07}_{-0.05}$  &  $ 2.86^{+0.11}_{-0.10}$  &  $ 2.55^{+0.68}_{-0.58 }$ \\ 
28 & $ 0.52^{+0.08}_{-0.10}$  &  $ 1.82^{+0.37}_{-0.57}$  &  $ 2.62^{+0.06}_{-0.09}$  &  $ 0.60^{+0.26}_{-0.11}$  &  $ 3.20^{+0.09}_{-0.13}$  &  $ 4.50^{+0.39}_{-0.59}$  &  $ 3.52^{+0.35}_{-0.53 }$ \\ 
29 & $ 0.15^{+0.03}_{-0.04}$  &  $ 0.88^{+0.21}_{-0.41}$  &  $ 1.25^{+0.02}_{-0.02}$  &  $ 0.65^{+0.37}_{-0.15}$  &  $ 1.46^{+0.03}_{-0.03}$  &  $ 2.20^{+0.21}_{-0.39}$  &  $ 6.03^{+1.38}_{-1.72 }$ \\ 
30 & $ 0.31^{+0.06}_{-0.03}$  &  $ 1.20^{+0.27}_{-0.13}$  &  $ 0.62^{+0.02}_{-0.03}$  &  $ 0.85^{+0.24}_{-0.25}$  &  $ 1.02^{+0.05}_{-0.03}$  &  $ 1.91^{+0.26}_{-0.13}$  &  $ 3.86^{+0.15}_{-0.07 }$ \\ 
31 & $ 0.36^{+0.03}_{-0.06}$  &  $ 3.34^{+0.62}_{-0.76}$  &  $ 0.73^{+0.04}_{-0.02}$  &  $ 0.81^{+0.17}_{-0.24}$  &  $ 1.17^{+0.02}_{-0.03}$  &  $ 4.16^{+0.61}_{-0.75}$  &  $ 9.42^{+2.13}_{-2.03 }$ \\ 
32 & $ 0.13^{+0.01}_{-0.02}$  &  $ 0.29^{+0.03}_{-0.05}$  &  $ 0.56^{+0.01}_{-0.01}$  &  $ 0.44^{+0.06}_{-0.09}$  &  $ 0.74^{+0.02}_{-0.03}$  &  $ 0.89^{+0.03}_{-0.05}$  &  $ 2.19^{+0.08}_{-0.13 }$ \\ 
33 & $ 0.34^{+0.05}_{-0.04}$  &  $ 1.25^{+0.30}_{-0.12}$  &  $ 1.08^{+0.01}_{-0.03}$  &  $ 0.23^{+0.05}_{-0.04}$  &  $ 1.44^{+0.03}_{-0.04}$  &  $ 2.35^{+0.28}_{-0.12}$  &  $ 3.72^{+0.42}_{-0.01 }$ \\ 
34 & $ 0.57^{+0.12}_{-0.17}$  &  $ 4.08^{+1.76}_{-1.94}$  &  $ 1.12^{+0.14}_{-0.08}$  &  $ 3.43^{+0.74}_{-0.86}$  &  $ 2.04^{+0.10}_{-0.11}$  &  $ 5.54^{+1.74}_{-1.88}$  &  $ 7.11^{+1.80}_{-1.88 }$ \\ 
35 & $ 0.26^{+0.07}_{-0.03}$  &  $ 1.17^{+0.73}_{-0.28}$  &  $ 1.14^{+0.01}_{-0.04}$  &  $ 0.38^{+0.05}_{-0.03}$  &  $ 1.44^{+0.04}_{-0.03}$  &  $ 2.35^{+0.70}_{-0.27}$  &  $ 4.42^{+1.50}_{-0.65 }$ \\ 
36 & $ 0.09^{+0.02}_{-0.01}$  &  $ 0.24^{+0.07}_{-0.05}$  &  $ 0.85^{+0.03}_{-0.01}$  &  $ 0.94^{+0.13}_{-0.09}$  &  $ 1.03^{+0.03}_{-0.01}$  &  $ 1.18^{+0.08}_{-0.05}$  &  $ 2.68^{+0.51}_{-0.47 }$ \\ 
37 & $ 1.46^{+0.17}_{-0.13}$  &  $ 3.49^{+0.46}_{-0.35}$  &  $ 1.89^{+0.14}_{-0.12}$  &  $ 1.04^{+0.14}_{-0.10}$  &  $ 3.46^{+0.21}_{-0.14}$  &  $ 5.48^{+0.46}_{-0.29}$  &  $ 2.38^{+0.08}_{-0.07 }$ \\ 
38 & $ 0.17^{+0.04}_{-0.03}$  &  $ 0.56^{+0.36}_{-0.19}$  &  $ 1.09^{+0.03}_{-0.02}$  &  $ 0.36^{+0.12}_{-0.04}$  &  $ 1.30^{+0.04}_{-0.03}$  &  $ 1.69^{+0.37}_{-0.19}$  &  $ 3.23^{+1.11}_{-0.71 }$ \\ 
39 & $ 0.50^{+0.06}_{-0.04}$  &  $ 0.41^{+0.07}_{-0.03}$  &  $ 0.23^{+0.02}_{-0.02}$  &  $ 0.39^{+0.10}_{-0.14}$  &  $ 0.76^{+0.06}_{-0.05}$  &  $ 0.68^{+0.07}_{-0.04}$  &  $ 0.82^{+0.05}_{-0.02 }$ \\ 
40 & $ 0.12^{+0.08}_{-0.01}$  &  $ 0.68^{+0.89}_{-0.09}$  &  $ 0.90^{+0.01}_{-0.03}$  &  $ 0.62^{+0.13}_{-0.04}$  &  $ 1.08^{+0.07}_{-0.01}$  &  $ 1.64^{+0.88}_{-0.09}$  &  $ 5.85^{+2.32}_{-0.43 }$ \\ 
41 & $ 0.12^{+0.02}_{-0.03}$  &  $ 0.36^{+0.04}_{-0.13}$  &  $ 0.42^{+0.04}_{-0.03}$  &  $ 1.39^{+0.44}_{-0.56}$  &  $ 0.68^{+0.03}_{-0.03}$  &  $ 0.91^{+0.05}_{-0.14}$  &  $ 2.90^{+0.01}_{-0.61 }$ \\ 
42 & $ 0.80^{+0.11}_{-0.09}$  &  $ 0.42^{+0.07}_{-0.06}$  &  $ 2.44^{+0.08}_{-0.05}$  &  $ 1.85^{+0.36}_{-0.59}$  &  $ 3.42^{+0.11}_{-0.09}$  &  $ 3.04^{+0.07}_{-0.06}$  &  $ 0.53^{+0.03}_{-0.03 }$ \\ 
\hline	
\end{tabular}
\caption{The AGN, starburst, spheroid, and total infrared luminosities (integrated over $1-1000\mu m$), derived from the CYGNUS model fits. For the AGN and total luminosities both the observed and anisotropy corrected luminosities are given, as well as the anisotropy factor $A(\theta_i)$ (see equation 1) used to perform the correction. The equivalent bolometric luminosities are given in Table B1 in the Appendix. The equivalent luminosities from the FR06, SKIRTOR and Siebenmorgen15 model fits are given in Tables  B2, B3 and B4 in the Appendix.
}
\end{center}
\end{table*}

This is the first of two papers which present a new look at local ULIRGs which is based on detailed models of their SEDs from the ultraviolet to the submillimetre. In this paper (Paper A) we assemble the SEDs of the 42 local ULIRGs that constitute the HERUS sample and fit them with multi-component radiative transfer models. In paper B (Farrah et al. 2021) we present a detailed analysis of the results. 

Our approach has three novel features compared to previous studies. The first novelty of this work is that we fit the ultraviolet to submillimetre SEDs of ULIRGs exclusively with radiative transfer models. The models constitute three libraries which describe the starburst, AGN and host galaxy components. These three libraries are part of the collection of radiative transfer models named CYprus models for Galaxies and their NUclear Spectra (CYGNUS)\footnote{The models are publicly available at https://arc.euc.ac.cy/cygnus/}. We have used extensively the starburst and AGN libraries for almost three decades  to fit the SEDs of a broad range of galaxies \citep{mrr93,mrr93b,efstathiour95,efs05,efstathiou06,efstathiou13,efs21,mrr97,hug98,ale99,mrr00,ruiz01,alon01,alon03,ver02,far02,far03,far12,far17,matt12,matt18,lon15,herr17,pitch19}. To these two well-tested libraries we added a third library of `spheroidal' models which represent the host galaxy in which the starburst and AGN reside \citep{efs21}. This library which is an evolution of the `cirrus' models of \citet{efs03} self-consistently takes into account the absorption of starlight and reemission by interstellar dust in a spheroidal geometry. As we discuss below our approach to use a library for the spheroidal component instead of running the model `on the fly' during the fitting (e.g. as with GRASIL, see also discussion by \citealt{john13}) speeds up considerably the fitting.

\begin{table*}
\label{tab:cyg:der}
\begin{center}
\begin{tabular}{lcccccccccc}
\hline
ID & $\theta_{i}$ & $\theta_{o}$ & $\tau_{uv}$ & $r_{2}/r_{1}$ & M$_{*}$              & $\dot{M}^{50}_{*}$    & $\dot{M}^{age}_{*}$  & $\dot{M}^{Sph}_{*}$  & Age        & $\tau_{V}$ \\ 
   & \multicolumn{2}{c}{$\degr$} &             &               & $10^{11}$M$_{\odot}$ & \multicolumn{3}{c}{M$_{\odot}$yr$^{-1}$}                            & $10^{7}$yr &            \\ 	
\hline
 1 & $79.1^{+2.3}_{-0.4}$ & $47.0^{+1.4}_{-1.3}$ & $389.5^{+18.8}_{-39.3}$ & $67.4^{+12.8}_{-8.0}$ & $ 2.2^{+0.3}_{-0.5}$ & $352^{+30}_{-69}$ & $598^{+58}_{-126}$ & $2^{+1}_{-0}$ & $ 2.9^{+0.1}_{-0.2}$ & $188.9^{+7.1}_{-8.8}$   \\ 
 2 & $86.4^{+0.6}_{-1.2}$ & $73.5^{+0.0}_{-0.4}$ & $300.4^{+44.5}_{-26.5}$ & $94.7^{+4.3}_{-12.8}$ & $ 1.2^{+0.3}_{-0.3}$ & $228^{+71}_{-11}$ & $2192^{+1132}_{-120}$ & $11^{+3}_{-3}$ & $ 0.5^{+0.2}_{-0.0}$ & $93.8^{+6.9}_{-3.4}$   \\ 
 3 & $57.1^{+1.9}_{-0.4}$ & $36.6^{+1.4}_{-0.6}$ & $1489.9^{+0.1}_{-71.1}$ & $26.4^{+71.6}_{-3.3}$ & $ 0.3^{+0.1}_{-0.0}$ & $75^{+16}_{-18}$ & $195^{+75}_{-60}$ & $2^{+0}_{-0}$ & $ 1.9^{+0.6}_{-0.4}$ & $51.5^{+9.0}_{-0.5}$   \\ 
 4 & $78.9^{+1.0}_{-0.0}$ & $37.9^{+1.2}_{-1.9}$ & $1070.5^{+70.2}_{-72.5}$& $78.8^{+2.9}_{-3.5}$  & $ 2.6^{+0.2}_{-0.4}$ & $330^{+77}_{-43}$ & $660^{+172}_{-118}$ & $0^{+0}_{-0}$ & $ 2.5^{+0.3}_{-0.3}$ & $244.9^{+4.0}_{-15.5}$   \\ 
 5 & $67.0^{+0.2}_{-0.1}$ & $44.3^{+3.4}_{-6.0}$ & $1488.1^{+1.9}_{-32.5}$ & $23.1^{+5.6}_{-2.1}$  & $ 0.7^{+0.2}_{-0.1}$ & $556^{+3}_{-27}$ & $798^{+5}_{-40}$ & $0^{+0}_{-0}$ & $ 3.5^{+0.0}_{-0.0}$ & $233.6^{+15.4}_{-19.8}$   \\ 
 6 & $68.4^{+0.6}_{-1.8}$ & $60.0^{+0.4}_{-1.6}$ & $1487.3^{+2.4}_{-177.2}$& $60.3^{+8.3}_{-10.0}$ & $ 1.5^{+0.3}_{-0.2}$ & $68^{+21}_{-9}$ & $170^{+57}_{-30}$ & $0^{+1}_{-0}$ & $ 2.0^{+0.3}_{-0.2}$ & $119.8^{+28.1}_{-19.1}$   \\ 
 7 & $72.3^{+1.6}_{-0.8}$ & $37.5^{+25.5}_{-1.5}$& $405.5^{+61.7}_{-41.1}$ & $95.4^{+3.5}_{-68.5}$ & $ 0.9^{+0.1}_{-0.3}$ & $45^{+177}_{-7}$ & $315^{+1733}_{-104}$ & $11^{+0}_{-4}$ & $ 0.7^{+2.8}_{-0.2}$ & $143.7^{+55.0}_{-4.2}$   \\ 
 8 & $71.5^{+0.0}_{-2.4}$ & $41.4^{+1.4}_{-3.8}$ & $756.7^{+120.1}_{-24.2}$& $70.8^{+1.4}_{-5.1}$  & $ 1.6^{+0.5}_{-0.4}$ & $228^{+57}_{-5}$ & $378^{+112}_{-8}$ & $3^{+2}_{-1}$ & $ 3.0^{+0.5}_{-0.0}$ & $231.3^{+17.7}_{-0.1}$   \\ 
 9 & $24.7^{+0.3}_{-1.2}$ & $56.4^{+2.0}_{-1.5}$ & $321.6^{+41.7}_{-18.9}$ & $90.3^{+2.0}_{-2.4}$  & $ 0.3^{+0.1}_{-0.0}$ & $356^{+27}_{-17}$ & $520^{+40}_{-27}$ & $0^{+0}_{-0}$ & $ 3.4^{+0.0}_{-0.1}$ & $52.1^{+5.6}_{-1.1}$   \\ 
10 & $61.7^{+1.8}_{-0.1}$ & $51.0^{+1.8}_{-0.7}$ & $1292.9^{+6.4}_{-27.6}$ & $26.7^{+0.4}_{-3.5}$  & $ 1.2^{+0.1}_{-0.1}$ & $422^{+3}_{-69}$ & $683^{+7}_{-122}$ & $11^{+1}_{-0}$ & $ 3.1^{+0.0}_{-0.2}$ & $57.8^{+6.0}_{-0.7}$   \\ 
11 & $85.5^{+1.4}_{-2.6}$ & $69.3^{+2.4}_{-1.1}$ & $519.1^{+37.7}_{-20.5}$ & $92.0^{+2.2}_{-5.5}$  & $ 0.2^{+0.0}_{-0.0}$ & $22^{+5}_{-3}$ & $206^{+76}_{-33}$ & $1^{+0}_{-0}$ & $ 0.5^{+0.2}_{-0.0}$ & $84.3^{+13.7}_{-10.8}$   \\ 
12 & $75.0^{+0.2}_{-0.3}$ & $73.3^{+0.2}_{-0.4}$ & $1111.8^{+0.6}_{-1.5}$  & $25.8^{+1.1}_{-1.3}$  & $ 1.4^{+0.2}_{-0.2}$ & $333^{+14}_{-9}$ & $482^{+20}_{-13}$ & $0^{+0}_{-0}$ & $ 3.5^{+0.0}_{-0.0}$ & $111.6^{+3.0}_{-0.6}$   \\ 
13 & $78.2^{+0.8}_{-1.7}$ & $60.1^{+2.5}_{-6.1}$ & $1488.0^{+2.0}_{-23.3}$ & $42.4^{+1.9}_{-3.1}$ & $ 0.9^{+0.4}_{-0.1}$ & $122^{+62}_{-55}$ & $245^{+134}_{-132}$ & $5^{+2}_{-3}$ & $ 2.5^{+0.5}_{-0.7}$ & $90.4^{+12.5}_{-9.0}$   \\ 
14 & $64.1^{+0.0}_{-0.3}$ & $39.0^{+2.6}_{-0.6}$ & $806.5^{+149.3}_{-125.2}$ & $36.8^{+3.6}_{-2.2}$ & $ 0.6^{+0.1}_{-0.0}$ & $220^{+12}_{-14}$ & $322^{+18}_{-23}$ & $5^{+0}_{-1}$ & $ 3.4^{+0.0}_{-0.1}$ & $169.8^{+2.7}_{-5.2}$   \\ 
15 & $77.3^{+1.6}_{-0.8}$ & $65.4^{+3.0}_{-1.6}$ & $989.2^{+52.5}_{-33.5}$ & $44.5^{+2.2}_{-4.4}$ & $ 0.2^{+0.0}_{-0.0}$ & $104^{+9}_{-21}$ & $151^{+14}_{-33}$ & $1^{+0}_{-0}$ & $ 3.4^{+0.1}_{-0.3}$ & $135.0^{+2.9}_{-5.6}$   \\ 
16 & $69.0^{+0.9}_{-0.2}$ & $38.9^{+3.4}_{-1.7}$ & $1480.5^{+8.3}_{-74.2}$ & $25.4^{+11.6}_{-3.0}$ & $ 1.0^{+0.2}_{-0.2}$ & $189^{+38}_{-121}$ & $298^{+66}_{-260}$ & $10^{+3}_{-1}$ & $ 3.2^{+0.3}_{-1.9}$ & $56.8^{+4.9}_{-1.4}$   \\ 
17 & $70.1^{+0.3}_{-0.2}$ & $38.5^{+6.6}_{-2.4}$ & $786.0^{+309.0}_{-57.5}$& $76.5^{+22.4}_{-12.2}$ & $ 0.8^{+0.1}_{-0.1}$ & $383^{+6}_{-8}$ & $549^{+9}_{-12}$ & $8^{+1}_{-1}$ & $ 3.5^{+0.0}_{-0.0}$ & $243.5^{+0.9}_{-0.9}$   \\ 
18 & $71.2^{+0.3}_{-1.7}$ & $60.0^{+0.5}_{-1.9}$ & $1488.3^{+1.7}_{-45.3}$ & $40.3^{+5.4}_{-4.9}$ & $ 2.5^{+1.8}_{-0.6}$ & $50^{+17}_{-25}$ & $87^{+34}_{-54}$ & $12^{+3}_{-6}$ & $ 2.9^{+0.6}_{-1.1}$ & $241.2^{+7.8}_{-74.4}$   \\ 
19 & $76.8^{+0.0}_{-2.0}$ & $45.8^{+0.1}_{-0.9}$ & $1455.9^{+14.8}_{-42.2}$& $47.4^{+0.2}_{-1.3}$ & $ 1.4^{+0.0}_{-0.3}$ & $240^{+25}_{-4}$ & $378^{+40}_{-7}$ & $16^{+0}_{-3}$ & $ 3.2^{+0.1}_{-0.0}$ & $225.9^{+3.5}_{-4.1}$   \\ 
20 & $71.2^{+0.3}_{-2.0}$ & $48.9^{+2.4}_{-3.1}$ & $1160.1^{+98.3}_{-44.2}$& $80.9^{+4.4}_{-3.6}$ & $ 0.4^{+0.1}_{-0.1}$ & $242^{+4}_{-26}$ & $347^{+6}_{-38}$ & $2^{+0}_{-0}$ & $ 3.5^{+0.0}_{-0.0}$ & $238.3^{+10.7}_{-6.7}$   \\ 
21 & $76.6^{+1.6}_{-0.2}$ & $56.5^{+1.3}_{-0.6}$ & $920.3^{+54.6}_{-42.9}$ & $31.0^{+2.6}_{-4.9}$ & $ 0.3^{+0.0}_{-0.0}$ & $90^{+13}_{-9}$ & $160^{+23}_{-19}$ & $2^{+0}_{-0}$ & $ 2.8^{+0.1}_{-0.2}$ & $161.5^{+12.5}_{-14.9}$   \\ 
22 & $50.3^{+1.5}_{-1.0}$ & $45.0^{+1.4}_{-1.5}$ & $577.1^{+51.8}_{-301.8}$& $29.3^{+18.8}_{-8.3}$ & $ 0.4^{+0.3}_{-0.1}$ & $128^{+42}_{-18}$ & $255^{+92}_{-49}$ & $3^{+3}_{-0}$ & $ 2.5^{+0.3}_{-0.3}$ & $75.1^{+18.1}_{-24.1}$   \\ 
23 & $80.4^{+1.3}_{-0.6}$ & $38.8^{+0.3}_{-1.0}$ & $517.7^{+39.0}_{-33.0}$ & $37.4^{+14.6}_{-4.1}$ & $ 0.9^{+0.1}_{-0.1}$ & $595^{+35}_{-62}$ & $855^{+50}_{-91}$ & $4^{+1}_{-0}$ & $ 3.5^{+0.0}_{-0.1}$ & $143.7^{+21.2}_{-13.5}$   \\ 
24 & $74.0^{+0.1}_{-1.0}$ & $36.1^{+5.2}_{-0.1}$ & $640.6^{+88.7}_{-54.9}$ & $78.5^{+8.2}_{-13.3}$ & $ 0.6^{+0.2}_{-0.1}$ & $536^{+6}_{-90}$ & $768^{+9}_{-129}$ & $3^{+1}_{-0}$ & $ 3.5^{+0.0}_{-0.0}$ & $248.9^{+0.1}_{-8.5}$   \\ 
25 & $68.9^{+2.2}_{-1.3}$ & $48.4^{+1.6}_{-0.9}$ & $1448.0^{+37.3}_{-805.2}$ & $77.4^{+0.3}_{-51.2}$ & $ 6.3^{+2.5}_{-1.3}$ & $22^{+34}_{-1}$ & $218^{+625}_{-16}$ & $2^{+0}_{-0}$ & $ 0.5^{+1.2}_{-0.0}$ & $51.3^{+34.2}_{-0.3}$   \\ 
26 & $73.7^{+2.8}_{-2.1}$ & $57.0^{+10.5}_{-5.2}$& $1343.2^{+146.0}_{-156.2}$ & $80.1^{+18.8}_{-16.2}$ & $ 1.2^{+0.3}_{-0.2}$ & $435^{+82}_{-75}$ & $625^{+118}_{-108}$ & $12^{+4}_{-3}$ & $ 3.5^{+0.0}_{-0.1}$ & $234.0^{+15.0}_{-39.1}$   \\ 
27 & $78.9^{+1.0}_{-0.2}$ & $48.1^{+5.9}_{-9.2}$ & $352.9^{+59.0}_{-101.8}$ & $94.5^{+4.4}_{-37.2}$ & $ 1.8^{+0.4}_{-0.6}$ & $644^{+22}_{-104}$ & $925^{+32}_{-149}$ & $2^{+2}_{-2}$ & $ 3.5^{+0.0}_{-0.0}$ & $197.5^{+17.6}_{-25.8}$   \\ 
28 & $81.2^{+0.2}_{-0.6}$ & $44.9^{+1.0}_{-8.9}$ & $623.6^{+105.0}_{-55.9}$ & $49.9^{+31.9}_{-7.3}$ & $ 0.9^{+0.5}_{-0.1}$ & $464^{+75}_{-178}$ & $772^{+137}_{-379}$ & $5^{+1}_{-1}$ & $ 3.0^{+0.2}_{-0.9}$ & $233.0^{+15.8}_{-34.2}$   \\ 
29 & $73.9^{+0.0}_{-0.4}$ & $54.0^{+5.3}_{-7.6}$ & $1196.7^{+285.8}_{-159.2}$ & $42.2^{+17.3}_{-15.6}$ & $ 0.5^{+0.1}_{-0.1}$ & $321^{+2}_{-47}$ & $460^{+3}_{-69}$ & $4^{+1}_{-0}$ & $ 3.5^{+0.0}_{-0.1}$ & $248.7^{+0.3}_{-8.1}$   \\ 
30 & $80.3^{+1.0}_{-1.4}$ & $41.1^{+1.9}_{-0.3}$ & $766.1^{+11.5}_{-23.2}$ & $75.1^{+5.1}_{-1.6}$ & $ 0.5^{+0.1}_{-0.1}$ & $55^{+26}_{-9}$ & $150^{+89}_{-34}$ & $6^{+1}_{-1}$ & $ 1.8^{+0.6}_{-0.3}$ & $231.7^{+12.9}_{-17.2}$   \\ 
31 & $76.4^{+0.1}_{-0.3}$ & $55.8^{+3.8}_{-4.3}$ & $1489.3^{+0.7}_{-64.0}$ & $25.7^{+2.6}_{-4.7}$ & $ 0.6^{+0.1}_{-0.1}$ & $164^{+8}_{-12}$ & $235^{+12}_{-18}$ & $6^{+0}_{-1}$ & $ 3.5^{+0.0}_{-0.1}$ & $226.8^{+12.1}_{-15.4}$   \\ 
32 & $61.0^{+0.7}_{-1.8}$ & $38.6^{+2.1}_{-1.0}$ & $1489.1^{+0.9}_{-174.4}$ & $21.0^{+6.3}_{-0.0}$ & $ 0.5^{+0.0}_{-0.1}$ & $114^{+12}_{-18}$ & $165^{+17}_{-29}$ & $4^{+0}_{-1}$ & $ 3.5^{+0.0}_{-0.3}$ & $117.6^{+10.7}_{-6.0}$   \\ 
33 & $74.0^{+1.3}_{-0.0}$ & $36.8^{+2.1}_{-0.4}$ & $1362.1^{+91.4}_{-92.7}$ & $34.8^{+5.5}_{-5.9}$ & $ 0.5^{+0.1}_{-0.1}$ & $254^{+8}_{-30}$ & $364^{+11}_{-44}$ & $1^{+0}_{-0}$ & $ 3.5^{+0.0}_{-0.1}$ & $125.4^{+6.4}_{-4.7}$   \\ 
34 & $75.8^{+0.7}_{-1.8}$ & $50.1^{+6.5}_{-5.9}$ & $1436.6^{+43.0}_{-76.1}$ & $40.3^{+7.4}_{-1.8}$ & $ 1.8^{+0.4}_{-0.4}$ & $137^{+60}_{-42}$ & $275^{+154}_{-105}$ & $20^{+4}_{-6}$ & $ 2.5^{+0.9}_{-0.6}$ & $243.6^{+5.4}_{-37.6}$   \\ 
35 & $74.9^{+1.5}_{-0.7}$ & $41.7^{+5.7}_{-3.8}$ & $1486.1^{+3.9}_{-65.2}$ & $23.3^{+9.2}_{-2.3}$ & $ 1.4^{+0.2}_{-0.1}$ & $223^{+13}_{-41}$ & $375^{+27}_{-78}$ & $0^{+0}_{-0}$ & $ 3.0^{+0.1}_{-0.3}$ & $244.1^{+4.9}_{-12.5}$   \\ 
36 & $79.3^{+5.2}_{-2.9}$ & $48.7^{+7.6}_{-8.1}$ & $328.3^{+42.8}_{-37.3}$ & $41.1^{+9.5}_{-12.0}$ & $ 1.9^{+0.3}_{-0.2}$ & $176^{+12}_{-18}$ & $254^{+18}_{-27}$ & $1^{+0}_{-0}$ & $ 3.5^{+0.0}_{-0.1}$ & $245.3^{+3.7}_{-51.9}$   \\ 
37 & $64.6^{+2.0}_{-0.5}$ & $55.2^{+1.5}_{-1.4}$ & $791.0^{+94.2}_{-77.1}$ & $89.4^{+6.7}_{-8.5}$ & $ 1.6^{+1.2}_{-0.4}$ & $309^{+53}_{-65}$ & $518^{+111}_{-131}$ & $7^{+3}_{-4}$ & $ 3.0^{+0.4}_{-0.4}$ & $161.6^{+52.1}_{-33.7}$   \\ 
38 & $73.8^{+0.2}_{-1.9}$ & $44.5^{+7.9}_{-8.5}$ & $844.4^{+279.9}_{-91.4}$ & $46.9^{+5.7}_{-12.0}$ & $ 1.3^{+0.1}_{-0.1}$ & $259^{+19}_{-49}$ & $373^{+28}_{-72}$ & $0^{+0}_{-0}$ & $ 3.5^{+0.0}_{-0.1}$ & $188.4^{+15.0}_{-16.2}$   \\ 
39 & $51.7^{+0.1}_{-1.3}$ & $47.2^{+0.3}_{-1.4}$ & $514.8^{+212.0}_{-92.3}$ & $27.3^{+2.5}_{-6.2}$ & $ 0.4^{+0.4}_{-0.1}$ & $30^{+10}_{-3}$ & $54^{+23}_{-6}$ & $4^{+1}_{-3}$ & $ 2.8^{+0.7}_{-0.0}$ & $52.9^{+19.0}_{-1.9}$   \\ 
40 & $80.2^{+1.2}_{-0.3}$ & $45.9^{+5.5}_{-2.0}$ & $1449.6^{+40.4}_{-39.7}$ & $22.2^{+4.6}_{-1.2}$ & $ 1.3^{+0.3}_{-0.1}$ & $140^{+38}_{-14}$ & $274^{+88}_{-29}$ & $0^{+0}_{-0}$ & $ 2.6^{+0.4}_{-0.1}$ & $248.9^{+0.1}_{-2.6}$   \\ 
41 & $73.9^{+0.1}_{-3.1}$ & $40.5^{+1.1}_{-4.4}$ & $895.4^{+144.1}_{-153.9}$ & $38.1^{+16.2}_{-13.9}$ & $ 0.9^{+0.2}_{-0.2}$ & $71^{+43}_{-24}$ & $107^{+65}_{-44}$ & $7^{+2}_{-3}$ & $ 3.3^{+0.2}_{-0.7}$ & $165.7^{+15.6}_{-13.9}$   \\ 
42 & $ 6.2^{+3.9}_{-1.2}$ & $45.0^{+3.6}_{-8.8}$ & $299.4^{+113.8}_{-48.3}$ & $61.8^{+37.1}_{-7.4}$ & $ 1.9^{+0.4}_{-0.5}$ & $348^{+31}_{-66}$ & $711^{+77}_{-177}$ & $23^{+5}_{-7}$ & $ 2.5^{+0.1}_{-0.4}$ & $51.1^{+4.5}_{-0.1}$   \\ 
\hline	
\end{tabular}
\caption{Fitted parameters and derived physical properties from the CYGNUS models.  $\theta_{o}$ is the half-opening angle of the torus, $\theta_{i} $ is the inclination of the torus,  $r_2/r_1$ is the ratio of outer to inner disc radius  and $\tau_{uv} $ is the torus equatorial optical depth at 1000\AA. M$_{*}$ is the total stellar mass, $\dot{M}^{50}_{*}$ is the star formation rate averaged over 50Myr, $\dot{M}^{age}_{*}$ is the star formation rate averaged over the age of the starburst episode, $\dot{M}^{Sph}_{*}$ is the star formation rate of the spheroidal component, $t_*$ and $\tau_{V}$ are the age of the starburst episode and the initial optical depth of the giant molecular clouds (GMCs) that constitute it.
}
\end{center}
\end{table*}

The second novelty of our approach is that we feed the three libraries into the Markov Chain Monte Carlo (MCMC) code SATMC \citep{john13} to get the best fit.  The output of SATMC is then post-processed by our own routines to get the luminosities of all components, star formation rates (SFRs) and stellar masses for the starburst and spheroid separately, AGN fraction, Core-collapse supernova rate and their errors.  The fit of a single galaxy takes about 10-15 minutes. This is comparable to the time needed by popular energy balance methods such as MAGPHYS \citep{dac08} and CIGALE \citep{noll09,bpq19}. The advantage of our method is that the model is more physically motivated and versatile as additional components (e.g. polar dust, a second AGN or a second starburst) can very easily be added to the model. We give examples of this in the paper.  We therefore foresee numerous applications of this method for the analysis of the rich datasets of ULIRGs and LIRGs at all redshifts which have been accumulated by multi-wavelength surveys such as H-ATLAS \citep{eal10} and  HerMES \citep{oli12}, projects such as the Herschel Extragalactic Legacy Project  (HELP) \citep{shi19}, and projects which will be made possible in the near future with facilities such as ALMA, JWST and Euclid.

\begin{figure*}
\centering
\includegraphics[width=16cm]{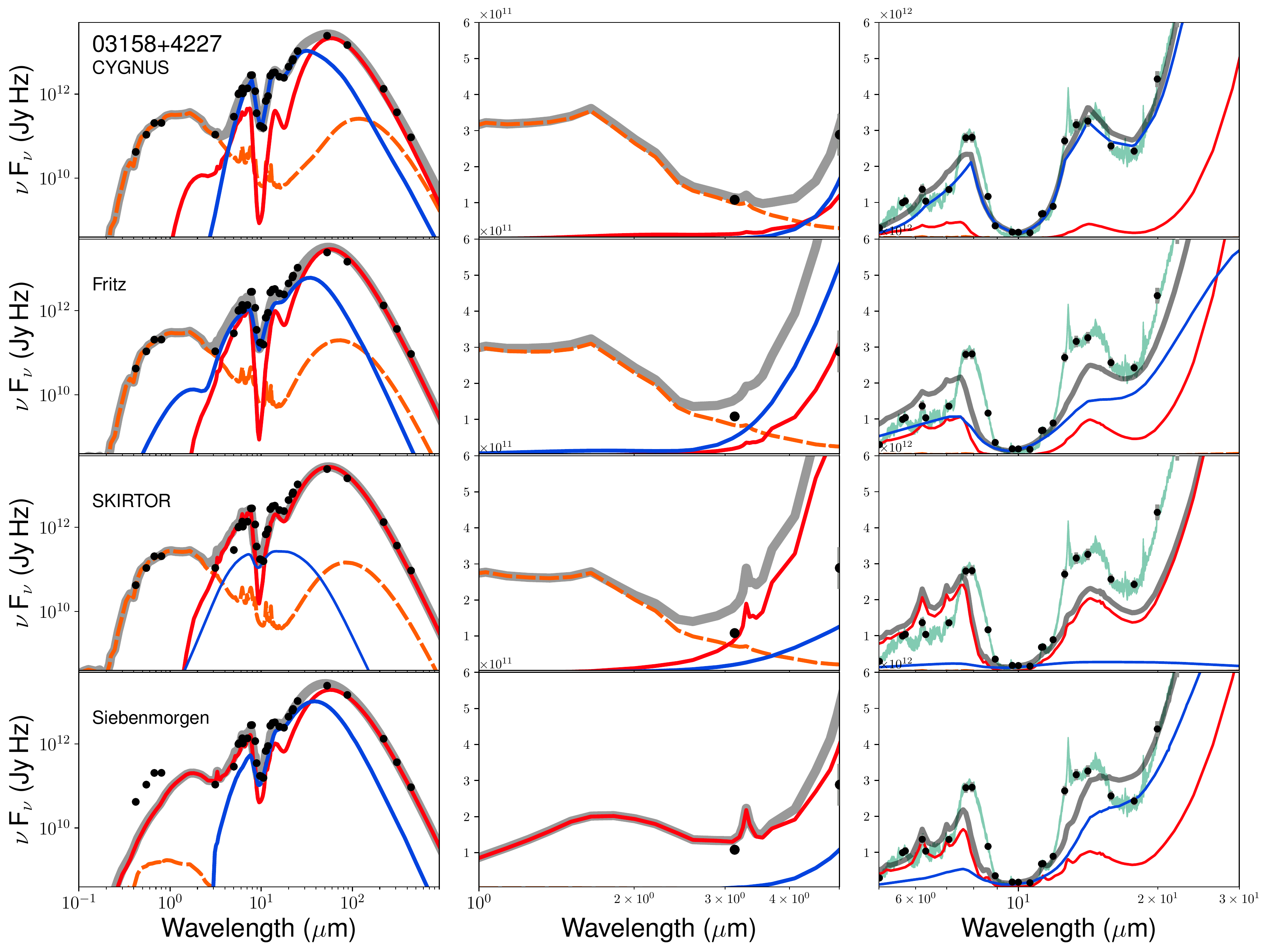}
		\caption{{ Comparison SED fit plots for IRAS~03158+4227. The top row shows fits with the CYGNUS combination of models: spheroidal host (orange), starburst (red), AGN torus (blue) and total (grey). The left panel shows the fit over the 0.1-1000$\mu m$ range, the middle panel the near-infrared range, and the right panel the 5-30$\mu m$ range. The second row shows fits with the CYGNUS AGN model replaced by the FR06 model, the third row replaces the CYGNUS AGN model with the SKIRTOR model, while the bottom row replaces the CYGNUS AGN model with the Siebenmorgen15 model. The full IRS data are plotted with green.}}
\label{fig:sedsrep1}
\end{figure*}

\begin{figure*}
\centering
\includegraphics[width=16cm]{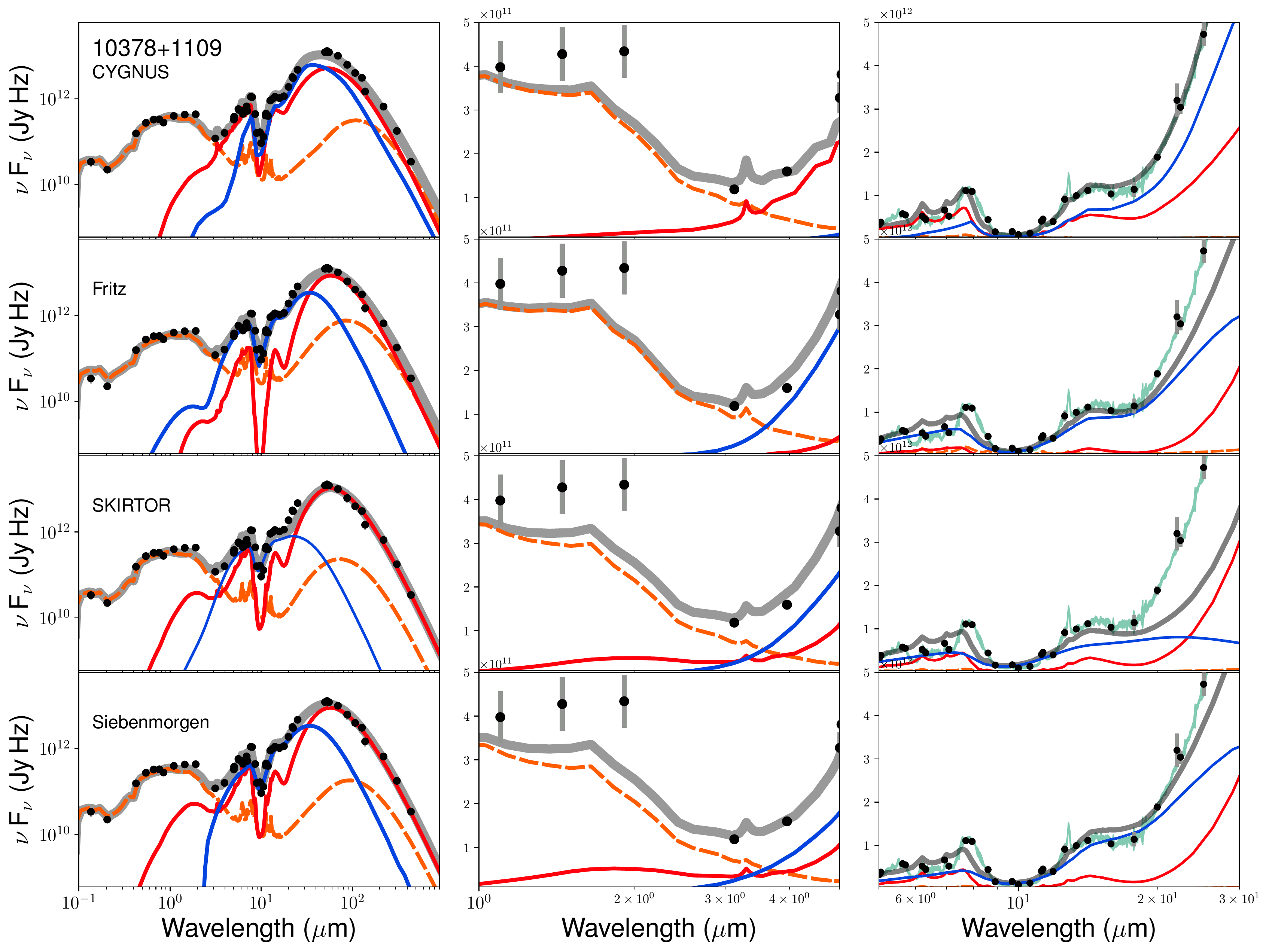}
		\caption{Comparison SED fit plots for { IRAS~10378+1109}. Other details are as in Figure \ref{fig:sedsrep1}.}
\label{fig:sedsrep2}
\end{figure*}

\begin{figure*}
\centering
\includegraphics[width=16cm]{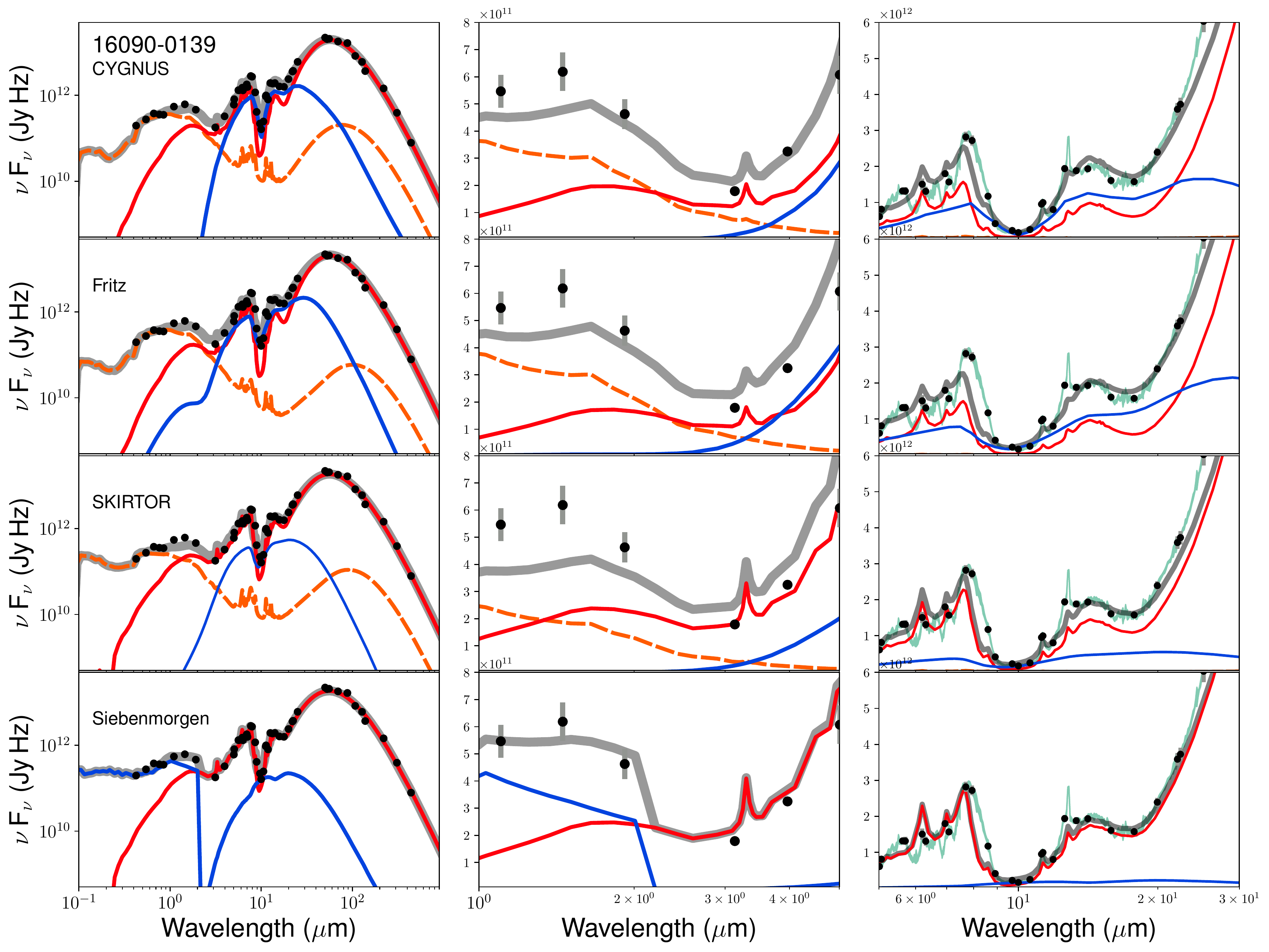}
		\caption{Comparison SED fit plots for IRAS~16090-0139.  Other details are as in Figure \ref{fig:sedsrep1}.}
\label{fig:sedsrep3}
\end{figure*}

\begin{figure*}
\centering
\includegraphics[width=16cm]{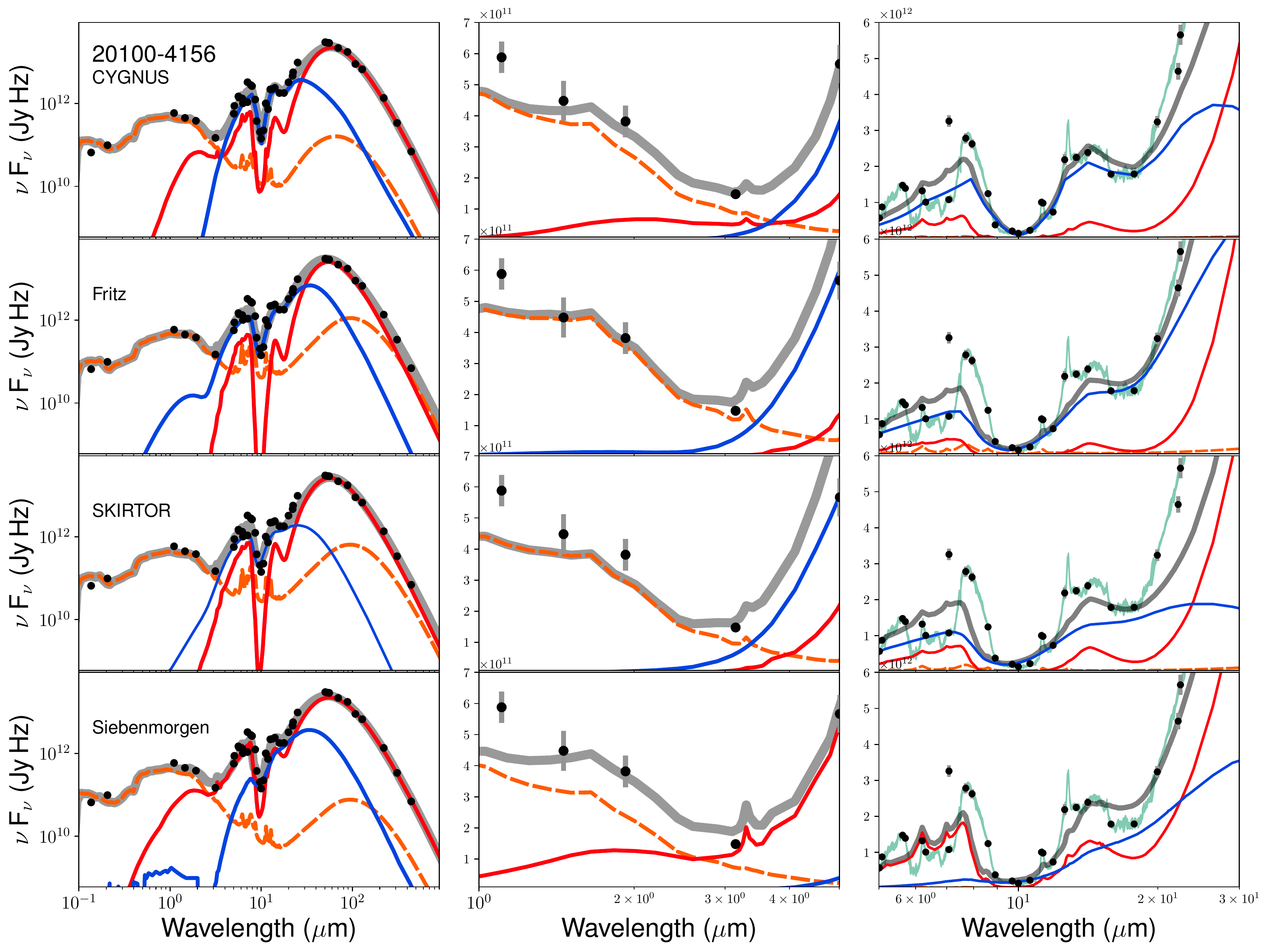}
		\caption{Comparison SED fit plots for IRAS~20100-4156.  Other details are as in Figure \ref{fig:sedsrep1}.}
\label{fig:sedsrep4}
\end{figure*}

\begin{figure*}
\centering
\includegraphics[width=16cm]{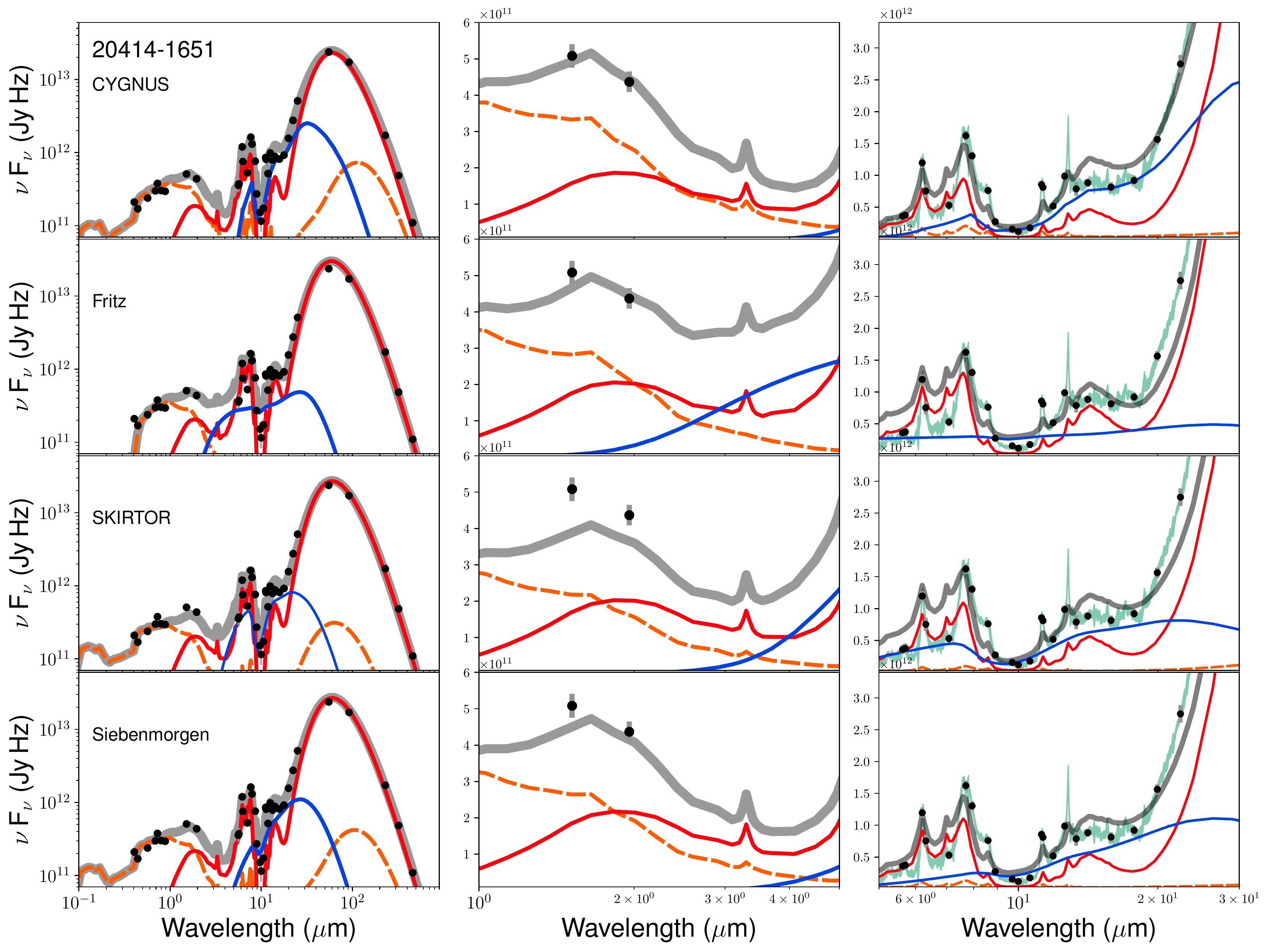}
		\caption{Comparison SED fit plots for IRAS~20414-1651.  Other details are as in Figure \ref{fig:sedsrep1}.}
\label{fig:sedsrep5}
\end{figure*}

\begin{figure*}
\centering
\includegraphics[width=16cm]{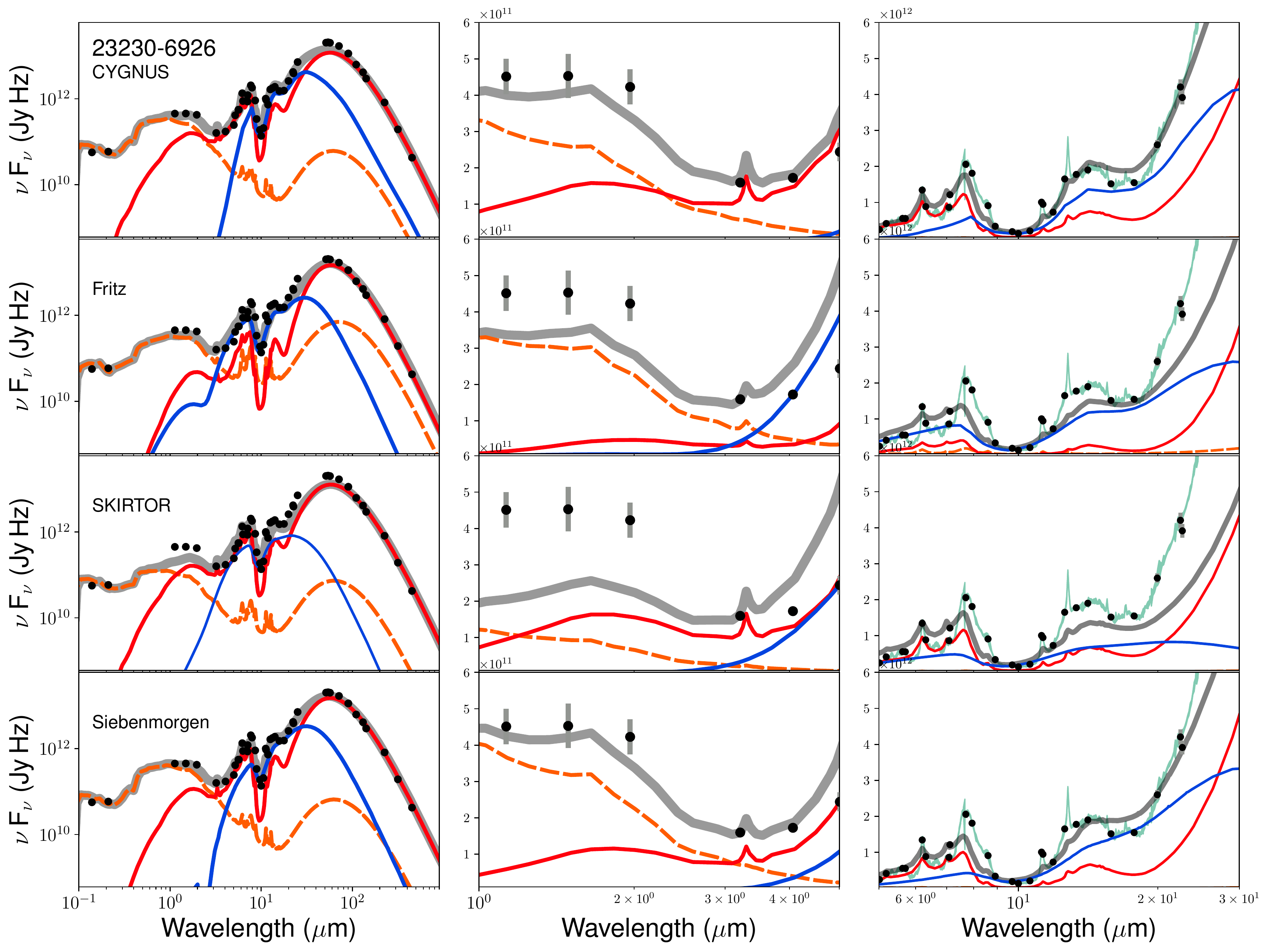}
		\caption{Comparison SED fit plots for IRAS~23230-6926.  Other details are as in Figure \ref{fig:sedsrep1}.}
\label{fig:sedsrep6}
\end{figure*}

The third novelty of the method used in this paper is that we include in the SED de-composition the Spitzer IRS spectroscopy at a resolution which is matched to the resolution of the radiative transfer models. PAH features give an indication of star formation activity \citep{pee02} whereas silicate absorption features constrain the degree of obscuration in either the starburst or AGN torus \citep{spo07}. According to most torus models silicate emission and absorption also constrain the inclination of the torus which as we show in the paper is one of the main factors that determine the intrinsic luminosity of the AGN. Including the spectroscopy in the SED fitting is therefore particularly important for breaking the degeneracy in the fits and constraining the starburst, host and AGN properties. 

This paper is organized as follows: In section 2 we describe the sample and in section 3 the data we assembled. In section 4 we describe the models and the SED de-composition method. In section 5 we present our results with the CYGNUS models and compare results for six representative ULIRGs from { four} different torus models, CYGNUS, the models from \citet{fritz06}, SKIRTOR \citep{stal16} and \citet{sie15}. In sections 6, 7 and 8 we discuss further our results and finally in section 9 we present our conclusions.

Throughout this work, we assume \mbox{$H_0 = 70$\,km\,s$^{-1}$\,Mpc$^{-1}$}, \mbox{$\Omega = 1$}, and \mbox{$\Omega_{\Lambda} = 0.7$}.

\section{Sample Selection}\label{sampleselect}
We start with the sample of ULIRGs observed by the HERschel Ultra Luminous Infrared Galaxy Survey (HERUS) carried out by the Herschel Space Observatory. This sample comprises all 40 ULIRGs from the IRAS PSC-z survey \citep{sau00} with 60$\mu$m fluxes greater than 2Jy, together with three randomly selected ULIRGs with lower 60$\mu$m fluxes; IRAS~00397-1312 (1.8 Jy), IRAS~07598+6508 (1.7 Jy) and IRAS~13451+1232 (1.9 Jy). We then exclude 3C~273 as it is a Blazar, to give a sample of 42 objects (Table \ref{tab:basic}). Strictly speaking this sample is not complete, but it includes nearly all known ULIRGs at $z<0.27$, and so should give an almost unbiased benchmark of local ULIRGs. All 42 objects were observed by the Infrared Spectrograph (IRS, \citealt{houck04}) on board Spitzer and by Herschel as part of both HERUS and the SHINING survey \citep{fisch10,sturm11,hail12,gonz13}.

\section{Data}
To assemble the SEDs of the galaxies fitted in this paper we combined data available from the literature and various archives. For all the HERUS galaxies there are Spitzer/IRS data which we downloaded from CASSIS \citep{leb11}. We do not use the IRS data in their full resolution but we reduce their spectral resolution so that they are matched to the spectral resolution of the radiative transfer models. In particular, the IRS data included in the fit with SATMC have a wavelength grid which is separated in steps of 0.05 in the log of the rest wavelength. We also add additional points around the 9.7$\mu m$ silicate feature and the PAH features to the equally spaced wavelength grid. We find that with this approach we maintain the constraining power of the data but make the fitting of the multi-wavelength SED feasible with current methods.   

To extend the wavelength coverage of the SEDs from the ultraviolet to the submillimetre we added to the IRS data the HERUS SPIRE photometry \citep{pea16,cle18} and IRAS data for all the galaxies and where available, GALEX, Pan-STARRS, 2MASS, Spitzer/IRAC and SCUBA or other infrared and submillimetre data. Apart from the Pan-STARRS data all of these data were extracted from the NASA Extragalactic Database (NED) { or directly from the 2MASS archive}. For IRAS~03158+4227 we used optical data from SDSS. Where available we added to the SEDs the PACS continuum data from \citet{far13}.

For seven of the ULIRGs that are included in the GOALS \citep{arm09} sample (IRAS~05189-2524,  IRAS~14348-1447, IRAS~15250+3608, IRAS~22491-1808, Arp~220, Mrk~231, Mrk~273, UGC~5101) we also added the optical photometry from \citet{u12}. For Arp~220, IRAS~08572+3915 and IRAS~10565+2448, we find that the optical photometry is more homogeneous if we use the SDSS data.

\begin{figure}
	\begin{center}
		\includegraphics[width=8cm]{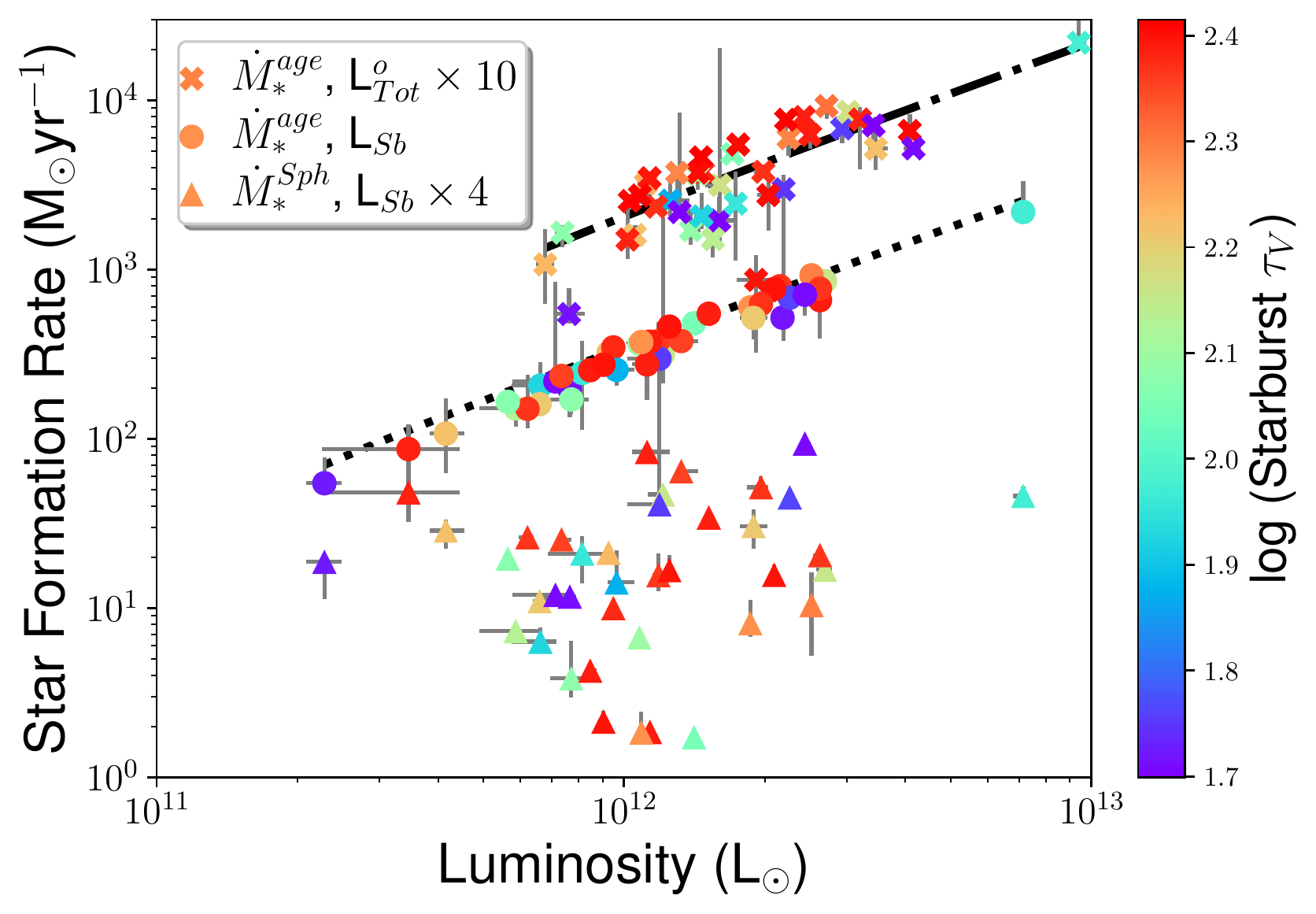}
		\caption{Star formation rates versus infrared luminosities (starburst or total). The dotted and dot-dashed lines show the fits that give rise to the relationships given in section 5.1 { (equations 2 and 4 respectively).}}
\label{fig:StarFormation}
\end{center}
\end{figure}

\begin{figure}
	\begin{center}
		\includegraphics[width=8cm]{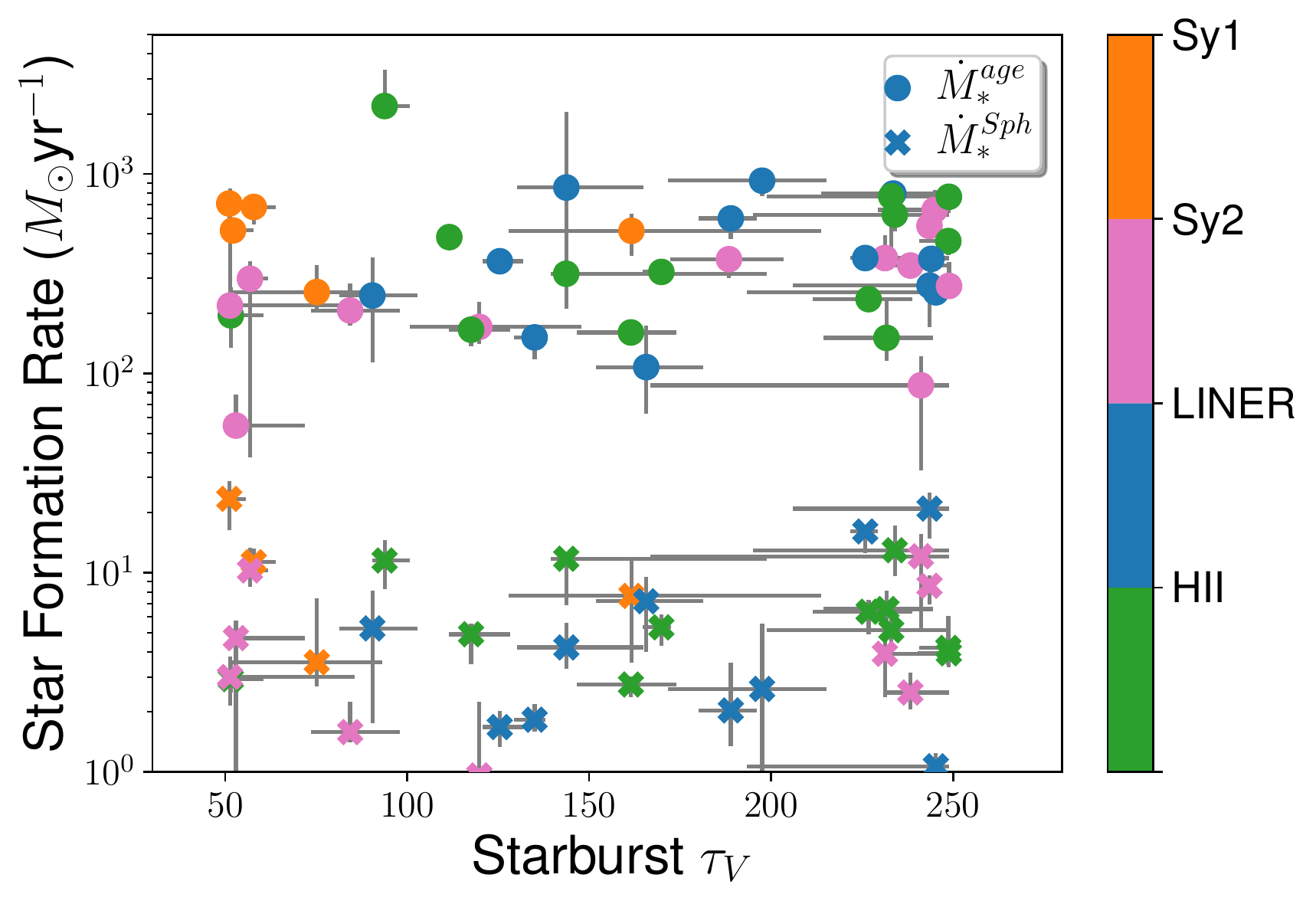}
		\caption{Star formation rates versus starburst obscuration which is parameterized with the initial optical depth of the molecular clouds $\tau_V$.}
\label{fig:StarFormation2}
\end{center}
\end{figure}

\section{Modelling}

\subsection{The CYGNUS models}

The available photometry and spectroscopy has been fitted with the MCMC code SATMC \citep{john13} using libraries of radiative transfer models for star formation, AGN activity, and host galaxy emission. These are part of the CYprus models for Galaxies and their NUclear Spectra (CYGNUS) project. CYGNUS includes models for starbursts as described in \citet{efstathiou00} and \citet{efstathiou09}. Models for massive star formation and starbursts were also presented and discussed in \citet{mrr89,mrr93b,efs94,kru94,sil98,tak03,dop05,sie07}. The starburst model parameters and their assumed ranges are: the age of the starburst ($t_{*} = 5-35$Myr), the giant molecular clouds' (GMC's) initial optical depth $\tau_{V}=50-250$, the time constant of the exponentially decaying star formation rate (SFR) $\tau_{*} =15-35$Myr.

The method employed for computing the libraries of spheroidal host galaxy models of CYGNUS is an evolution of the `cirrus' model of \citet{efs03} and is described in more detail in \citet{efs21} and Efstathiou (in preparation). The models of \citet{bru93,bru03} are used in combination with an assumed star formation history (SFH) to compute the spectrum of starlight which is illuminating the dust throughout the model galaxy. The spectrum of starlight is assumed to be constant throughout the galaxy but its intensity varies throughout the galaxy according to a S\'ersic profile with $n=4$ which is equivalent to de Vaucouleurs's law. \citet{efs03}  assumed an exponentially decaying SFH whereas here we assume a delayed exponential ($\dot{M}_{\ast} \propto t \times e^{-t/\tau^s}$), where $\tau^s$ is the e-folding time of the exponential.

The parameters assumed by the spheroidal model and their assumed ranges are: the e-folding time of the delayed exponential $\tau^s= 0.125-8$Gyr, the optical depth of the spherical cloud from its centre to its surface $\tau_v^s= 0.1-15$ and $\psi^s= 1-17$ which controls the bolometric intensity of stellar emission relative to that of the bolometric intensity of starlight in the solar neighborhood. The library used in this paper was computed assuming all the galaxies have an age equal to the age of the Universe at a redshift of $z=0.1$. We assumed that all the stars in the galaxy formed with a Salpeter IMF out of gas with a metallicity of 40\% of solar.

There exist a wide variety of AGN obscurer models in the literature with different assumptions about the obscurer geometry \citep{pier93,mrr93,gra94,efstathiour95,nen02,dull05,fritz06,hon06,nen08,sch08,hey12,sta12,stal16,efstathiou13,sie15,hk17}. We therefore explored the impact of { four} different AGN models:

\begin{enumerate}

\item The CYGNUS AGN torus model. More details of the implementation of this combination of models within SATMC are given in \citet{efs21}. A number of results with this combination of models in the MCMC code SATMC have previously been presented \citep{herr17,kool18,matt18,pitch19,efs21,kan21}.

\item The AGN torus model of \citet{fritz06}, hereafter FR06. 

\item The two-phase AGN torus model SKIRTOR of \citet{stal16}.

\item The two-phase AGN torus model of \citet{sie15}.

\end{enumerate}

The CYGNUS AGN model parameters and their assumed ranges are the half-opening angle of the torus ($\theta_{o} = 30\degr - 75\degr$), the inclination of the torus ($\theta_{i} = 0\degr - 90\degr$), the ratio of outer to inner disc radius ($r_2/r_1 = 20 - 100$)  and the equatorial optical depth at 1000\AA ~~($\tau_{uv} = 250 - 1450$; for the dust model used in \citet{efstathiour95} this translates to an $A_V\approx \tau_{uv}/5$ and $\tau_{9.7\mu m} \approx \tau_{uv}/61$). All the models assume that the density distribution in the tapered disc falls off with distance from the supermassive black hole $r$ as $r^{-1}$. In addition we explore for some objects the impact of adding a component of polar dust in the fits. This is discussed further in section 5.5.

For the CYGNUS combination we therefore have a total of 14 free parameters in the fits:  $\tau^s$, $\tau_v^s$, $\psi^s$, $\theta_{o}$, $\theta_{i}$, $r_2/r_1$, $\tau_{uv}$, $t_{*}$, $\tau_V$, $\tau_{*}$, $f_{SB}$, $f_{AGN}$, $f_s$, $f_{p}$. The last four are scaling factors that determine the luminosities of the starburst, AGN, spheroidal and polar dust components respectively. The parameters of all models used in this work are also listed in Table 2 together with a summary of other useful information.

\subsection{The Fritz et al. and SKIRTOR AGN torus models}

The FR06 model assumes a smooth torus as in \citet{efstathiour95}. The models are available in a public library\footnote{https://www.irya.unam.mx/gente/j.fritz/JFhp/AGN\_models.html}. The FR06 model has two more parameters than the CYGNUS model. We use the subset of the library that results by fixing the two parameters that determine the density distribution within the torus in the radial direction ($\beta=0$) and polar direction ($\gamma=4$). This gives the same number of free parameters as the CYGNUS AGN torus model. The four parameters are the equatorial optical depth at 9.7$\mu m$, the ratio of outer to inner radius,  the half-opening angle of the torus and the inclination plus the scaling factor. For these parameters we use the full range available in the library. For the FR06 combination we also have a total of 14 free parameters in the fits. It is important to note that the FR06 models have a lower resolution in inclination compared to the CYGNUS model, with 10 angles covering the range $0-90\degr$ degrees. We linearly interpolate the models between these values to obtain the same resolution we have in CYGNUS.

The SKIRTOR model assumes a two-phase geometry, i.e. it assumes that the torus dust lies in discrete clouds that are embedded in a smooth distribution. The SKIRTOR models are also available in a public library\footnote{https://sites.google.com/site/skirtorus/}. As in the case of the FR06 models we selected the subset of the library by fixing the parameters that describe the density distribution in the torus in the radial direction ($p=1$) and polar direction ($q=1$). The parameter $p$ takes four discrete values in the library (0, 0.5, 1 and 1.5). As we discuss later in the paper the main problem with these models is that they do not produce deep enough silicate absorption features to fit the SEDs of the ULIRGs in this sample. Assuming a value for $p < 1$ leads to AGN torus spectra with shallower silicate absorption features and therefore worse fits with these models. We have four remaining free parameters in the SKIRTOR model which are the same as those in the FR06 model.  For these parameters we also use the full range available in the library. For the SKIRTOR combination we also have a total of 14 free parameters in the fits. The SKIRTOR models also has 10 inclinations that cover the range $0-90\degr$ so again we linearly interpolate the templates to have the same resolution as in the CYGNUS fits. 

\subsection{The Siebenmorgen et al. models}

As in the case of the SKIRTOR models, the \cite{sie15} models also assume a two-phase geometry and they are computed with a Monte Carlo radiative transfer code. The models are available in a public library\footnote{http://www.eso.org/\textasciitilde{}rsiebenm/agn\_models/index.html}. A particular feature of the Siebenmorgen models is that the opening angle of the torus is not a parameter of the model. This model is in that respect similar to the smooth `anisotropic sphere' model of \cite{efstathiour95} which was used to fit the SED of IRASF~10214+4724 \citep{mrr93}. Another feature of the model is that it assumes that the dust grains are fluffy and have higher emissivity in the far-infrared and submillimeter compared to normal interstellar grains. The model assumes the following parameters: The cloud volume filling factor, the optical depth (in the V band) of the individual clouds, the optical depth (in the V band) of the disk mid-plane and the inclination which takes 9 values corresponding to bins at 86, 80, 73, 67, 60, 52, 43, 33, and 19 degrees measured from the pole. As in the case of the SKIRTOR and Fritz models we interpolate the models to have the same resolution in inclination as the CYGNUS model. The model also has as a parameter the inner radius of the dusty torus. We select the value of $1000 \times 10^{15}$cm which for an AGN luminosity of $10^{11} L_\odot$ gives a temperature at the inner torus radius of about 1000K. Our fits with the Siebenmorgen models also have a total of 14 free parameters.

\begin{figure*}
	\begin{center}
		\includegraphics[width=15cm]{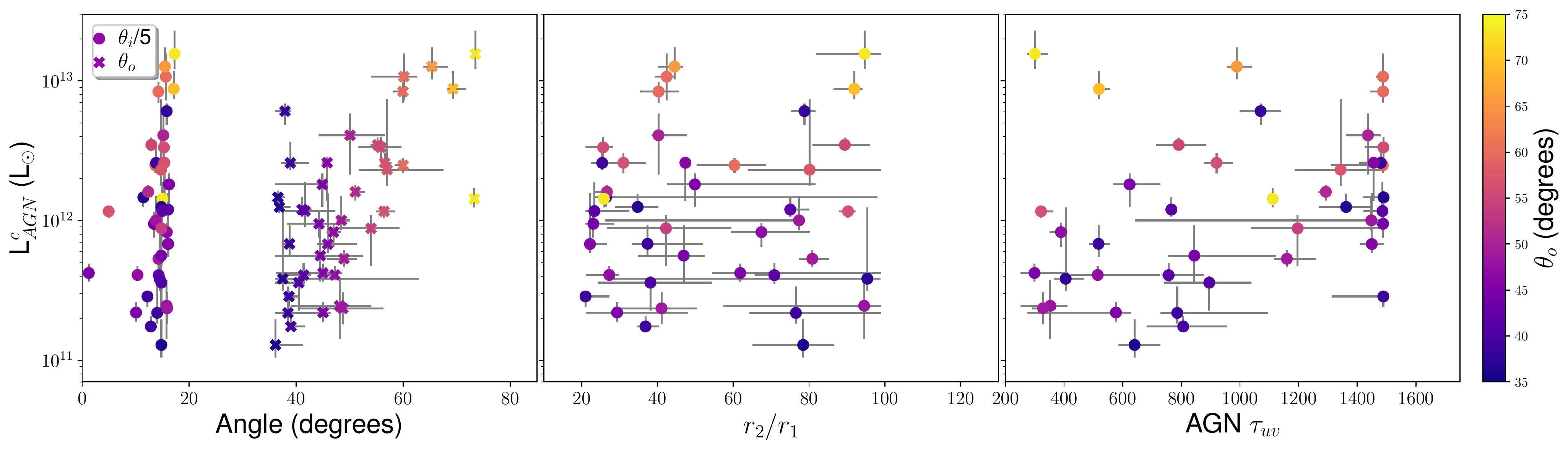}
		\caption{Anisotropy-corrected infrared AGN luminosity as a function of three obscurer parameters: {\itshape Left:} inclination and half-opening angles, {\itshape Middle:} outer to inner torus radius,  {\itshape Right:} torus obscuration.}
\label{fig:AGNlumsvsprops}
\end{center}
\end{figure*}	

\begin{figure}
	\begin{center}
		\includegraphics[width=8cm]{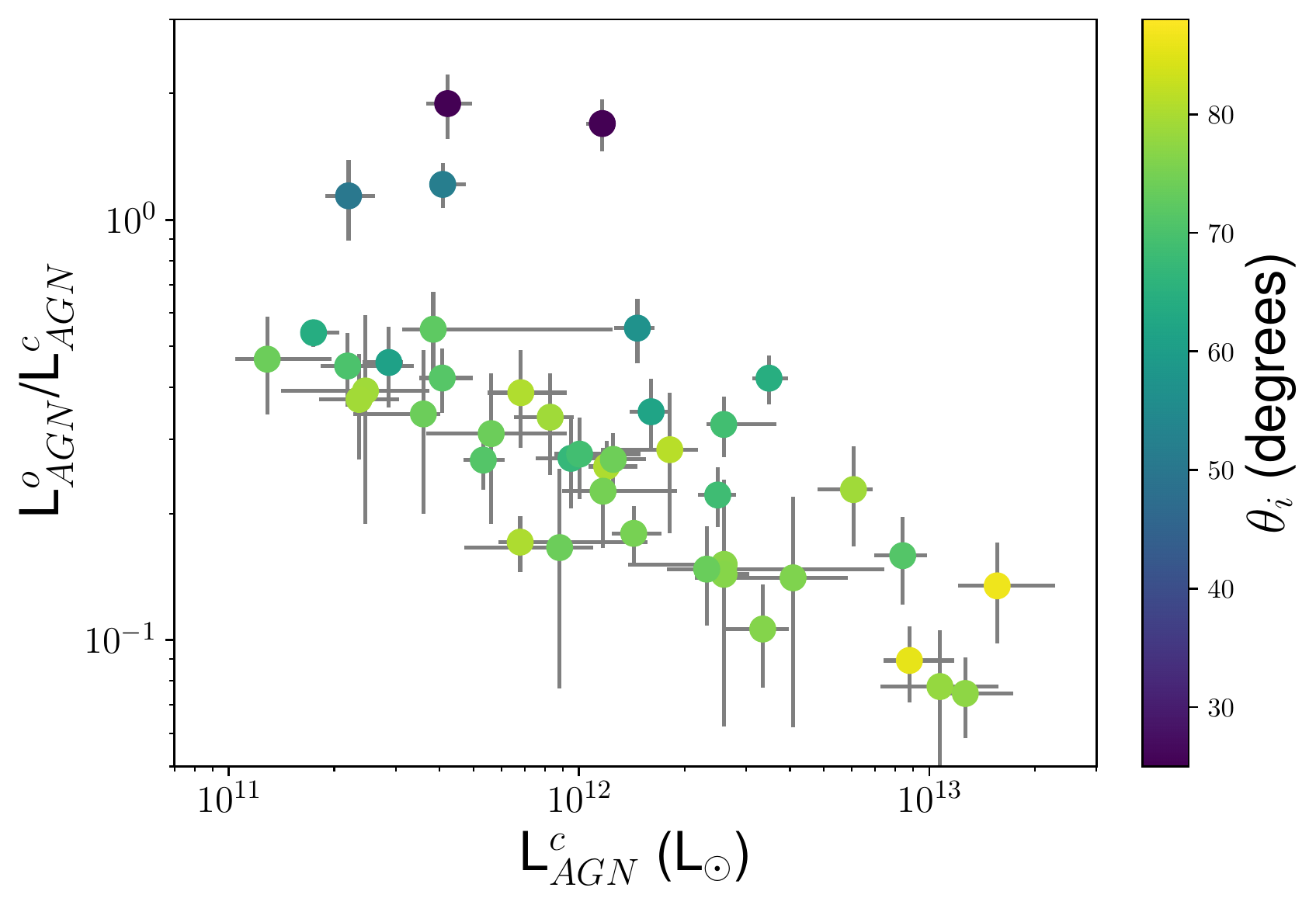}
		\caption{Observed to corrected luminosity ratio vs anisotropy-corrected infrared AGN luminosity. The scatter in this relation appears smaller for larger values of $\theta_{i} - \theta_{o}$, suggesting that the relation is tighter when the viewing angle is `well into' the torus.}
\label{fig:AGNaniso}
\end{center}
\end{figure}

\begin{figure*}
	\begin{center}
		\includegraphics[width=15cm]{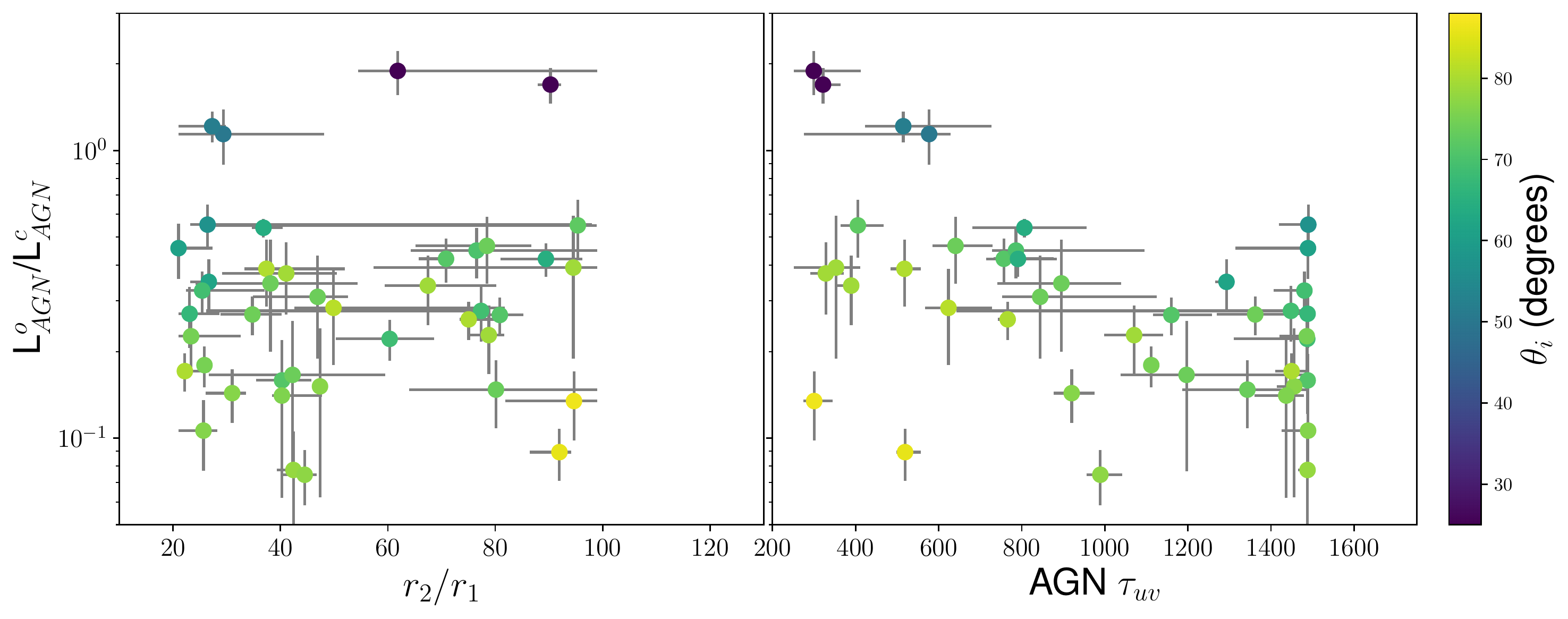}
		\caption{Observed to corrected AGN luminosity ratio versus AGN obscurer parameters.}
\label{fig:AGNanisovsprops}
\end{center}
\end{figure*}

\begin{figure*}
	\begin{center}
		\includegraphics[width=15cm]{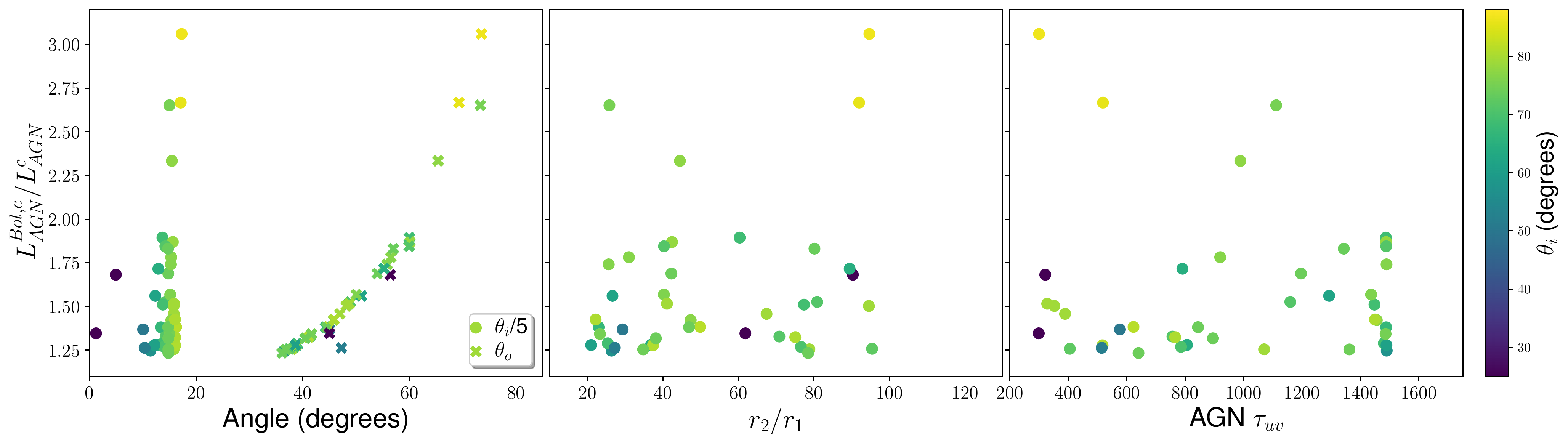}
		\caption{Anisotropy-corrected bolometric to infrared AGN luminosity ratio versus obscurer parameters.}
\label{fig:AGNbolanisovsprops}
\end{center}
\end{figure*}

\section{Results} \label{sec:res}
The infrared luminosities and derived parameters and physical quantities from the SED fits are presented in Tables 4 and 5. We also give additional derived physical quantities in Tables B1, B2, B3 and B4 in the Appendix. A description of all the derived physical quantities is given in Table 3. 
To give an impression of the results we selected six representative objects from the sample,  IRAS~03158+4227, IRAS~10378+1109, IRAS~16090-0139, IRAS~20100-4156, IRAS~20414-1651 and IRAS~23230-6926. We plot the SEDs of these six galaxies in Figures \ref{fig:sedsrep1} to \ref{fig:sedsrep6}. The full suite of SED fits with the CYGNUS models are presented in Figures A1 to A7 in the Appendix.

It is clear from the SED fits shown in Figures \ref{fig:sedsrep1} to \ref{fig:sedsrep6}, and also from the log-likelihood values listed in Table~1, that the CYGNUS combination of models provide good fits to the data including the IRS spectroscopy. The fits with the combination that includes the SKIRTOR and Fritz models are usually worse especially in the range covered by the IRS data. The SKIRTOR models generally do not produce deep enough silicate absorption features to fit the mid-infrared spectroscopy.
The Siebenmorgen15 models give better fits than SKIRTOR and Fritz but usually worse than CYGNUS.

An important feature of the results from the SED fitting is that due to the anisotropy of the emission from the AGN torus,  which is usually optically thick to its own radiation \citep{pier93,efstathiour95,efs14}, the AGN luminosity needs to be corrected by the anisotropy correction factor $A(\theta_i)$ defined in \citet{efstathiou06}, 

\begin{equation}
A(\theta_i) = {{\int_0^{\pi/2} ~~S(\theta_i') ~~sin ~~\theta_i' ~~d \theta_i' } \over {S(\theta_i) }}
\end{equation}

\noindent where $S(\theta_i)$ is the bolometric emission over the relevant wavelength range.  $A(\theta_i)$ is generally different for the infrared and bolometric luminosities and is necessary and significant for all the AGN torus models considered in this paper. In Table \ref{tab:cyg:irlums} we give both the `observed' or uncorrected AGN luminosities as well as their `corrected' counterparts.

\subsection{Star formation}\label{ssectsfr}
In this work we discuss two estimates of the star formation rate of the starburst episodes in the ULIRGs: the star formation rate averaged over the age of the starburst as determined from the fits $\dot{M}_{Sb}^{age}$ and the star formation rate averaged over a flat timescale of 50Myr $\dot{M}_{Sb}^{50}$.
We find that all objects harbor high rates of star formation, between 22 and 644$M_{\odot}$ yr$^{-1}$ when the SFR is averaged over 50Myr and between 59 and 2203$M_{\odot}$ yr$^{-1}$ when averaged over the age of the starburst. The starburst component always dominates the total SFR. Host star formation rates are in most cases at least an order of magnitude lower. 

We plot SFR against $L_{Sb}$ and $L^{o}_{Tot}$ in Figure \ref{fig:StarFormation}. There is a strong correlation between $L_{Sb}$ and $\dot{M}_{Sb}^{age}$ (Kendall's $\tau = 0.88$), of which an acceptable linear parameterization is:

\begin{equation}
\frac{{\dot{M}_{Sb}^{age}}}{M_{\odot}yr^{-1}} = (3.14\pm0.08)\times10^{-10}\frac{L_{Sb}}{L_{\odot}} - 5\pm16
\end{equation}

\noindent or:

\begin{equation}
\frac{{\dot{M}_{Sb}^{age}}}{M_{\odot}yr^{-1}} = (3.11\pm0.05)\times10^{-10}\frac{L_{Sb}}{L_{\odot}}
\end{equation}

\noindent with a zero intercept, with a scatter of about 0.1 dex\footnote{The starburst models assume a Salpeter IMF. With a Chabrier IMF the SFRs would be about a factor of 0.6 lower.}. A power-law is not an appreciably better fit. The relation does not appear to depend on $t_*$ but there is a dependence on $\tau_{V}$, with a flatter slope for more extincted starbursts. There is no correlation between $L_{Sb}$ and $\dot{M}_{sph}$. 

Since a more commonly measured value for luminous infrared galaxies are their observed (that is, uncorrected for anisotropic emission) total infrared luminosities, we also compare $\dot{M}_{Sb}^{age}$ to $L^{o}_{Tot}$. We see a correlation, albeit with a wider dispersion:

\begin{equation}
\frac{{\dot{M}_{Sb}^{age}}}{M_{\odot}yr^{-1}} = (2.24\pm0.17)\times10^{-10}\frac{L^{o}_{Tot}}{L_{\odot}} - 13\pm42
\end{equation}

\noindent or:

\begin{equation}
\frac{{\dot{M}_{Sb}^{age}}}{M_{\odot}yr^{-1}} = (2.22\pm0.10)\times10^{-10}\frac{L^{o}_{Tot}}{L_{\odot}}
\end{equation}

\noindent with a zero intercept. More extincted starbursts are more likely to scatter below this relation. 

Previous calibrations of the conversion between infrared luminosity and star formation rate generally derive 

\begin{equation}
\frac{SFR}{M_{\odot}yr^{-1}} = (0.6 - 3.2)\times10^{-10}\frac{L_{IR}}{L_{\odot}}
\end{equation}

\noindent for 0.1-100$M_{\odot}$, a Kroupa IMF, and SFR timescales of 10Gyr to 2Myr \citep{calz07,ken12}. Our calibration for ULIRGs is consistent with this range, with a slope consistent with star formation timescales of $\lesssim100$Myr. This agrees with the ages we derive for the starburst episode. Previous values are closer to our $L^{o}_{Tot}$ as that is what is typically measured in other studies. 
We plot SFR against $\tau_{V}$ in Figure \ref{fig:StarFormation2}. There is a weak positive correlation between $\tau_{V}$ and $\dot{M}_{Sb}^{age}$ { (Kendall's $\tau = 0.11$)} , consistent with more luminous starbursts being systematically more extincted. The correlation does not appear to depend on starburst age. This is reasonable as $\tau_{V}$ is a measure of the obscuration of the starburst episode over its whole duration. There is no dependence between $\tau_{V}$ and $\dot{M}_{sph}$, suggesting that more extincted starbursts do not straightforwardly imply more extincted host galaxies.

\subsection{AGN activity}
We find AGN in all objects, with anisotropy-corrected infrared luminosities spanning $1\times10^{11} - 1.5\times10^{13}$L$_{\odot}$. There is a correlation between $L^{c}_{AGN}$ and $\theta_{o}$, with more luminous AGN having larger values of $\theta_{o}$ (Figure \ref{fig:AGNlumsvsprops}). Conversely, there is no relation between $L^{c}_{AGN}$ and $\theta_{i}$. Instead plotting the obscurer parameters against $L^{o}_{AGN}$ gives conceptually identical results. This is unlikely to be a selection effect unless there exists a population of low redshift ULIRGs with luminous AGN that have both high covering fractions and very cold SEDs. Thus, more infrared-luminous AGN in ULIRGs seem to be associated with smaller covering fractions. This is also consistent with the general class of `receding torus' models \citep{law91} according to which the covering factor of the torus decreases with increasing luminosity due to dust sublimation. 

Conversely, there is no relation between corrected AGN luminosity and either $r_{2}/r_{1}$, or $\tau_{uv}$ (Figure \ref{fig:AGNlumsvsprops}). There is also no dependence between $r_{2}/r_{1}$ and $\theta_{o}$. It is notable that more than half of the objects have a relatively compact obscurer, with $r_{2}/r_{1}$  $\sim 20-40$, but there does not appear to be anything unusual in the other properties of the systems with compact obscurers. 

The obscurer geometry affects the observed AGN luminosity $L^{o}_{AGN}$, obtained by integrating the observed SED over 4$\pi$ steradians. The correction of the AGN luminosity depends on the intrinsic AGN luminosity, with more intrinsically luminous AGN having a larger correction (Figure \ref{fig:AGNaniso}). The mean $L^{o}_{AGN}$/$L^{c}_{AGN}$ ratio of the sample is 0.40, but more luminous AGN have a greater difference between $L^{o}_{AGN}$ and $L^{c}_{AGN}$, with $L^{o}_{AGN}$/$L^{c}_{AGN}=0.51$ at $L^{c}_{AGN} < 2\times10^{12}$L$_{\odot}$ and $L^{o}_{AGN}$/$L^{c}_{AGN}=0.17$ at $L^{c}_{AGN} > 2\times10^{12}$L$_{\odot}$. Moreover, the relation appears to depend on $\theta_{i}$, with a clearer relation for objects viewed closer to edge-on, consistent with a higher optical depth in the equatorial plane. The observed to corrected luminosity ratio does not however seem to depend as strongly on the other torus parameters (Figure \ref{fig:AGNanisovsprops}). Whether or not the viewing angle intersects the torus is the primary driver of the size of the anisotropy correction. 

Finally, we examine the bolometric AGN luminosity correction as a function of obscurer geometry (Figure \ref{fig:AGNbolanisovsprops}). The bolometric corrections range from factors of 1.25 to 3.10. As expected, there is a strong dependence on torus half-opening angle, especially when the line-of sight intersects the torus. There is however no significant dependence on the other AGN parameters. The magnitude of the bolometric correction also does not seem to depend on AGN luminosity.

\subsection{The Starburst-AGN Connection}
\begin{figure*}
	\begin{center}
		\includegraphics[width=15cm]{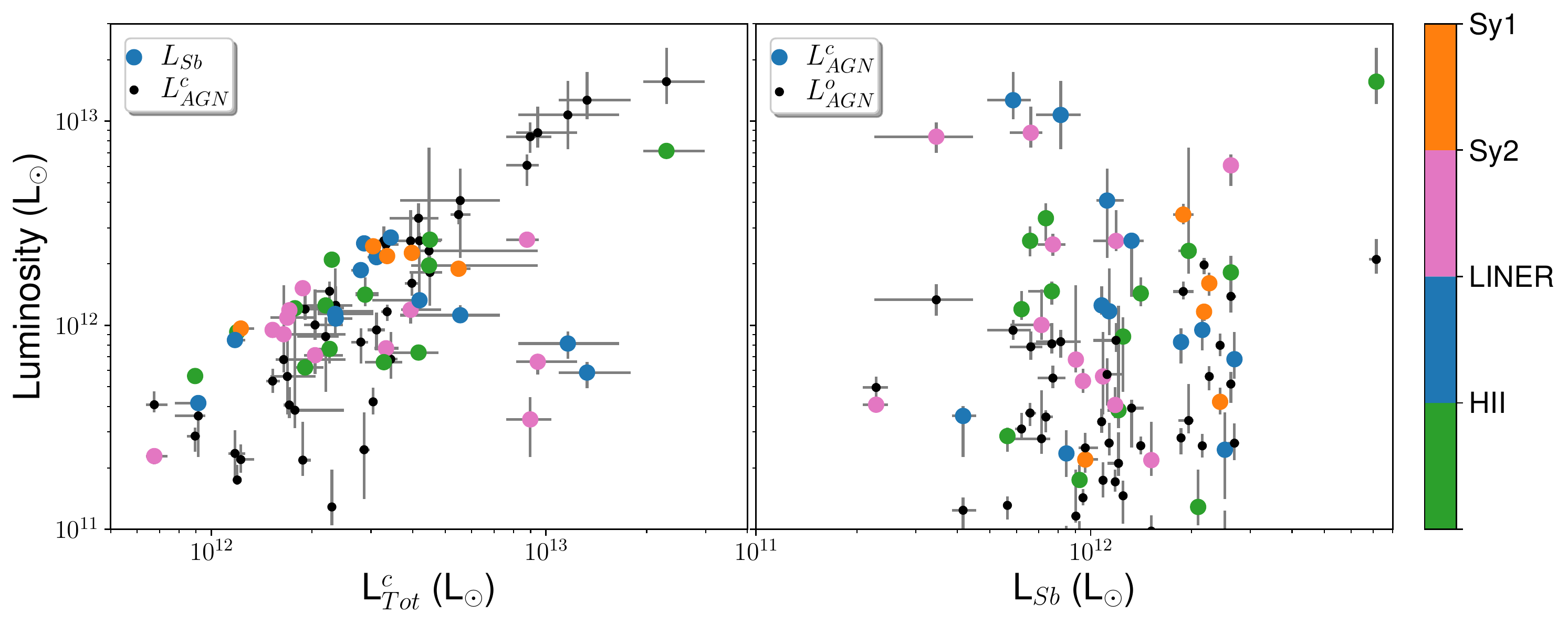}
		\caption{{\itshape Left:} Starburst and anisotropy-corrected AGN luminosity versus total corrected infrared luminosity. The starburst usually contributes the majority of infrared emission except at $L^{c}_{tot}\gtrsim 3\times10^{12} L_\odot$ when AGN start to dominate. {\itshape Right:}  Observed and anisotropy-corrected AGN luminosity versus Starburst luminosity. There is no observable relation between $L_{Sb}$ and $L_{AGN}$. In both cases we see no dependence on optical spectral type.}
\label{fig:sbconnbasic1}
\end{center}
\end{figure*}

\begin{figure*}
	\begin{center}
        \includegraphics[width=15cm]{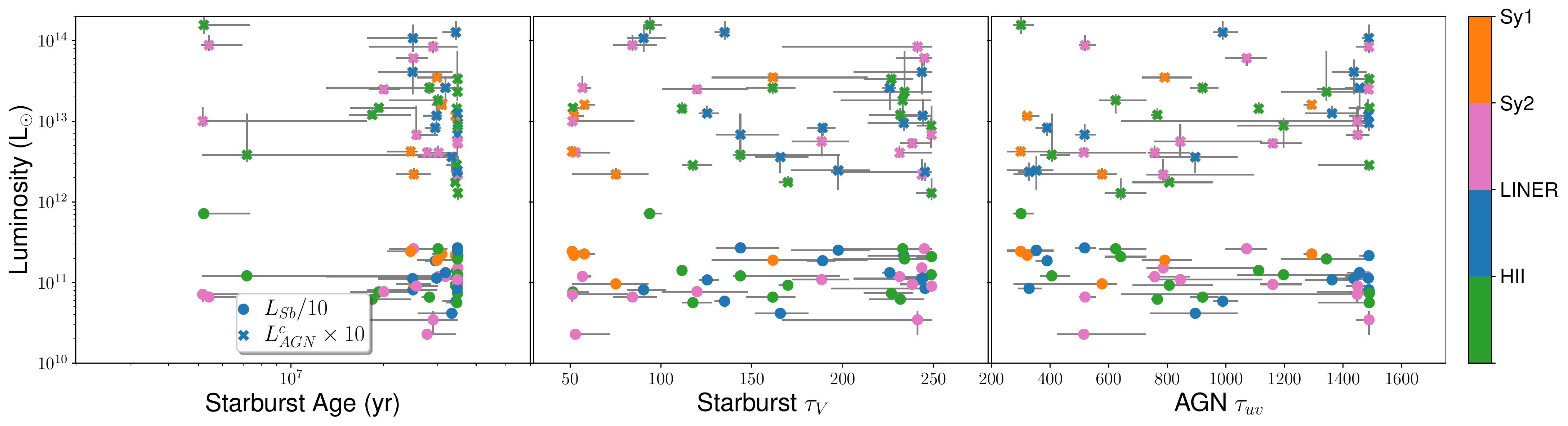}
		\includegraphics[width=15cm]{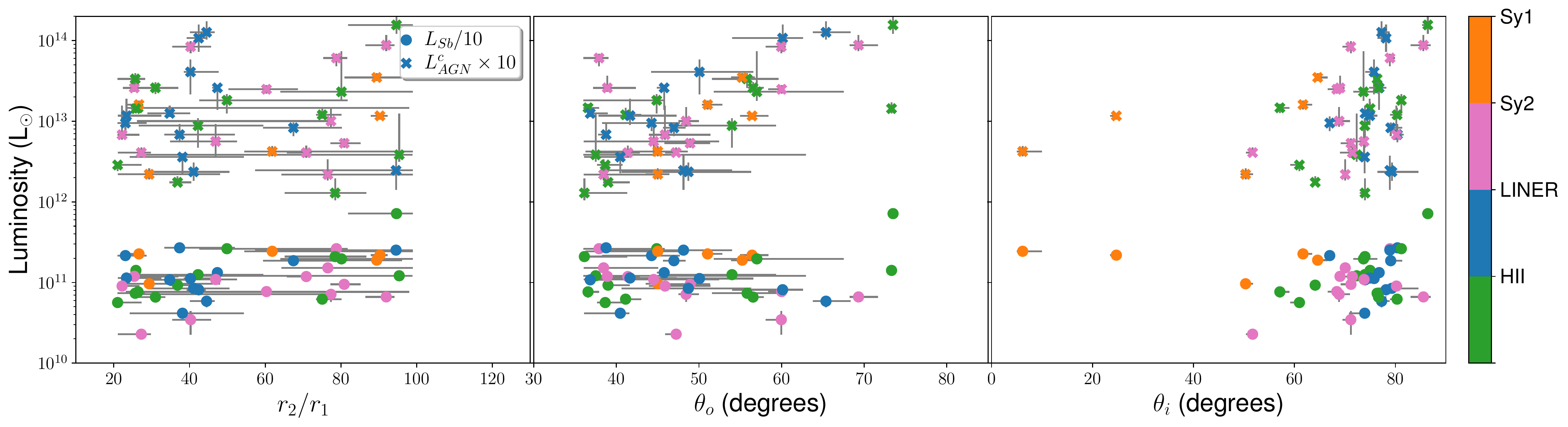}
		\caption{Starburst and AGN luminosity versus Starburst and AGN properties.}
\label{fig:sac_lumsprops}
\end{center}
\end{figure*}

\begin{figure}
	\begin{center}
		\includegraphics[width=8cm]{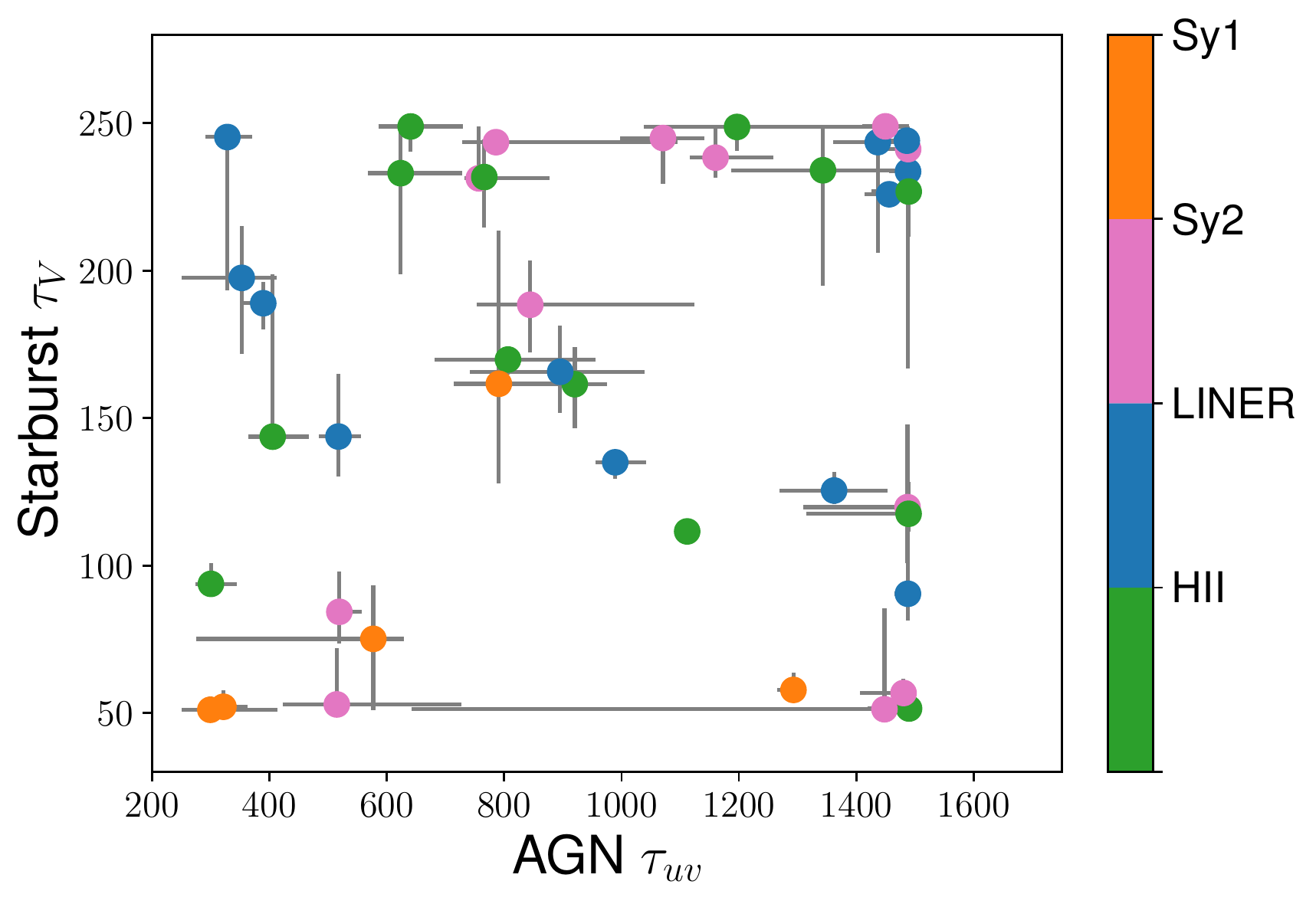}
		\caption{$\tau_{V}$ against $\tau_{uv}$. There is no evidence for a relation, even after accounting for e.g. optical spectral type.}
\label{fig:sbagnobs}
\end{center}
\end{figure}

Both star formation and AGN contribute significantly in most ULIRGs. The starburst usually contributes the majority of infrared emission except at $L^{c}_{tot}\gtrsim3\times10^{12}$L$_{\odot}$ when AGN start to dominate. There is no observable relation between $L_{Sb}$ and $L^{c}_{AGN}$ or $L^{o}_{AGN}$ (Figure \ref{fig:sbconnbasic1}). Objects with high $L^{c}_{AGN}$ do not have low $L_{Sb}$ in absolute terms. Moreover, there is no observed relation between any of $L^{c}_{tot}$, $L_{Sb}$, $L^{c}_{AGN}$, or starburst fraction with starburst age. We also see no trends in any of $L^{c}_{tot}$, $L_{Sb}$, $L^{c}_{AGN}$ with optical spectral classification. A conceptually similar disconnect between starburst and AGN luminosity has also recently been found in other classes of object, including Sloan Digital Sky Survey quasars \citep{maya15,pitch16}, though see also \citet{harr16}. Neither do we find evidence that $\tau_{V}$, $\tau_{uv}$, $\theta_{o}$ or $r_{2}/r_{1}$ depend on luminosity (Figure \ref{fig:sac_lumsprops}), or evidence for a relation between $\tau_{V}$ and $\tau_{uv}$ (Figure \ref{fig:sbagnobs}). 

The overall impression is an at best weak {\itshape direct} connection between starburst and AGN activity. The lack of trends between starburst and AGN luminosities suggests that they do not strongly affect each other. The lack of trends with optical spectral class is also expected if optical spectral class is primarily a function of orientation.  Moreover, there is no evidence that this (lack of) connection evolves over the lifetime of the starburst. 

\begin{figure*}
	\begin{center}
		\includegraphics[width=15cm]{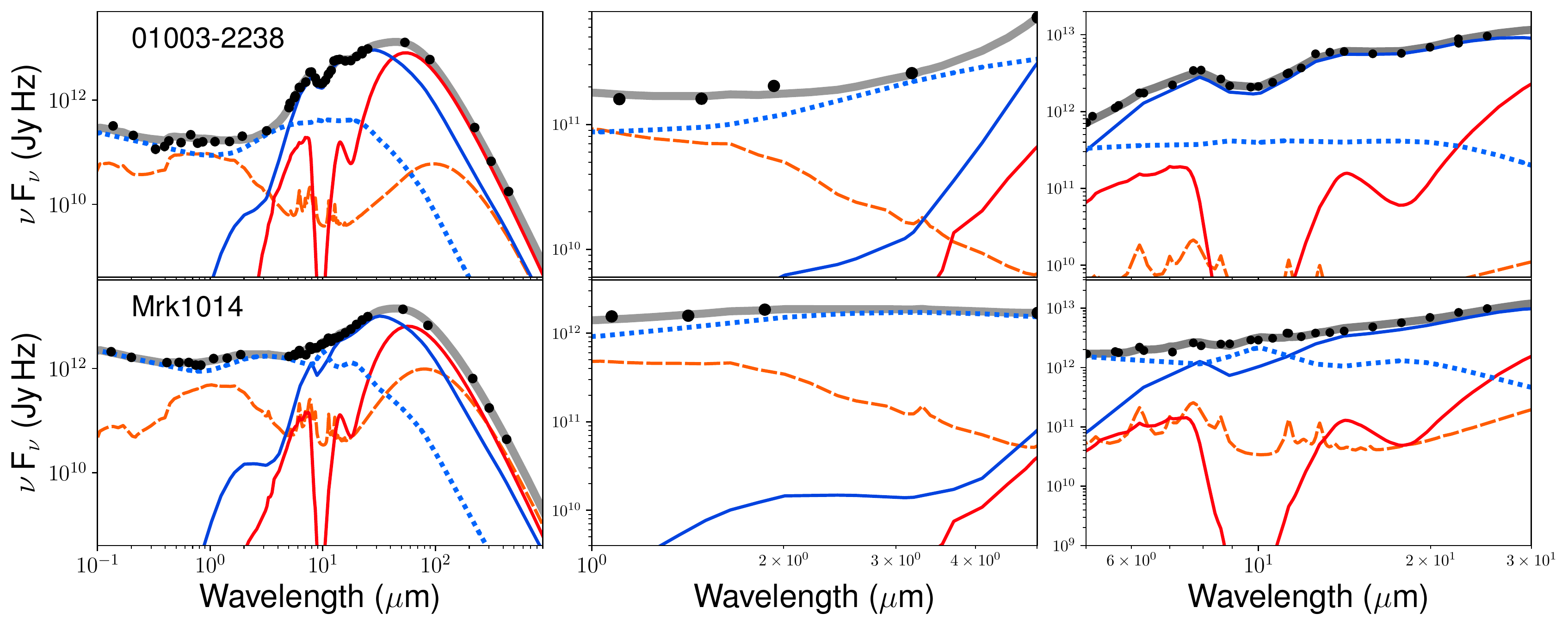}
		\caption{SED fit plots of the ULIRGs with dual AGN: spheroidal host (orange), starburst (red), edge-on AGN torus (solid blue), face-on AGN torus (dotted blue), total (grey).}
\label{fig:sedsdual}
\end{center}
\end{figure*}

\subsection{Dual AGN}
Two ULIRGs in the sample - IRAS~01003-2238 and Mrk~1014 - show evidence for dual AGN in their SED fits (Figure \ref{fig:sedsdual} and Table \ref{tab:dualagnprop}). At face value, this implies a dual AGN fraction of about 5 percent in local ULIRGs, but this value is unlikely to be correct. There is independent corroborating evidence for at least two other objects, see below. \citet{tad17} discovered a tidal disruption event (TDE) in IRAS~01003-2238 and conclude that TDEs are much more common in ULIRGs compared to the general galaxy population probably because of the presence of dual AGN in the process of merging. However, the fits are likely biased. The fits are only sensitive to dual AGN where the second AGN is sufficiently different in one or more physical parameters. We thus can not rule out the possibility that there may be other dual AGN with both AGN being broadly similar except, potentially, in luminosity. There is for example evidence for a triple AGN in NGC~6240 \citep{kollatschny20} and a dual AGN in Mrk~231 \citep{yan15}. However, we do not see any evidence for more than one AGN from the SEDs of NGC~6240 and Mrk~231. So we interpret the 5 percent of ULIRGs harboring a dual AGN as a lower limit.

\begin{table*}
\begin{center}
\begin{tabular}{lcccccccccc}
\hline
ID & $L^{o}_{AGN}$           & $L^{c}_{AGN}$            &$\theta_{o}$         & $\theta_{i}$         & $\tau_{uv}$         & $r_{2}/r_{1}$    & $L_{Sb}$                 & $\tau_{V}$              & $t_{*}$    & $M_{*}$               \\
   & $10^{12}$L$_{\odot}$    & $10^{12}$L$_{\odot}$     & $\degr$             & $\degr$              &                     &                  & $10^{12}$L$_{\odot}$     &                         & $10^{7}$yr & $10^{10}$M$_{\odot}$   \\ 	

\hline	
3  & $1.12^{+0.20}_{-0.21}$  & $2.47^{+0.44}_{-0.46}$   &$39.6^{+1.9}_{-3.5}$ & $61.8^{+2.2}_{-0.1}$ & $1408^{+82}_{-172}$ & $54^{+5}_{-11}$  & $0.73^{+0.05}_{-0.07}$  &  $204^{+45}_{-10}$ & $0.76^{+0.39}_{-0.25}$  &  $1.18^{+5.72}_{-0.13}$ \\
   & $0.09^{+0.06}_{-0.01}$  & $0.04^{+0.02}_{-0.005}$  &$40.7^{+6.2}_{-4.7}$ & $\sim10.0$           & $\geq 693$          & $37^{+34}_{-16}$ &  &   & & \\ 
42 & $2.21^{+0.47}_{-0.28}$  & $12.8^{+2.8}_{-1.6}$     & $52.5^{+3.3}_{-2.0}$& $71.0^{+0.5}_{-1.1}$ & $1489^{+1}_{-66}$   & $37^{+6}_{-9}$   & $1.17^{+0.20}_{-0.27}$  &  $236^{+13}_{-44}$ & $2.19^{+0.66}_{-1.69}$  &  $16.1^{+7.9}_{-5.8}$ \\ 
   & $0.78^{+0.15}_{-0.11}$  & $0.46^{+0.09}_{-0.04}$   & $59.9^{+3.8}_{-3.8}$& $\sim10.0$           & $327^{+88}_{-76}$   & $91^{+8}_{-8}$   &  &   & & \\  
\hline	
\end{tabular}
\caption{Fitted parameters and derived physical quantities for the dual AGN ULIRGs.}
\label{tab:dualagnprop}
\end{center}
\end{table*}
	
\subsection{Polar Dust}
For a few objects the emission from AGN polar dust is suggested from the SEDs. To model the emission of polar dust we use a library of spherical optically thick polar dust clouds for which we carry out a full radiative transfer calculation with the exception that we don't solve for the temperature of the dust but we assume a constant temperature for all grain species. We also assume that the clouds have a uniform density. There is therefore only one parameter in this model, the temperature of the polar dust clouds which we assume to be 900K. In addition we have the scaling factor $f_p$.

In a number of AGN there is clear evidence by high resolution imaging in the near- and mid-infrared that some of the nuclear emission is not coming from the torus but from the ionization cones. This has been observed in NGC~1068 since the early 1990s by \citet{braa93} and \citet{cam93}, in Circinus by \citet{tris07} and other AGN by \citet{hon13} and more recently in \citet{asm19}. The model of \citet{efs95} for NGC~1068 explored the presence of polar dust in the nucleus. \citet{efstathiou06} and \citet{efstathiou13} also discussed the idea that the mid-infrared emission of IRASF~10214+4724 is due to polar dust. \citet{matt18} also interpreted the near-infrared emission arising from the dust-enshrouded TDE in Arp~299 as arising from polar dust which was illuminated by the TDE. A similar phenomenon was also observed in the LIRG IRAS~23436+5257 by \citet{kool20}.

It is not clear how widespread polar dust is in ULIRGs. In the HERUS sample we do not find much evidence for it (three ULIRGs or about 8\% of the sample) but this may be due to the large optical depths to the nuclear region or because the covering factor is too low to detect with SED fitting. We see evidence for polar dust in the following ULIRGs: IRAS~05189-2524, IRAS~07598+6508 and  IRAS~13451+1232. IRAS~07598+6508 is the only unobscured quasar in the sample whereas IRAS~05189-2524 shows broad lines in polarized flux \citep{young96} which also requires a relatively unobscured view to the nucleus. The fraction of AGN luminosity due to polar dust ranges from $\sim 1-10$\% which is also to a good approximation the covering factor of polar dust.

\begin{figure}
	\begin{center}
		\includegraphics[width=7cm, angle=0]{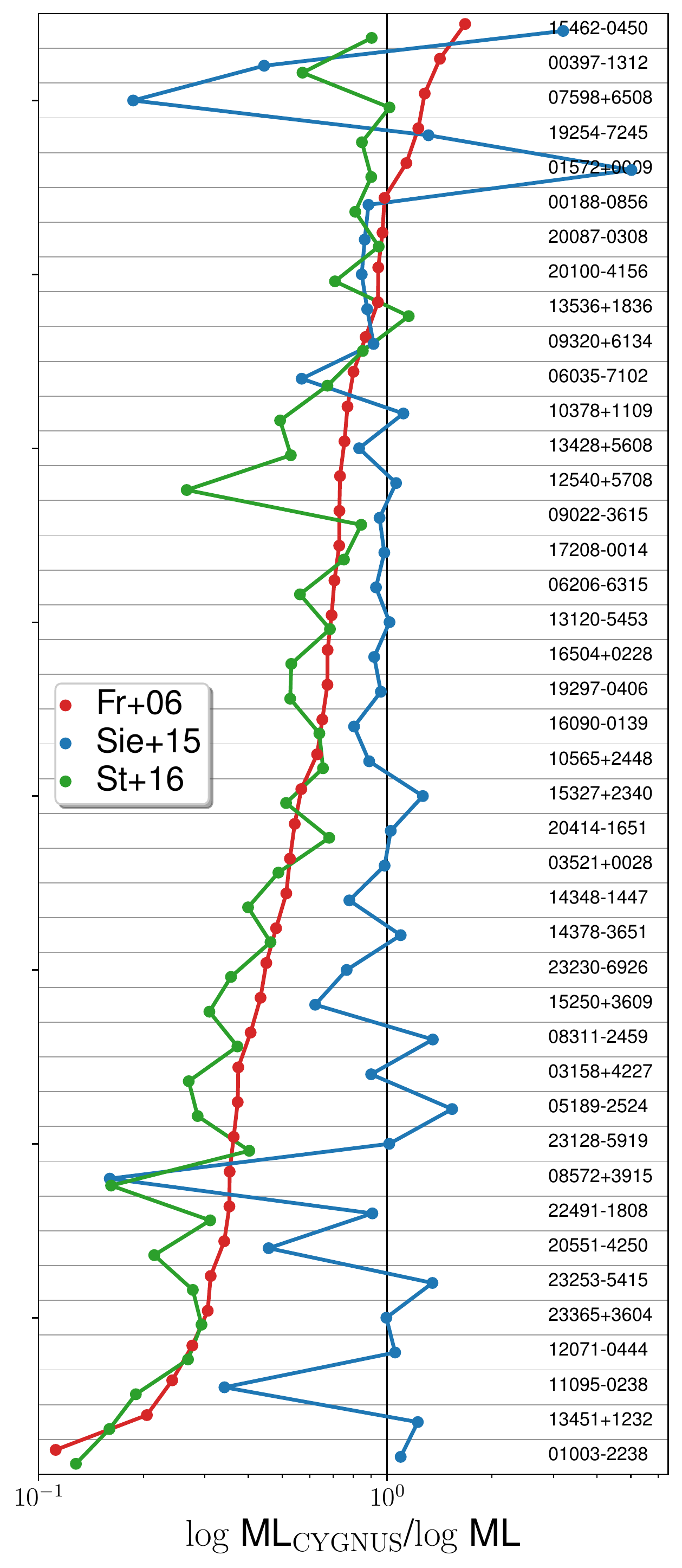}
		\caption{Comparison of the likelihood ratios of all objects fitted with the CYGNUS, FR06, SKIRTOR { and Siebenmorgen15} models. The plot shows the ratio of the FR06, SKIRTOR { and Siebenmorgen15} likelihood ratios, relative to the CYGNUS values. The CYGNUS and { Siebenmorgen15} models give systematically better fits than FR06 and  SKIRTOR. { CYGNUS is usually better overall.}}
\label{fig:lum_likes}
\end{center}
\end{figure}

\begin{figure*}
	\begin{center}
		\includegraphics[width=16cm]{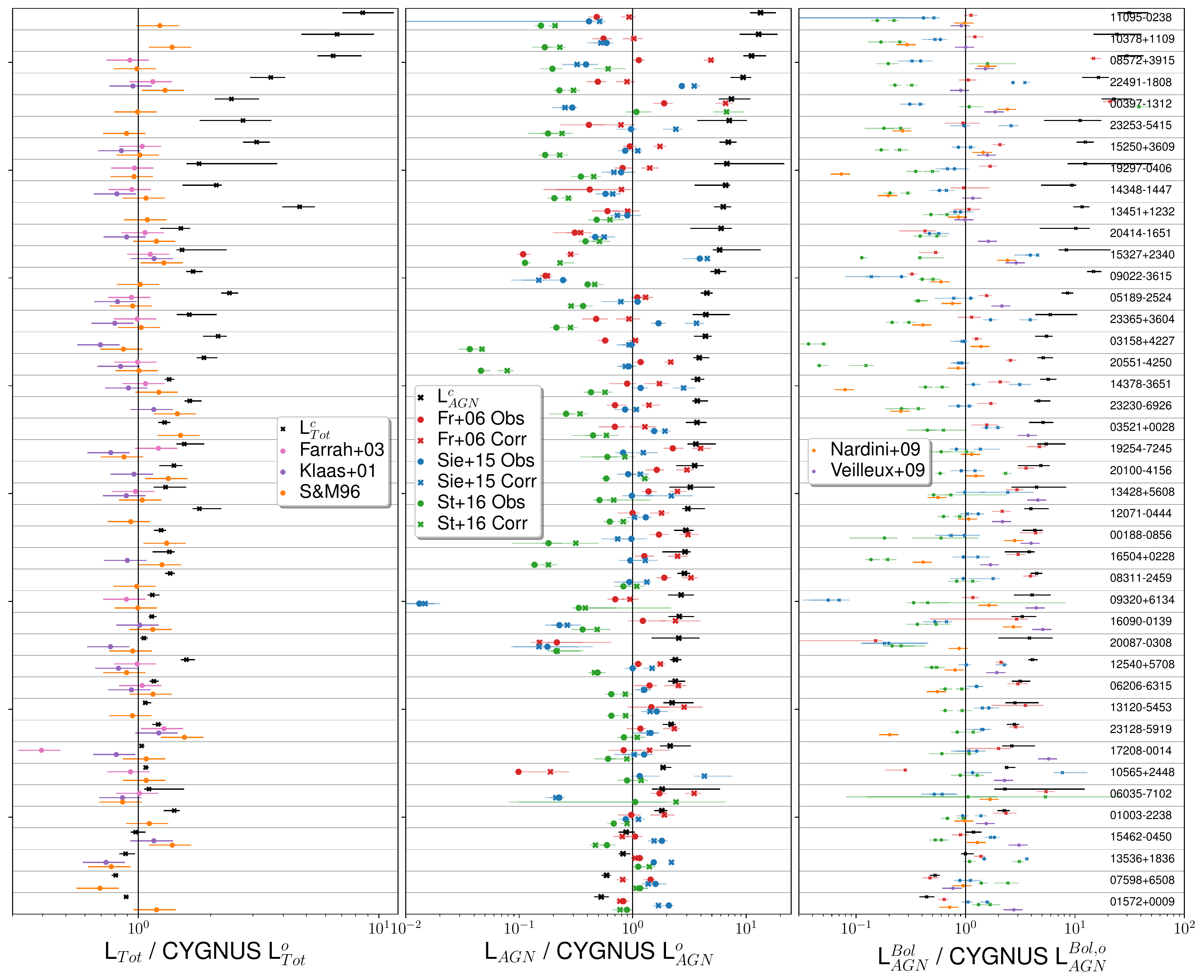}
		\caption{Comparison of luminosities. {\itshape Left:} The total infrared luminosities (observed and corrected) from the CYGNUS fits, compared against those derived by \citet{san96,klaas01,far03}. The values are presented as the ratios to the observed CYGNUS luminosities. {\itshape Middle:} The AGN infrared luminosities (observed and corrected) from the CYGNUS fits, compared against those from the FR06, SKIRTOR { and Siebenmorgen15} model fits. The values are presented as the ratios to the observed CYGNUS luminosities. {\itshape Right:} The corrected AGN bolometric luminosities from the CYGNUS fits, compared to those from the FR06 and SKIRTOR fits, and to those derived by \citet{nard09} and \citet{vei09}.}
\label{fig:lum_comp2}
\end{center}
\end{figure*}

\section{Comparison with other luminosities}
In order to assess the robustness of our SED decomposition results and our estimate of the correction of the AGN luminosity due to anisotropic torus emission we compare our results to the following:

\begin{itemize}
\item Total infrared luminosities derived from the IRAS fluxes \citep{san96}, and those presented by \citet{klaas01} from fitting a modified blackbody dust model. 

\item AGN infrared luminosities from FR06, SKIRTOR and { Siebenmorgen15}.

\item AGN bolometric luminosities from FR06, SKIRTOR and { Siebenmorgen15}, as well as \citet{nard09} and \citet{vei09}.

\end{itemize}

A comparison of the likelihoods from the fits with all { four} combinations of models is given in Figure \ref{fig:lum_likes}, and of the luminosities in Figure \ref{fig:lum_comp2}. It is clear that the CYGNUS models nearly always give better fits than the FR06 and SKIRTOR models, and usually better fits than the Siebenmorgen15 models. We discuss possible reasons for this and their effect on the predicted AGN luminosities in section 8.

The uncorrected total infrared luminosities from CYGNUS are consistent with prior estimates \citep{san96} though slightly higher that those based on pure modified blackbody dust models, likely due to lack of PAH emission in these models \citep{klaas01}.

The infrared AGN luminosities from CYGNUS and FR06 are in general consistent, though with significant dispersion. The infrared AGN luminosities from SKIRTOR and Siebenmorgen15 are however systematically lower than those from either FR06 or CYGNUS. As we also discuss in section 8 this may be related to the fact that the SKIRTOR fits are generally poorer than those with CYGNUS. As the same effect is also observed with the Siebenmorgen15 models, this is most probably also related to the two-phase geometry which is the common characteristic of SKIRTOR and Siebenmorgen15.

With all combinations of models significant correction of the AGN luminosities due to anisotropic AGN emission is needed. In the most extreme objects, IRAS~00397-1312 and IRAS~08572+3915 the corrected luminosities predicted by the CYGNUS, FR06 and SKIRTOR combinations agree very well. This is not the case for the Siebenmorgen15 models but the fits with this combination are not very good. The CYGNUS corrections are usually but not always the highest. The SKIRTOR and Siebenmorgen15 corrections do not bring their luminosities in line with either CYGNUS or FR06.

The bolometric AGN luminosities from CYGNUS and FR06 are also in general consistent, and usually higher than those from SKIRTOR and Siebenmorgen15. All are higher than the bolometric AGN luminosities from \citet{nard09} though their bolometric luminosities are in reality infrared luminosities, so this is expected. The comparison with the \citet{vei09} bolometric luminosities is more interesting. The \citet{vei09} luminosities are not based on SED fits, but instead are averages of five different approaches. The CYGNUS uncorrected luminosities are systematically lower than those from \citet{vei09}  but the corrected ones are often consistent. We conclude that empirical calibrations such as those in \citet{vei09} do capture some of the anisotropy correction.

\begin{figure}
	\begin{center}
		\includegraphics[width=8cm, angle=0]{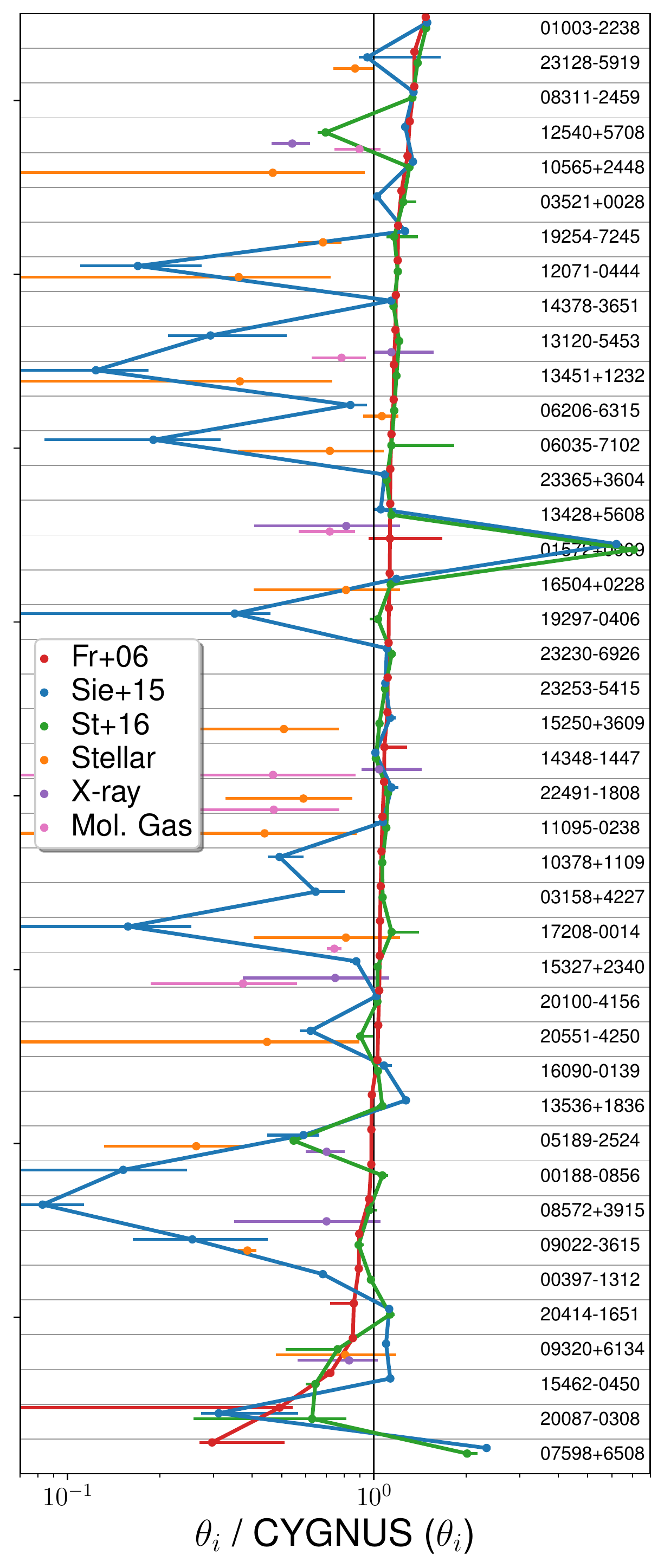}
		\caption{Comparison of inclination angles derived with various methods. The values are given as the ratio to the CYGNUS inclination angle. Inclination angles for the X-ray obscurer are taken from \citet{feru15,oda17,xu17,yamada21}. Stellar disk inclination angles are from \citet{das06,rou07,bel13,med14}. 
		Molecular or atomic gas disk inclination angles are taken from \citet{car98,car00,sco17,pri17,per18}.}
\label{fig:comp_angle}
\end{center}
\end{figure}

\section{Comparison with other inclination angles}
The inclination angle of the AGN obscurer relative to the line of sight plays an important role in determining AGN luminosities. We therefore compare the inclination angles inferred from the CYGNUS models to those derived from other approaches. 

It is possible to observationally infer inclination angles of several galactic components - the molecular or atomic gas disk, the X-ray obscurer, the infrared obscurer, and the stellar disk. These estimates are challenging and usually carry large uncertainties. Few such measures are available, nevertheless, such comparisons can be illuminating, so we plot the { values derived by all combinations of models} with all available literature values in  Figure \ref{fig:comp_angle}. 

There is no {\itshape a priori} reason to expect alignment of different components, but it is reasonable to expect some level of systematic consistency. In general however there is consistency between the CYGNUS-derived inclination angles and those obtained from other methods. The CYGNUS values may be slightly biased towards edge-on compared to the stellar disk derived values, but in most cases there is consistency well within $2\sigma$. Moreover, as discussed by \citet{das06}, stellar disk inclination angles carry substantial uncertainty, so it is plausible that the bias is in these measures rather than the CYGNUS values. Compared to the molecular disk values there is consistency, except in one case: the molecular gas disk in the appropriate nucleus of Arp~220 \citep{sco17}. It is inclined at $30\degr$ while the  CYGNUS value is $80.2\degr$ (both from pole-on). \citet{sco17} note however (their \S5.2) that their inclination angle is inconsistent with the supernova remnant distribution in \citet{lon06b}. Moreover, the inclination angle inferred by \citet{lon06b} is consistent with the CYGNUS value. Resolving this contradiction is beyond the scope of this paper, though it does suggest that the CYGNUS value is plausible. Finally, compared to the X-ray derived values there is consistency in one case but not in another. It is (perhaps) unreasonable to expect consistency here, though consistency between CYGNUS and X-ray derived inclination angles has been noted previously \citep{far16}. We here simply note the result as a starting point for further work. 

\section{Discussion}
As we can see in Figure 17, the fits with SKIRTOR and FR06 are consistently worse than with CYGNUS or Siebenmorgen15. SKIRTOR is a two-phase torus model whereas CYGNUS and FR06 are both smooth models. There are also differences in the assumed `shape' of the torus. Both SKIRTOR and FR06 assume a flared disc geometry whereas in CYGNUS the torus has the tapered disc geometry suggested by \citet{efstathiour95}. This may be the reason why CYGNUS is giving overall better fits than FR06. So overall the conclusion we can draw from our analysis is that a smooth tapered disc is the best approximation for the distribution of dust in the torus among the four distributions considered by the models.

The fundamental difference between a smooth and clumpy or two-phase geometry for the torus is that in the latter case we can see through gaps in the cloud distribution to the inner torus. This has the effect of making the emission from a clumpy torus be generally more isotropic and appear less obscured. A clumpy or two-phase torus also can not produce deep enough silicate absorption features in order to fit well most of the galaxies in our sample. In cases where the SKIRTOR fit is poor, such as for example in IRAS~11095-0238 or IRAS~03158+4227, this will usually manifest itself with a lower AGN luminosity as the model under predicts the emission at 20-30$\mu m$ where the torus emission peaks. 

The conclusion that a smooth tapered disc is a better approximation than the SKIRTOR Siebenmorgen15 geometry is not necessarily at odds with the detections of compact tori in nearby Seyferts like NGC~1068 with ALMA \citep{gar16,alon18,combs19}. It is important to note that even in NGC~1068 the compact torus of a few parsec diameter is surrounded by a circumnuclear disc of diameter of the order of 100 parsec which dominates the far-infrared and submillimetre emission. \citet{lyu21} also recently presented results from reverberation mapping of the nucleus of NGC~4151 at 1-40$\mu m$ which shows lack of variability at 20-40$\mu m$. This implies the 20-40$\mu m$ emitting region is more extended than that predicted by clumpy torus models and more consistent with smooth models. Because of the limited spatial resolution of the infrared data in this sample what we are modeling with the AGN torus models are structures similar to the circumnuclear disc in NGC~1068 or the extended structure in NGC~4151 inferred by \citet{lyu21}. This may explain why the smooth torus models provide better fits to the SEDs of the ULIRGs in our sample. 

It is also interesting to consider if the results presented here are consistent regarding the ratio of predicted type 1 and type 2 AGN. We have a total of 45 AGN in this sample (including the 2 type 1 AGN in the dual cases and 3C~273). Out of these, 5 are type 1 (IRAS~07598+6508, Mrk~231, 3C~273, and the 2 dual AGN). So we have a minimum type 1 fraction of 5/45=11.1\%. Assuming all the tori in these AGN have $\theta_{o} = 30\degr$ then we should expect a type 1 fraction of $\sim $13\%. So we conclude that our results are consistent.    

With the more detailed analysis presented in this paper, which includes the addition of the emission of the host galaxy, we confirm the result obtained by \citet{efs14} for IRAS~08572+3915 namely that it is intrinsically a hyperluminous infrared galaxy. The predicted higher luminosity is due to the anisotropic emission of the torus. The only way to avoid this conclusion is if the AGN torus were more `spherical' than predicted in our model but more spherical models approaching a torus covering factor of ~90\% are included in our AGN torus libraries. There is also clear evidence for large outflows in this ULIRG which suggest a non-spherical geometry \citep{spo13,rup13,gonz17}. Attempts to fit the SED of IRAS~08572+3915 with a foreground screen model have also not been successful \citep{levenson07}. IRAS~08572+3915 is one of the first ULIRGs which will be targets of JWST so we will be able to elucidate further the nature of this interesting system in the near future. We also find that there are three other ULIRGs which are predicted to be hyperluminous. These are IRAS~00397-1312, IRAS~10378+1109 and IRAS~11095-0238 which show very similar SEDs to IRAS~08572+3915. IRAS~00397-1312 belongs to class 3A of
\citet{spo07} which are associated with a deep silicate absorption feature and weak PAH emission. Another galaxy that belongs to class 3A of \citet{spo07} is IRAS~00188-0856. This ULIRG is also fitted with a similar model as the other objects in this class but is not predicted to be hyperluminous.

In Arp~220 we find evidence for an AGN which may be consistent with the prediction of \cite{gas19} for the presence of a luminous AGN with an infrared luminosity of $0.76-1.2 \times 10^{12} L_\odot$ in the western nucleus. \cite{sco17} also identified a compact Keplerian component in the western nucleus of Arp~220.  We also note that three ULIRGs have significant contribution in the submillimetre from cold dust in the host galaxy component. In two of the cases, IRAS~19254-7245 (SuperAntena) and IRAS~23253-5415 cold dust dominates the submillimetre emission. These two galaxies are known to have prominent tidal tails extending for tens of kpc. IRAS~13451+123 is a radio galaxy so there may be contribution from synchrotron emission in the submillimetre, which is not included in our models. These three galaxies would be interesting to study with ALMA, NOEMA or SMA.

\section{Conclusions}

We have presented a detailed analysis of the SEDs of the HERUS local ULIRGs using radiative transfer models and an MCMC SED fitting code.

\begin{enumerate}

\item We fit the SEDs of the ULIRGs in our sample with four combinations of radiative transfer models. Two of the combinations assume a smooth torus geometry \citep{efstathiour95,fritz06} and two combinations a two-phase geometry \citep{sie15,stal16}. We find the smooth CYGNUS models \citep{efstathiour95} provide better fits than the other combinations of models.

\item All objects harbor high rates of star formation. There is a strong correlation between $L_{Sb}$ and $\dot{M}_{Sb}^{age}$ and more luminous starbursts are unlikely to be systematically different in either age or extinction to less luminous starbursts.

\item We find bolometrically significant AGN in all objects. The obscurer geometry can substantially affect the observed AGN luminosity, by factors of up to $\sim 10$ in tori viewed nearly edge-on. This is due to the anisotropic nature of the emission of the torus which is to a large extent independent of the torus model. The correction may depend on luminosity; more infrared-luminous AGN in ULIRGs seem to be associated with both smaller covering fractions {\itshape and} viewing angles closer to equatorial. It is also consistent with the general class of `receding torus' models.

\item We see no relation between AGN luminosity and either $r_{2}/r_{1}$, or $\tau_{uv}$. Moreover, neither $r_{2}/r_{1}$ or $\tau_{uv}$ appear to depend on $\theta_{o}$. This is consistent with only the covering fraction having a relation with $L^{c}_{AGN}$, but other properties of the obscurer being independent of luminosity and covering fraction.

\item We find an observed dual AGN fraction of about 5\%. Due to observational biases this is likely a lower limit.

\item We find evidence for significant amounts of polar dust in three ULIRGs or about 8\% of the sample. This may be considered as a lower limit as in other objects polar dust emission may be obscured due to the large optical depths of ULIRGs.

\end{enumerate}

\section*{Acknowledgements}

We would like to thank the referee Ralf Siebenmorgen for his comments and suggestions which led to an improvement of the paper. The work leading to this paper has received funding from the European Union Seventh Framework Programme FP7/2007-2013/ under grant agreement no.607254. This publication reflects only the authors' view and the European Union is not responsible for any use that may be made of the information contained therein. AE, DF and VPL acknowledge support from the project EXCELLENCE/1216/0207/ GRATOS funded by the Cyprus Research \& Innovation Foundation. AE and VPL acknowledge support from the project CYGNUS funded by the European Space Agency. DR acknowledges support from STFC through grant ST/S000488/1. E.G-A is a Research Associate at the Harvard-Smithsonian Center for Astrophysics and thanks the Spanish Ministerio de Econom\'{\i}a y Competitividad for support under projects ESP2017-86582-C4-1-R and PID2019-105552RB-C41.

\section*{Data Availability Statement}

The data underlying this article are available in the article or are publicly available in databases like The Cornell Atlas of Spitzer/Infrared Spectrograph Sources (CASSIS) and the NASA Extragalactic Database (NED).

\clearpage




\appendix

\section*{Appendix A: Individual SED fits with the CYGNUS models}
The following figures present SED fits using the CYGNUS AGN torus models for all the objects in the sample.

\begin{figure*}
\centering
\includegraphics[width=16cm]{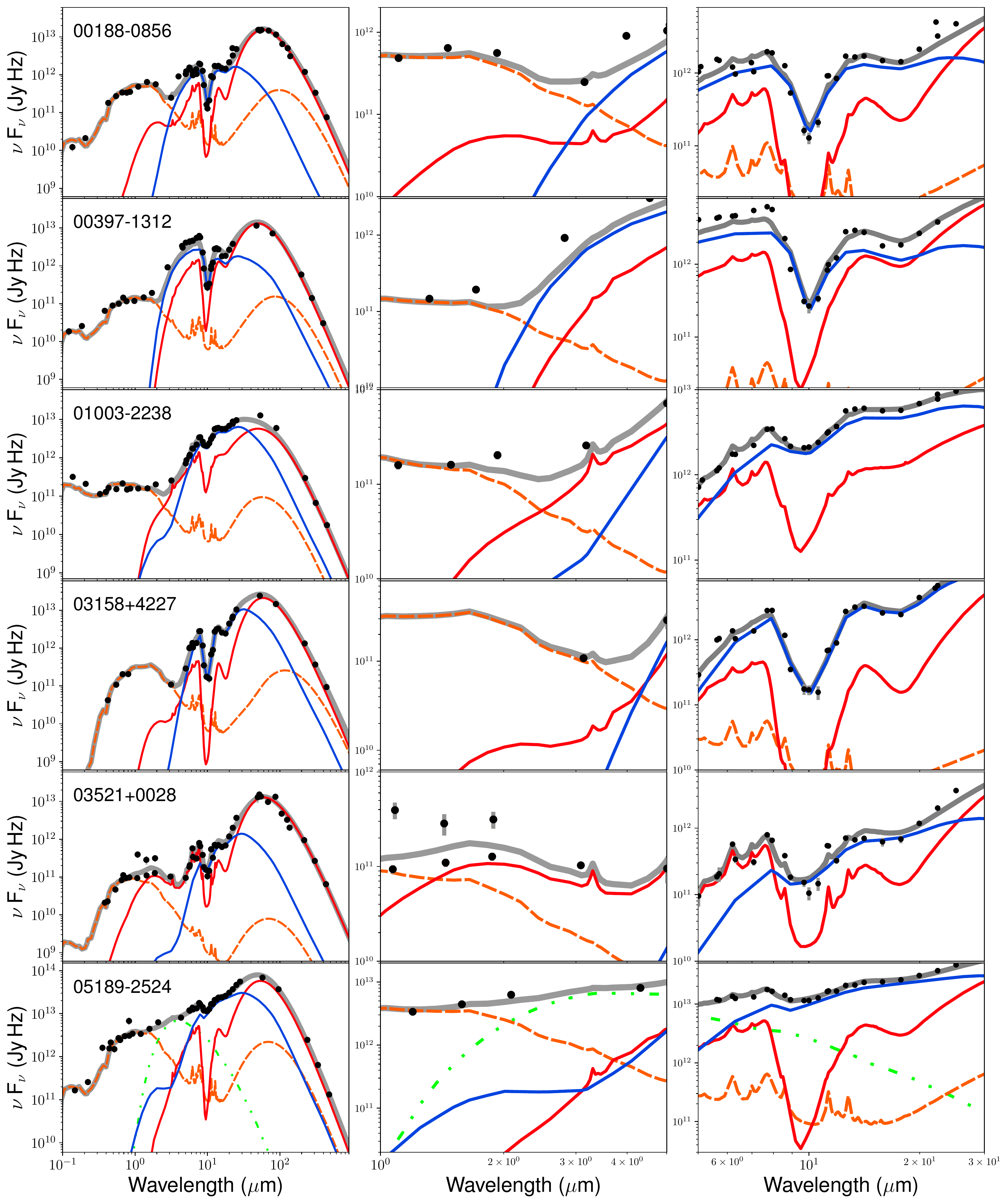}
        \par
		{{\bf Figure A1} SED fit plots of all objects, using the CYGNUS models: spheroidal host (orange), starburst (red), AGN torus (blue), polar dust (green) and total (grey). These fits assume a single AGN.}
\label{fig:seds1}
\end{figure*}

\begin{figure*}
\centering
\includegraphics[width=16cm]{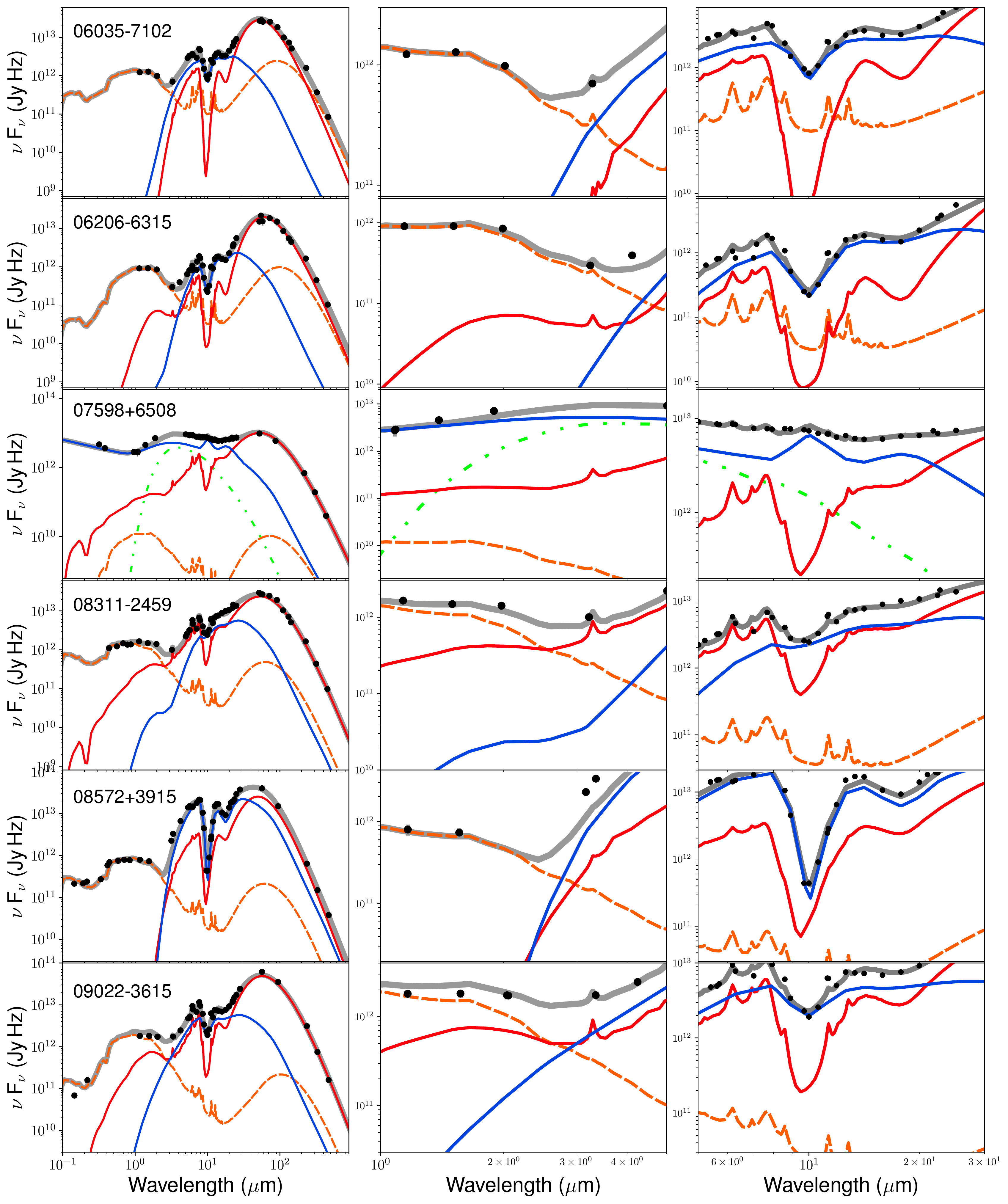}
        \par
		{{\bf Figure A2} SED fit plots of all objects, using the CYGNUS models.}
\label{fig:seds2}
\end{figure*}

\begin{figure*}
\centering
\includegraphics[width=16cm]{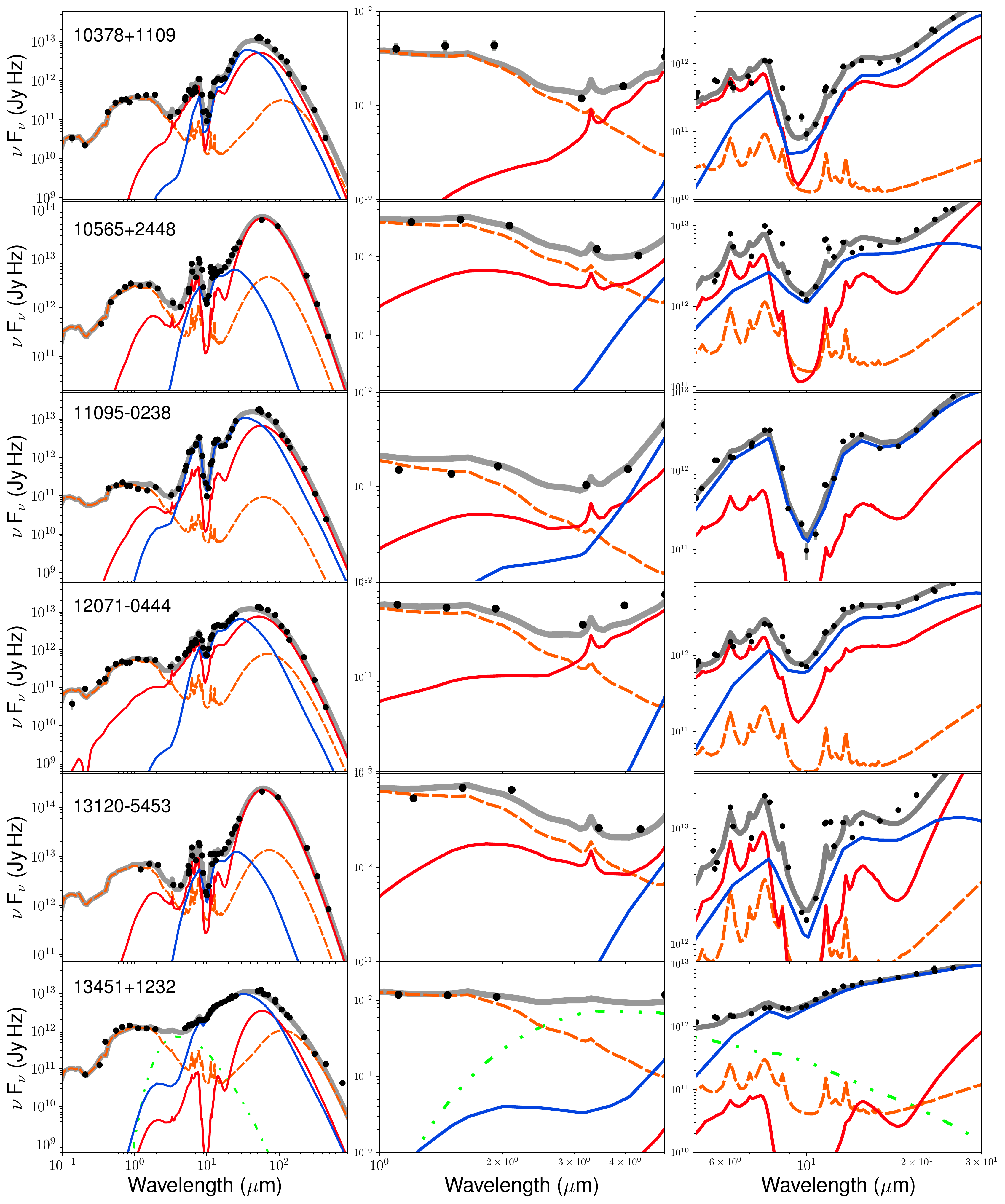}
        \par
		{{\bf Figure A3} SED fit plots of all objects, using the CYGNUS models.}
\label{fig:seds3}
\end{figure*}

\begin{figure*}
\centering
\includegraphics[width=16cm]{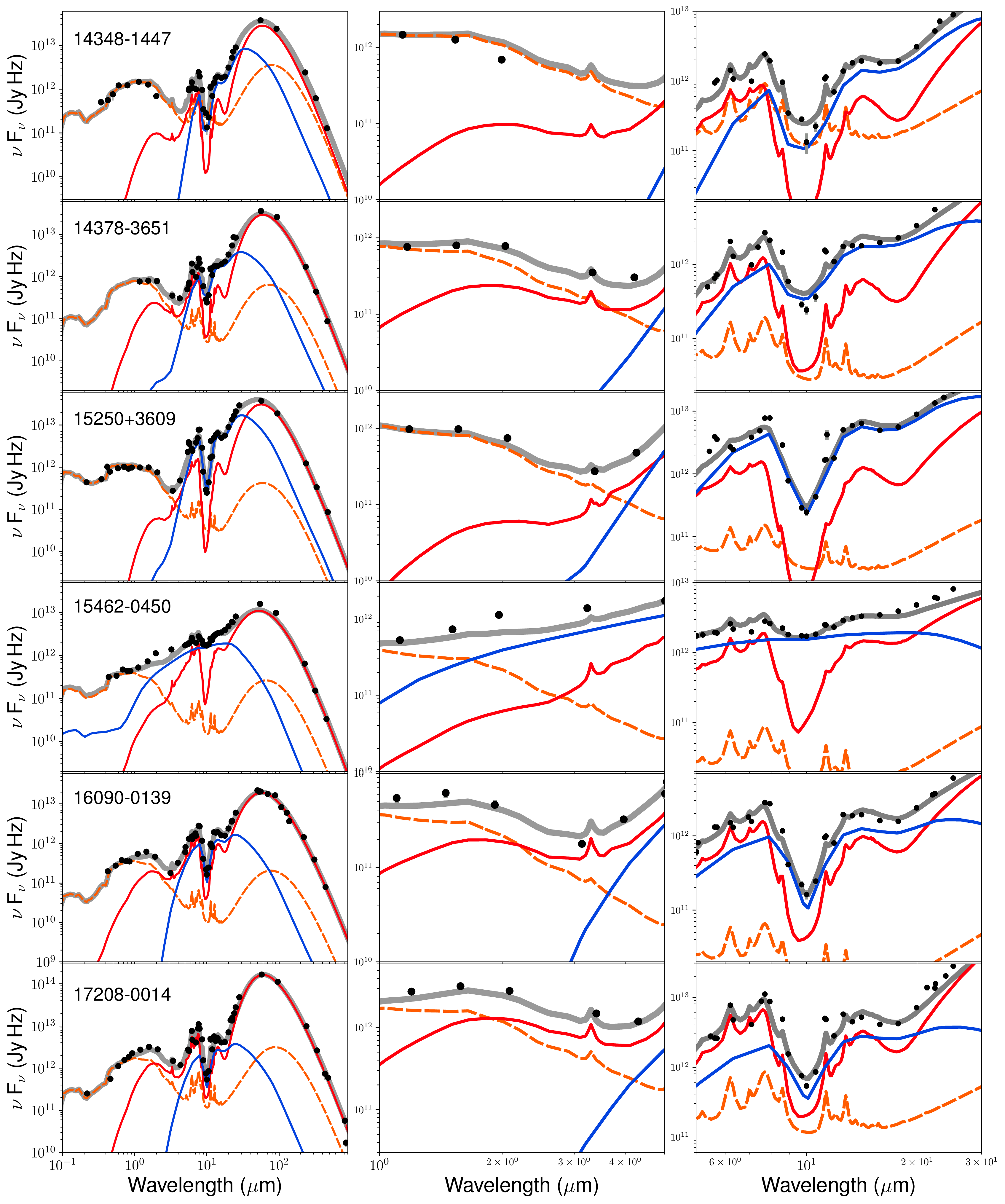}
        \par
		{{\bf Figure A4} SED fit plots of all objects, using the CYGNUS models.}
\label{fig:seds4}
\end{figure*}

\begin{figure*}
\centering
\includegraphics[width=16cm]{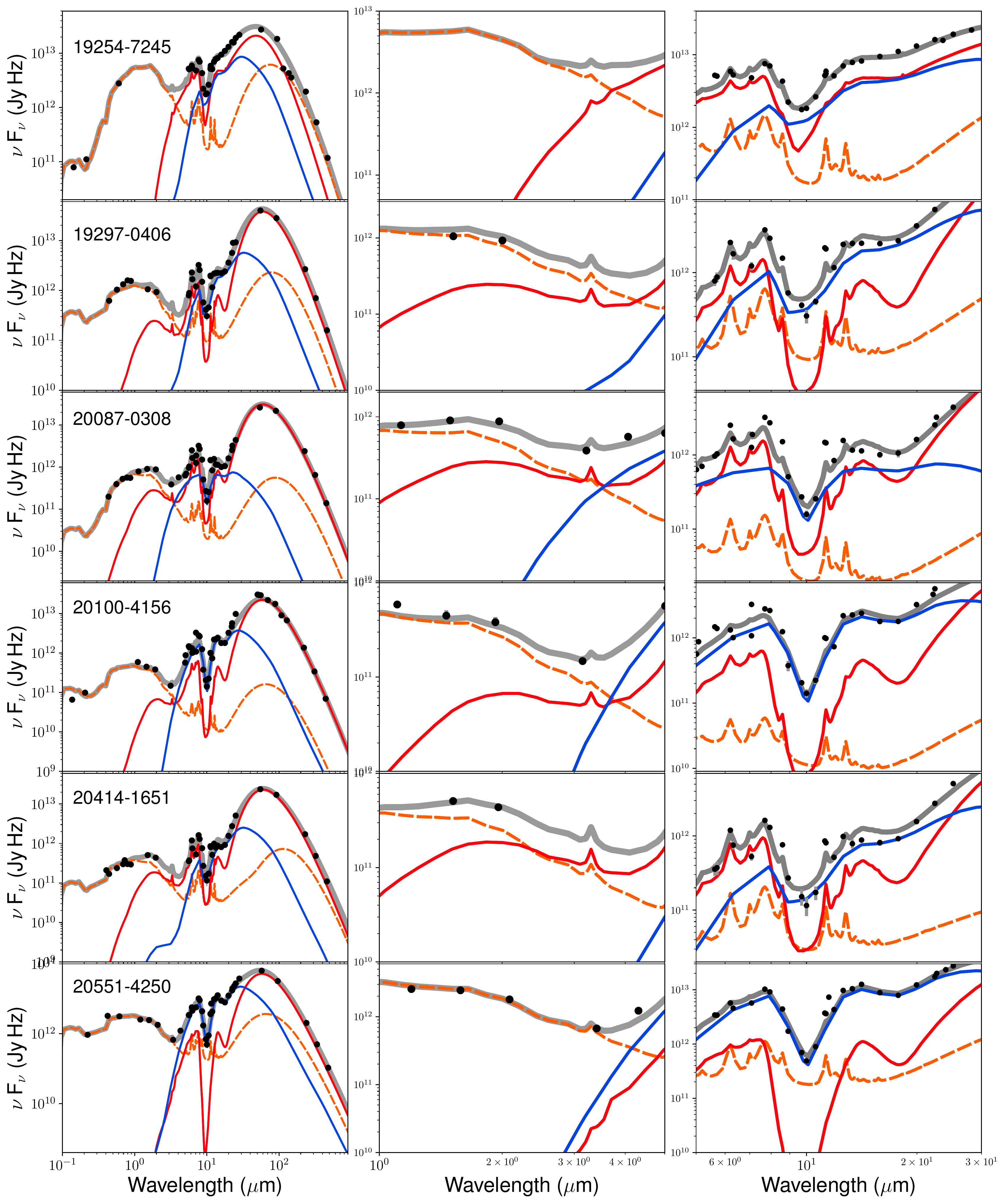}
        \par
		{{\bf Figure A5} SED fit plots of all objects, using the CYGNUS models.}
\label{fig:seds5}
\end{figure*}

\begin{figure*}
\centering
\includegraphics[width=16cm]{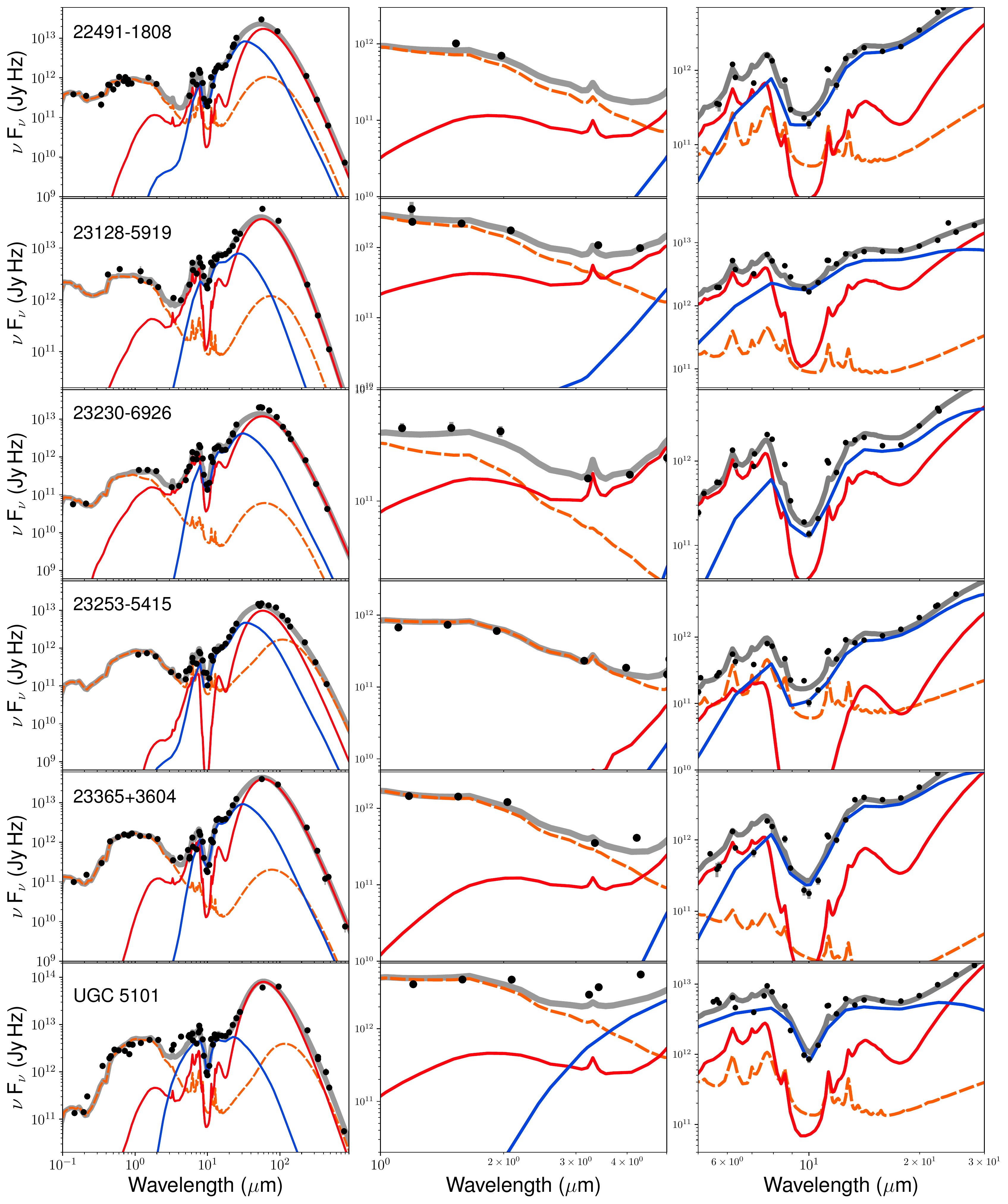}
        \par
		{{\bf Figure A6} SED fit plots of all objects, using the CYGNUS models.}
\label{fig:seds6}
\end{figure*}

\begin{figure*}
\centering
\includegraphics[width=16cm]{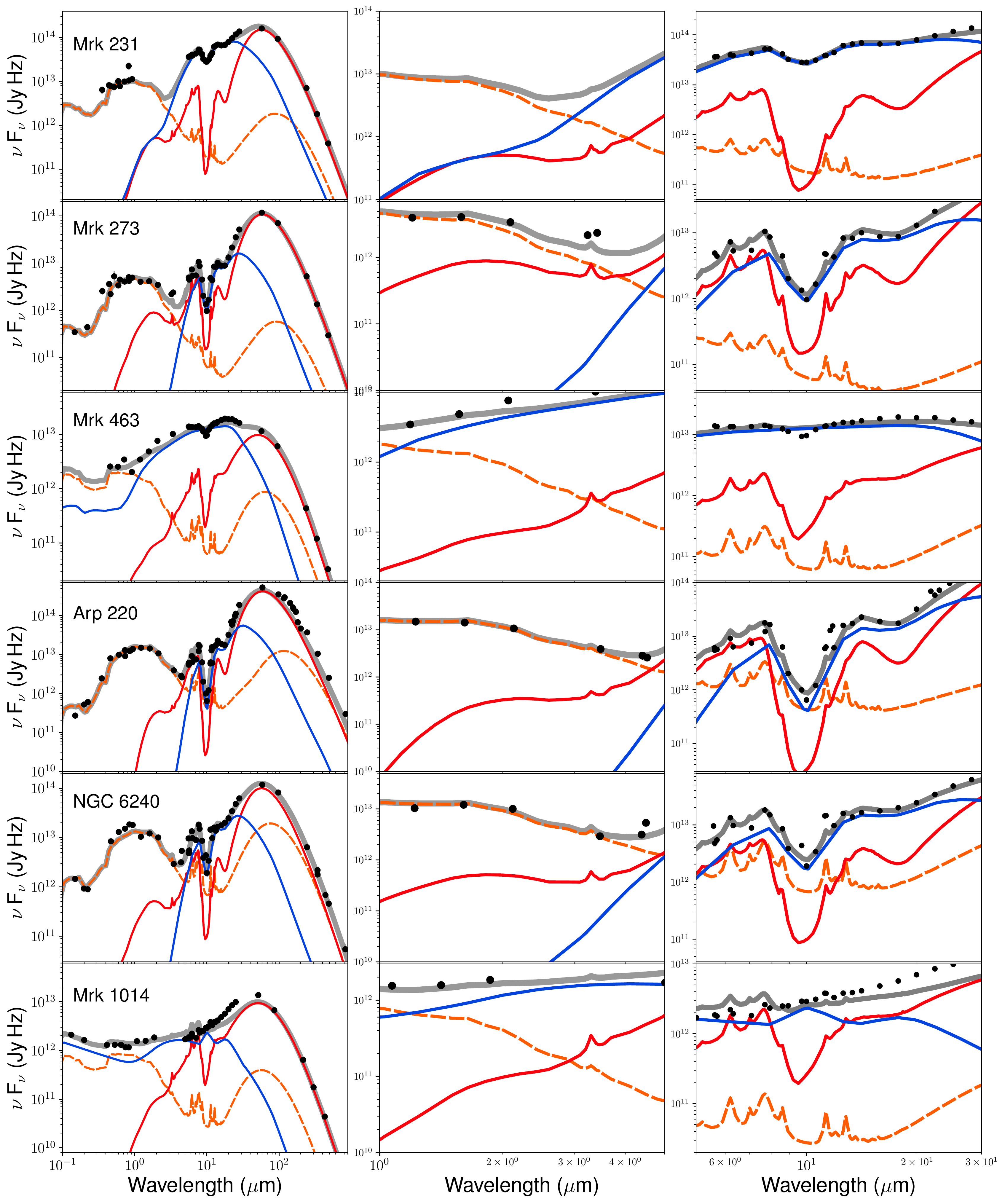}
        \par
		{{\bf Figure A7} SED fit plots of all objects, using the CYGNUS models.}
\label{fig:seds7}
\end{figure*}

\appendix

\section*{Appendix B: Ancillary Data}
The following tables present the full suite of luminosities together with luminosities from the literature. Where appropriate the luminosities have been scaled to our adopted cosmology.

\begin{table*}
\label{tab:bololums}
\begin{center}
\begin{tabular}{lcccccccccc}
\hline
   & \multicolumn{4}{c}{Other Luminosities}              & \multicolumn{6}{c}{CYGNUS Bolometric Luminosities}   \\
ID & L$_{Tot}$ & L$_{Tot}$ & L$_{AGN}$ & L$_{AGN}^{Bol}$ &  L$_{AGN}^{o}$        & L$_{AGN}^{c}$        & L$_{Sb}$             & L$_{host}$           & L$_{Tot}^{o}$        & L$_{Tot}^{c}$             \\ 
   & SM96      & K01       & N09       & V09             &                       &                      &                      &                      &                      &                       \\ 	

   & \multicolumn{4}{c}{$10^{12}$L$_{\odot}$}            & \multicolumn{3}{c}{$10^{12}$L$_{\odot}$}                            & $10^{11}$L$_{\odot}$ & \multicolumn{2}{c}{$10^{12}$L$_{\odot}$}   \\
\hline	
1 & $2.95$& --- & $0.79$& $1.11$& $0.28^{+0.03}_{-0.05}$ & $1.21^{+0.22}_{-0.27}$ & $1.86^{+0.03}_{-0.02}$ & $1.43^{+0.16}_{-0.27}$ & $2.29^{+0.05}_{-0.06}$ & $3.21^{+0.22}_{-0.29 }$\\ 
2 & $9.33$& --- & $5.08$& $3.92$& $2.10^{+0.55}_{-0.31}$ & $47.79^{+22.09}_{-10.94}$ & $7.14^{+0.18}_{-0.35}$ & $2.13^{+0.56}_{-0.60}$ & $9.46^{+0.34}_{-0.25}$ & $55.15^{+21.78}_{-10.77 }$\\ 
3 & $1.78$& --- & $0.80$& $1.26$& $0.81^{+0.22}_{-0.09}$ & $1.83^{+0.22}_{-0.25}$ & $0.77^{+0.02}_{-0.12}$ & $0.58^{+0.13}_{-0.08}$ & $1.63^{+0.12}_{-0.08}$ & $2.65^{+0.18}_{-0.27 }$\\ 
4 & $3.55$& $2.85$& $1.92$& $-0.00$& $1.39^{+0.18}_{-0.23}$ & $7.62^{+1.08}_{-1.66}$ & $2.63^{+0.10}_{-0.06}$ & $0.97^{+0.11}_{-0.11}$ & $4.11^{+0.13}_{-0.15}$ & $10.34^{+1.02}_{-1.57 }$\\ 
5 & $3.63$& --- & $0.00$& $0.96$& $0.26^{+0.04}_{-0.03}$ & $1.31^{+0.33}_{-0.34}$ & $2.16^{+0.03}_{-0.02}$ & $0.22^{+0.04}_{-0.02}$ & $2.44^{+0.03}_{-0.03}$ & $3.49^{+0.31}_{-0.33 }$\\ 
6 & $1.32$& $1.14$& $0.42$& $1.18$& $0.55^{+0.08}_{-0.06}$ & $4.71^{+0.60}_{-0.58}$ & $0.77^{+0.07}_{-0.08}$ & $1.00^{+0.34}_{-0.21}$ & $1.42^{+0.07}_{-0.05}$ & $5.58^{+0.58}_{-0.53 }$\\ 
7 & $1.38$& $1.38$& $0.35$& $-0.00$& $0.21^{+0.09}_{-0.03}$ & $0.48^{+2.11}_{-0.10}$ & $1.21^{+0.03}_{-0.09}$ & $2.41^{+0.20}_{-0.96}$ & $1.66^{+0.03}_{-0.10}$ & $1.93^{+1.98}_{-0.10 }$\\ 
8 & $1.70$& $1.38$& $0.09$& $-0.00$& $0.17^{+0.02}_{-0.02}$ & $0.54^{+0.13}_{-0.09}$ & $1.18^{+0.01}_{-0.03}$ & $1.52^{+0.30}_{-0.14}$ & $1.51^{+0.02}_{-0.03}$ & $1.88^{+0.12}_{-0.10 }$\\ 
9 & $2.88$& --- & $3.54$& $2.86$& $3.72^{+0.33}_{-0.37}$ & $1.96^{+0.21}_{-0.20}$ & $2.19^{+0.01}_{-0.06}$ & $0.05^{+0.04}_{-0.01}$ & $5.92^{+0.33}_{-0.37}$ & $4.16^{+0.21}_{-0.22 }$\\ 
10 & $2.88$& --- & --- & $-0.00$& $0.56^{+0.07}_{-0.08}$ & $2.51^{+0.32}_{-0.30}$ & $2.27^{+0.02}_{-0.06}$ & $2.40^{+0.39}_{-0.21}$ & $3.07^{+0.06}_{-0.09}$ & $5.02^{+0.30}_{-0.28 }$\\ 
11 & $1.45$& --- & $1.26$& $1.19$& $0.78^{+0.15}_{-0.11}$ & $23.39^{+9.73}_{-4.27}$ & $0.66^{+0.05}_{-0.09}$ & $0.41^{+0.11}_{-0.03}$ & $1.49^{+0.08}_{-0.05}$ & $24.09^{+9.69}_{-4.24 }$\\ 
12 & $1.74$& --- & $0.15$& $-0.00$& $0.26^{+0.03}_{-0.02}$ & $3.80^{+0.68}_{-0.48}$ & $1.42^{+0.01}_{-0.02}$ & $0.74^{+0.11}_{-0.09}$ & $1.75^{+0.03}_{-0.03}$ & $5.29^{+0.68}_{-0.48 }$\\ 
13 & $2.40$& --- & $0.24$& $0.83$& $0.83^{+0.12}_{-0.14}$ & $20.04^{+13.56}_{-7.75}$ & $0.81^{+0.12}_{-0.13}$ & $1.27^{+0.49}_{-0.38}$ & $1.77^{+0.04}_{-0.06}$ & $20.98^{+13.47}_{-7.69 }$\\ 
14 & $1.20$& --- & --- & $0.21$& $0.09^{+0.02}_{-0.01}$ & $0.22^{+0.05}_{-0.01}$ & $0.93^{+0.02}_{-0.01}$ & $1.21^{+0.16}_{-0.14}$ & $1.14^{+0.02}_{-0.01}$ & $1.27^{+0.05}_{-0.01 }$\\ 
15 & $1.91$& --- & $0.93$& $0.86$& $0.94^{+0.12}_{-0.10}$ & $29.55^{+14.51}_{-6.45}$ & $0.59^{+0.07}_{-0.10}$ & $0.38^{+0.07}_{-0.05}$ & $1.57^{+0.03}_{-0.04}$ & $30.18^{+14.42}_{-6.38 }$\\ 
16 & $2.04$& --- & $0.90$& $1.84$& $0.84^{+0.18}_{-0.10}$ & $3.34^{+1.53}_{-0.43}$ & $1.19^{+0.06}_{-0.18}$ & $2.12^{+0.58}_{-0.42}$ & $2.25^{+0.06}_{-0.09}$ & $4.74^{+1.42}_{-0.41 }$\\ 
17 & $1.66$& --- & $0.00$& $-0.00$& $0.10^{+0.02}_{-0.01}$ & $0.28^{+0.18}_{-0.05}$ & $1.52^{+0.02}_{-0.02}$ & $1.78^{+0.23}_{-0.33}$ & $1.80^{+0.02}_{-0.02}$ & $1.97^{+0.17}_{-0.06 }$\\ 
18 & $2.09$& --- & $1.15$& $1.31$& $1.33^{+0.25}_{-0.22}$ & $15.48^{+2.61}_{-2.60}$ & $0.35^{+0.10}_{-0.12}$ & $3.33^{+0.76}_{-0.74}$ & $2.01^{+0.20}_{-0.19}$ & $16.16^{+2.58}_{-2.57 }$\\ 
19 & $2.14$& $1.62$& $0.08$& $0.46$& $0.39^{+0.04}_{-0.14}$ & $3.69^{+0.35}_{-1.77}$ & $1.33^{+0.12}_{-0.01}$ & $3.28^{+0.13}_{-0.73}$ & $2.05^{+0.02}_{-0.09}$ & $5.34^{+0.33}_{-1.72 }$\\ 
20 & $1.38$& $1.03$& $0.01$& $-0.00$& $0.14^{+0.01}_{-0.01}$ & $0.81^{+0.16}_{-0.13}$ & $0.95^{+0.01}_{-0.01}$ & $0.63^{+0.14}_{-0.10}$ & $1.15^{+0.02}_{-0.02}$ & $1.83^{+0.16}_{-0.13 }$\\ 
21 & $1.07$& $0.90$& $0.54$& $0.59$& $0.37^{+0.04}_{-0.05}$ & $4.62^{+0.90}_{-0.78}$ & $0.66^{+0.03}_{-0.02}$ & $0.59^{+0.05}_{-0.07}$ & $1.09^{+0.04}_{-0.04}$ & $5.34^{+0.87}_{-0.76 }$\\ 
22 & $1.74$& $1.46$& $0.33$& $0.78$& $0.25^{+0.05}_{-0.04}$ & $0.30^{+0.06}_{-0.04}$ & $0.96^{+0.09}_{-0.04}$ & $0.74^{+0.79}_{-0.16}$ & $1.29^{+0.13}_{-0.06}$ & $1.34^{+0.14}_{-0.06 }$\\ 
23 & $3.47$& $3.06$& $0.72$& $1.34$& $0.26^{+0.07}_{-0.05}$ & $0.87^{+0.30}_{-0.18}$ & $2.69^{+0.09}_{-0.11}$ & $1.03^{+0.27}_{-0.19}$ & $3.06^{+0.06}_{-0.07}$ & $3.67^{+0.21}_{-0.13 }$\\ 
24 & $2.40$& $1.80$& $0.00$& $0.34$& $0.06^{+0.02}_{-0.01}$ & $0.16^{+0.10}_{-0.03}$ & $2.10^{+0.01}_{-0.02}$ & $0.85^{+0.33}_{-0.13}$ & $2.24^{+0.03}_{-0.01}$ & $2.34^{+0.10}_{-0.03 }$\\ 
25 & $1.15$& $1.01$& $0.32$& $-0.00$& $0.28^{+0.20}_{-0.04}$ & $1.52^{+0.77}_{-0.25}$ & $0.71^{+0.04}_{-0.14}$ & $4.02^{+1.18}_{-0.69}$ & $1.39^{+0.17}_{-0.05}$ & $2.63^{+0.73}_{-0.26 }$\\ 
26 & $2.40$& --- & $0.03$& $-0.00$& $0.34^{+0.17}_{-0.05}$ & $4.24^{+13.24}_{-1.32}$ & $1.96^{+0.07}_{-0.13}$ & $2.61^{+0.85}_{-0.59}$ & $2.57^{+0.09}_{-0.05}$ & $6.46^{+13.13}_{-1.31 }$\\ 
27 & $2.57$& $2.08$& $0.08$& $-0.00$& $0.10^{+0.03}_{-0.03}$ & $0.37^{+0.24}_{-0.18}$ & $2.52^{+0.07}_{-0.02}$ & $1.31^{+0.31}_{-0.75}$ & $2.75^{+0.08}_{-0.06}$ & $3.02^{+0.23}_{-0.17 }$\\ 
28 & $4.27$& $3.07$& $0.64$& $-0.00$& $0.52^{+0.08}_{-0.10}$ & $2.51^{+0.52}_{-0.96}$ & $2.63^{+0.06}_{-0.09}$ & $1.18^{+0.37}_{-0.20}$ & $3.26^{+0.10}_{-0.13}$ & $5.25^{+0.53}_{-0.98 }$\\ 
29 & $1.74$& $1.31$& $0.00$& $0.24$& $0.15^{+0.03}_{-0.04}$ & $1.49^{+0.51}_{-0.79}$ & $1.25^{+0.02}_{-0.02}$ & $0.85^{+0.38}_{-0.15}$ & $1.48^{+0.03}_{-0.03}$ & $2.82^{+0.50}_{-0.78 }$\\ 
30 & $1.02$& $0.86$& $0.26$& $-0.00$& $0.31^{+0.06}_{-0.03}$ & $1.59^{+0.37}_{-0.18}$ & $0.62^{+0.02}_{-0.03}$ & $1.37^{+0.30}_{-0.27}$ & $1.07^{+0.06}_{-0.03}$ & $2.35^{+0.36}_{-0.18 }$\\ 
31 & $1.51$& $1.11$& $0.00$& $0.32$& $0.36^{+0.03}_{-0.06}$ & $5.82^{+1.42}_{-1.71}$ & $0.74^{+0.04}_{-0.02}$ & $1.30^{+0.19}_{-0.28}$ & $1.22^{+0.02}_{-0.04}$ & $6.69^{+1.41}_{-1.71 }$\\ 
32 & $1.15$& $0.90$& $0.03$& $-0.00$& $0.13^{+0.01}_{-0.02}$ & $0.37^{+0.04}_{-0.06}$ & $0.56^{+0.01}_{-0.01}$ & $1.04^{+0.13}_{-0.24}$ & $0.80^{+0.03}_{-0.04}$ & $1.04^{+0.05}_{-0.07 }$\\ 
33 & $2.09$& $1.67$& $0.09$& $-0.00$& $0.34^{+0.05}_{-0.04}$ & $1.57^{+0.43}_{-0.14}$ & $1.08^{+0.01}_{-0.03}$ & $0.49^{+0.10}_{-0.09}$ & $1.47^{+0.04}_{-0.04}$ & $2.70^{+0.42}_{-0.13 }$\\ 
34 & $1.82$& --- & $0.15$& $-0.00$& $0.57^{+0.12}_{-0.17}$ & $6.40^{+3.64}_{-3.42}$ & $1.12^{+0.14}_{-0.08}$ & $4.22^{+0.88}_{-0.96}$ & $2.12^{+0.10}_{-0.10}$ & $7.94^{+3.62}_{-3.33 }$\\ 
35 & $1.48$& $1.15$& $0.11$& $-0.00$& $0.26^{+0.07}_{-0.03}$ & $1.57^{+1.21}_{-0.42}$ & $1.14^{+0.01}_{-0.04}$ & $0.77^{+0.10}_{-0.06}$ & $1.48^{+0.05}_{-0.03}$ & $2.79^{+1.18}_{-0.41 }$\\ 
36 & $1.02$& --- & $0.14$& $0.39$& $0.09^{+0.02}_{-0.01}$ & $0.36^{+0.17}_{-0.11}$ & $0.85^{+0.03}_{-0.01}$ & $1.26^{+0.18}_{-0.12}$ & $1.06^{+0.04}_{-0.02}$ & $1.33^{+0.18}_{-0.11 }$\\ 
37 & $3.09$& $2.86$& $1.18$& $2.82$& $1.46^{+0.17}_{-0.13}$ & $5.98^{+0.70}_{-0.50}$ & $1.89^{+0.14}_{-0.12}$ & $2.30^{+0.38}_{-0.27}$ & $3.59^{+0.22}_{-0.13}$ & $8.10^{+0.73}_{-0.52 }$\\ 
38 & $1.35$& $1.16$& $0.10$& $0.80$& $0.17^{+0.04}_{-0.03}$ & $0.78^{+0.67}_{-0.32}$ & $1.09^{+0.03}_{-0.02}$ & $0.73^{+0.13}_{-0.08}$ & $1.34^{+0.04}_{-0.03}$ & $1.94^{+0.68}_{-0.31 }$\\ 
39 & $0.59$& $0.56$& --- & $-0.00$& $0.52^{+0.07}_{-0.05}$ & $0.52^{+0.10}_{-0.05}$ & $0.23^{+0.02}_{-0.02}$ & $0.96^{+0.23}_{-0.46}$ & $0.84^{+0.07}_{-0.07}$ & $0.84^{+0.10}_{-0.07 }$\\ 
40 & $1.38$& $1.26$& $0.28$& $0.34$& $0.12^{+0.08}_{-0.01}$ & $0.97^{+1.49}_{-0.15}$ & $0.90^{+0.01}_{-0.03}$ & $0.83^{+0.17}_{-0.05}$ & $1.10^{+0.07}_{-0.01}$ & $1.95^{+1.48}_{-0.15 }$\\ 
41 & $0.85$& $0.61$& $0.05$& $0.21$& $0.12^{+0.02}_{-0.03}$ & $0.47^{+0.06}_{-0.19}$ & $0.42^{+0.04}_{-0.03}$ & $1.78^{+0.49}_{-0.58}$ & $0.72^{+0.03}_{-0.04}$ & $1.07^{+0.06}_{-0.20 }$\\ 
42 & $4.07$& --- & $0.92$& $3.56$& $1.29^{+0.20}_{-0.15}$ & $0.57^{+0.10}_{-0.08}$ & $2.44^{+0.07}_{-0.05}$ & $4.55^{+1.04}_{-1.39}$ & $4.18^{+0.18}_{-0.15}$ & $3.46^{+0.10}_{-0.10 }$\\ 
\hline	
\end{tabular}
\par
{{\bf Table B1} Comparison of luminosities from the literature. Also, the AGN, starburst, spheroid, and total bolometric luminosities, derived from the CYGNUS model fits. For the AGN and total luminosities both the observed and anisotropy-corrected luminosities are given.
}
\end{center}
\end{table*}

\begin{table*}
\label{tab:complums1}
\begin{center}
\begin{tabular}{lcccccccccc}
\hline
   & \multicolumn{5}{c}{Infrared Luminosities}   &  \multicolumn{5}{c}{Bolometric Luminosities}  \\         
ID & L$_{Sb}$             & L$_{AGN}^{o}$        & L$_{AGN}^{c}$            & L$_{host}$           & L$_{Tot}^{c}$        & L$_{Sb}$             & L$_{AGN}^{o}$        & L$_{AGN}^{c}$        & L$_{host}$           & L$_{Tot}^{c}$            \\
   & \multicolumn{3}{c}{$10^{12}$L$_{\odot}$}                               & $10^{11}$L$_{\odot}$ & $10^{12}$L$_{\odot}$ & \multicolumn{3}{c}{$10^{12}$L$_{\odot}$}                           & $10^{11}$L$_{\odot}$ & $10^{12}$L$_{\odot}$ \\
\hline	
1 & $ 1.53^{+0.07}_{-0.03}$ & $ 0.48^{+0.11}_{-0.09}$ & $ 0.86^{+0.21}_{-0.14}$ & $ 1.42^{+0.45}_{-0.34}$ & $ 2.54^{+0.24}_{-0.14}$ & $ 1.53^{+0.07}_{-0.03}$ & $ 0.48^{+0.11}_{-0.09}$ & $ 1.23^{+0.36}_{-0.21}$ & $ 1.76^{+0.53}_{-0.37}$ & $ 2.93^{+0.39}_{-0.18}$  \\ 
2 & $ 4.12^{+0.39}_{-0.28}$ & $ 3.99^{+0.76}_{-0.79}$ & $ 13.86^{+2.11}_{-2.60}$ & $ 3.06^{+0.61}_{-0.84}$ & $ 18.28^{+1.90}_{-2.29}$ & $ 4.12^{+0.39}_{-0.28}$ & $ 3.99^{+0.76}_{-0.79}$ & $ 43.96^{+6.28}_{-7.96}$ & $ 3.68^{+0.75}_{-0.92}$ & $ 48.45^{+6.21}_{-7.67}$  \\ 
3 & $ 0.73^{+0.04}_{-0.03}$ & $ 0.79^{+0.16}_{-0.19}$ & $ 1.55^{+0.35}_{-0.38}$ & $ 0.22^{+0.04}_{-0.05}$ & $ 2.29^{+0.32}_{-0.35}$ & $ 0.73^{+0.04}_{-0.03}$ & $ 0.79^{+0.16}_{-0.19}$ & $ 1.90^{+0.44}_{-0.47}$ & $ 0.56^{+0.11}_{-0.13}$ & $ 2.68^{+0.42}_{-0.45}$  \\ 
4 & $ 3.66^{+0.08}_{-0.01}$ & $ 0.79^{+0.07}_{-0.11}$ & $ 1.47^{+0.12}_{-0.19}$ & $ 0.62^{+0.03}_{-0.16}$ & $ 5.19^{+0.15}_{-0.18}$ & $ 3.66^{+0.08}_{-0.01}$ & $ 0.79^{+0.07}_{-0.11}$ & $ 1.74^{+0.15}_{-0.22}$ & $ 0.81^{+0.04}_{-0.17}$ & $ 5.49^{+0.18}_{-0.21}$  \\ 
5 & $ 1.88^{+0.03}_{-0.02}$ & $ 0.18^{+0.04}_{-0.05}$ & $ 0.33^{+0.08}_{-0.09}$ & $ 0.50^{+0.06}_{-0.18}$ & $ 2.26^{+0.07}_{-0.08}$ & $ 1.88^{+0.03}_{-0.02}$ & $ 0.18^{+0.04}_{-0.05}$ & $ 0.40^{+0.11}_{-0.11}$ & $ 0.63^{+0.07}_{-0.20}$ & $ 2.34^{+0.09}_{-0.10}$  \\ 
6 & $ 0.72^{+0.06}_{-0.00}$ & $ 0.61^{+0.09}_{-0.07}$ & $ 0.72^{+0.11}_{-0.09}$ & $ 0.70^{+0.04}_{-0.39}$ & $ 1.50^{+0.11}_{-0.09}$ & $ 0.72^{+0.06}_{-0.00}$ & $ 0.61^{+0.09}_{-0.07}$ & $ 0.86^{+0.14}_{-0.10}$ & $ 1.02^{+0.07}_{-0.41}$ & $ 1.68^{+0.14}_{-0.10}$  \\ 
7 & $ 1.03^{+0.17}_{-0.03}$ & $ 0.36^{+0.05}_{-0.06}$ & $ 0.73^{+0.11}_{-0.13}$ & $ 1.90^{+0.29}_{-1.06}$ & $ 1.96^{+0.13}_{-0.08}$ & $ 1.03^{+0.17}_{-0.03}$ & $ 0.36^{+0.05}_{-0.06}$ & $ 1.15^{+0.21}_{-0.23}$ & $ 2.47^{+0.26}_{-0.95}$ & $ 2.43^{+0.23}_{-0.20}$  \\ 
8 & $ 0.91^{+0.07}_{-0.02}$ & $ 0.24^{+0.04}_{-0.07}$ & $ 0.43^{+0.08}_{-0.11}$ & $ 2.27^{+0.22}_{-0.66}$ & $ 1.57^{+0.09}_{-0.10}$ & $ 0.91^{+0.07}_{-0.02}$ & $ 0.24^{+0.04}_{-0.07}$ & $ 0.51^{+0.10}_{-0.12}$ & $ 2.73^{+0.27}_{-0.73}$ & $ 1.70^{+0.13}_{-0.12}$  \\ 
9 & $ 1.65^{+0.06}_{-0.05}$ & $ 2.84^{+0.29}_{-0.38}$ & $ 1.62^{+0.19}_{-0.19}$ & $ 0.00^{+0.00}_{-0.00}$ & $ 3.27^{+0.18}_{-0.20}$ & $ 1.66^{+0.06}_{-0.05}$ & $ 4.52^{+0.60}_{-0.89}$ & $ 1.76^{+0.25}_{-0.35}$ & $ 0.00^{+0.00}_{-0.00}$ & $ 3.42^{+0.25}_{-0.36}$  \\ 
10 & $ 1.42^{+0.12}_{-0.14}$ & $ 1.07^{+0.16}_{-0.15}$ & $ 1.83^{+0.27}_{-0.29}$ & $ 3.12^{+1.34}_{-1.38}$ & $ 3.57^{+0.26}_{-0.30}$ & $ 1.42^{+0.12}_{-0.14}$ & $ 1.07^{+0.16}_{-0.15}$ & $ 2.20^{+0.32}_{-0.34}$ & $ 4.32^{+1.32}_{-1.59}$ & $ 4.06^{+0.31}_{-0.36}$  \\ 
11 & $ 0.67^{+0.06}_{-0.13}$ & $ 0.89^{+0.12}_{-0.09}$ & $ 3.84^{+0.32}_{-0.52}$ & $ 0.19^{+0.01}_{-0.03}$ & $ 4.54^{+0.24}_{-0.50}$ & $ 0.67^{+0.06}_{-0.13}$ & $ 0.89^{+0.12}_{-0.09}$ & $ 11.58^{+0.85}_{-1.92}$ & $ 0.41^{+0.02}_{-0.06}$ & $ 12.29^{+0.77}_{-1.92}$  \\ 
12 & $ 1.56^{+0.03}_{-0.02}$ & $ 0.04^{+0.00}_{-0.01}$ & $ 0.05^{+0.00}_{-0.01}$ & $ 0.00^{+0.00}_{-0.00}$ & $ 1.61^{+0.03}_{-0.02}$ & $ 1.57^{+0.03}_{-0.02}$ & $ 0.05^{+0.01}_{-0.01}$ & $ 0.08^{+0.01}_{-0.01}$ & $ 0.00^{+0.00}_{-0.00}$ & $ 1.66^{+0.03}_{-0.02}$  \\ 
13 & $ 1.07^{+0.13}_{-0.03}$ & $ 0.46^{+0.09}_{-0.09}$ & $ 0.85^{+0.17}_{-0.17}$ & $ 1.65^{+0.34}_{-0.88}$ & $ 2.09^{+0.17}_{-0.16}$ & $ 1.07^{+0.13}_{-0.03}$ & $ 0.46^{+0.09}_{-0.09}$ & $ 1.01^{+0.20}_{-0.20}$ & $ 2.00^{+0.37}_{-0.87}$ & $ 2.29^{+0.21}_{-0.19}$  \\ 
14 & $ 0.65^{+0.01}_{-0.02}$ & $ 0.01^{+0.01}_{-0.00}$ & $ 0.02^{+0.01}_{-0.00}$ & $ 0.53^{+0.13}_{-0.07}$ & $ 0.72^{+0.01}_{-0.01}$ & $ 0.65^{+0.01}_{-0.02}$ & $ 0.01^{+0.01}_{-0.00}$ & $ 0.03^{+0.01}_{-0.00}$ & $ 0.72^{+0.17}_{-0.08}$ & $ 0.75^{+0.02}_{-0.01}$  \\ 
15 & $ 1.10^{+0.02}_{-0.02}$ & $ 0.46^{+0.06}_{-0.07}$ & $ 0.88^{+0.12}_{-0.13}$ & $ 0.17^{+0.02}_{-0.04}$ & $ 1.99^{+0.10}_{-0.11}$ & $ 1.10^{+0.02}_{-0.02}$ & $ 0.46^{+0.06}_{-0.07}$ & $ 1.06^{+0.15}_{-0.16}$ & $ 0.42^{+0.05}_{-0.10}$ & $ 2.20^{+0.13}_{-0.14}$  \\ 
16 & $ 1.00^{+0.13}_{-0.11}$ & $ 0.84^{+0.13}_{-0.22}$ & $ 1.52^{+0.27}_{-0.31}$ & $ 1.56^{+1.59}_{-1.10}$ & $ 2.67^{+0.31}_{-0.27}$ & $ 1.00^{+0.13}_{-0.11}$ & $ 0.84^{+0.13}_{-0.22}$ & $ 1.82^{+0.34}_{-0.35}$ & $ 2.09^{+1.73}_{-1.12}$ & $ 3.03^{+0.40}_{-0.32}$  \\ 
17 & $ 1.12^{+0.07}_{-0.04}$ & $ 0.14^{+0.06}_{-0.05}$ & $ 0.28^{+0.13}_{-0.12}$ & $ 1.56^{+0.54}_{-0.95}$ & $ 1.56^{+0.14}_{-0.14}$ & $ 1.12^{+0.07}_{-0.04}$ & $ 0.14^{+0.06}_{-0.05}$ & $ 0.35^{+0.18}_{-0.16}$ & $ 1.87^{+0.64}_{-1.07}$ & $ 1.66^{+0.20}_{-0.18}$  \\ 
18 & $ 0.75^{+0.06}_{-0.06}$ & $ 0.80^{+0.22}_{-0.21}$ & $ 1.21^{+0.32}_{-0.32}$ & $ 2.44^{+0.98}_{-0.50}$ & $ 2.19^{+0.38}_{-0.29}$ & $ 0.75^{+0.06}_{-0.06}$ & $ 0.80^{+0.22}_{-0.22}$ & $ 1.43^{+0.40}_{-0.37}$ & $ 3.36^{+1.23}_{-0.64}$ & $ 2.51^{+0.48}_{-0.36}$  \\ 
19 & $ 1.65^{+0.08}_{-0.02}$ & $ 0.16^{+0.04}_{-0.10}$ & $ 0.31^{+0.07}_{-0.23}$ & $ 1.06^{+0.27}_{-0.19}$ & $ 2.07^{+0.08}_{-0.17}$ & $ 1.65^{+0.08}_{-0.02}$ & $ 0.16^{+0.04}_{-0.10}$ & $ 0.38^{+0.09}_{-0.27}$ & $ 1.62^{+0.32}_{-0.27}$ & $ 2.19^{+0.10}_{-0.22}$  \\ 
20 & $ 0.80^{+0.03}_{-0.02}$ & $ 0.13^{+0.03}_{-0.04}$ & $ 0.25^{+0.05}_{-0.07}$ & $ 0.75^{+0.16}_{-0.34}$ & $ 1.12^{+0.04}_{-0.06}$ & $ 0.80^{+0.03}_{-0.02}$ & $ 0.13^{+0.03}_{-0.04}$ & $ 0.30^{+0.07}_{-0.07}$ & $ 0.92^{+0.18}_{-0.35}$ & $ 1.19^{+0.05}_{-0.07}$  \\ 
21 & $ 0.68^{+0.03}_{-0.01}$ & $ 0.35^{+0.05}_{-0.05}$ & $ 0.65^{+0.09}_{-0.08}$ & $ 0.32^{+0.11}_{-0.10}$ & $ 1.36^{+0.08}_{-0.07}$ & $ 0.68^{+0.03}_{-0.01}$ & $ 0.35^{+0.05}_{-0.05}$ & $ 0.77^{+0.11}_{-0.10}$ & $ 0.63^{+0.10}_{-0.11}$ & $ 1.51^{+0.10}_{-0.08}$  \\ 
22 & $ 1.05^{+0.03}_{-0.04}$ & $ 0.26^{+0.04}_{-0.04}$ & $ 0.20^{+0.03}_{-0.03}$ & $ 0.23^{+0.05}_{-0.04}$ & $ 1.28^{+0.04}_{-0.06}$ & $ 1.05^{+0.03}_{-0.04}$ & $ 0.27^{+0.04}_{-0.04}$ & $ 0.23^{+0.04}_{-0.04}$ & $ 0.45^{+0.09}_{-0.08}$ & $ 1.33^{+0.04}_{-0.06}$  \\ 
23 & $ 2.81^{+0.15}_{-0.17}$ & $ 0.33^{+0.19}_{-0.09}$ & $ 0.63^{+0.42}_{-0.17}$ & $ 0.40^{+0.07}_{-0.09}$ & $ 3.48^{+0.28}_{-0.09}$ & $ 2.81^{+0.15}_{-0.17}$ & $ 0.33^{+0.19}_{-0.09}$ & $ 0.78^{+0.65}_{-0.21}$ & $ 0.84^{+0.12}_{-0.20}$ & $ 3.67^{+0.50}_{-0.12}$  \\ 
24 & $ 2.33^{+0.01}_{-0.01}$ & $ 0.05^{+0.02}_{-0.01}$ & $ 0.08^{+0.04}_{-0.02}$ & $ 0.60^{+0.09}_{-0.15}$ & $ 2.47^{+0.04}_{-0.02}$ & $ 2.33^{+0.01}_{-0.01}$ & $ 0.05^{+0.02}_{-0.01}$ & $ 0.12^{+0.06}_{-0.04}$ & $ 0.74^{+0.12}_{-0.15}$ & $ 2.53^{+0.06}_{-0.03}$  \\ 
25 & $ 0.32^{+0.07}_{-0.11}$ & $ 0.63^{+0.15}_{-0.11}$ & $ 1.10^{+0.25}_{-0.17}$ & $ 2.57^{+1.21}_{-0.80}$ & $ 1.68^{+0.26}_{-0.16}$ & $ 0.32^{+0.07}_{-0.11}$ & $ 0.63^{+0.15}_{-0.11}$ & $ 1.32^{+0.30}_{-0.20}$ & $ 3.27^{+1.41}_{-0.91}$ & $ 1.96^{+0.34}_{-0.20}$  \\ 
26 & $ 2.20^{+0.01}_{-0.01}$ & $ 0.28^{+0.06}_{-0.04}$ & $ 0.48^{+0.10}_{-0.07}$ & $ 0.49^{+0.09}_{-0.10}$ & $ 2.73^{+0.09}_{-0.07}$ & $ 2.20^{+0.01}_{-0.01}$ & $ 0.28^{+0.06}_{-0.04}$ & $ 0.57^{+0.12}_{-0.09}$ & $ 1.03^{+0.21}_{-0.26}$ & $ 2.87^{+0.12}_{-0.09}$  \\ 
27 & $ 1.65^{+0.11}_{-0.08}$ & $ 0.02^{+0.01}_{-0.00}$ & $ 0.01^{+0.05}_{-0.00}$ & $ 1.88^{+0.62}_{-0.52}$ & $ 1.85^{+0.18}_{-0.10}$ & $ 1.65^{+0.11}_{-0.08}$ & $ 0.02^{+0.01}_{-0.00}$ & $ 0.01^{+0.10}_{-0.00}$ & $ 2.25^{+0.75}_{-0.58}$ & $ 1.89^{+0.25}_{-0.10}$  \\ 
28 & $ 2.72^{+0.10}_{-0.08}$ & $ 0.84^{+0.13}_{-0.17}$ & $ 1.55^{+0.24}_{-0.30}$ & $ 2.17^{+0.64}_{-0.49}$ & $ 4.49^{+0.23}_{-0.23}$ & $ 2.72^{+0.10}_{-0.08}$ & $ 0.84^{+0.13}_{-0.17}$ & $ 1.84^{+0.29}_{-0.35}$ & $ 2.66^{+0.75}_{-0.47}$ & $ 4.83^{+0.28}_{-0.28}$  \\ 
29 & $ 1.63^{+0.03}_{-0.02}$ & $ 0.05^{+0.01}_{-0.02}$ & $ 0.05^{+0.01}_{-0.01}$ & $ 0.14^{+0.03}_{-0.02}$ & $ 1.69^{+0.03}_{-0.02}$ & $ 1.63^{+0.03}_{-0.02}$ & $ 0.05^{+0.01}_{-0.02}$ & $ 0.06^{+0.03}_{-0.02}$ & $ 0.26^{+0.06}_{-0.04}$ & $ 1.72^{+0.04}_{-0.02}$  \\ 
30 & $ 0.55^{+0.01}_{-0.03}$ & $ 0.37^{+0.03}_{-0.05}$ & $ 0.67^{+0.06}_{-0.08}$ & $ 0.36^{+0.08}_{-0.03}$ & $ 1.26^{+0.05}_{-0.09}$ & $ 0.55^{+0.01}_{-0.03}$ & $ 0.37^{+0.03}_{-0.05}$ & $ 0.80^{+0.07}_{-0.10}$ & $ 0.84^{+0.14}_{-0.08}$ & $ 1.43^{+0.07}_{-0.11}$  \\ 
31 & $ 0.73^{+0.02}_{-0.01}$ & $ 0.18^{+0.03}_{-0.04}$ & $ 0.32^{+0.05}_{-0.06}$ & $ 0.92^{+0.12}_{-0.27}$ & $ 1.14^{+0.05}_{-0.06}$ & $ 0.73^{+0.02}_{-0.01}$ & $ 0.18^{+0.03}_{-0.04}$ & $ 0.38^{+0.07}_{-0.07}$ & $ 1.41^{+0.13}_{-0.32}$ & $ 1.25^{+0.06}_{-0.07}$  \\ 
32 & $ 0.56^{+0.02}_{-0.04}$ & $ 0.15^{+0.02}_{-0.04}$ & $ 0.31^{+0.04}_{-0.06}$ & $ 0.86^{+0.38}_{-0.16}$ & $ 0.95^{+0.04}_{-0.06}$ & $ 0.56^{+0.02}_{-0.04}$ & $ 0.15^{+0.02}_{-0.04}$ & $ 0.38^{+0.05}_{-0.08}$ & $ 1.36^{+0.29}_{-0.19}$ & $ 1.07^{+0.05}_{-0.08}$  \\ 
33 & $ 1.16^{+0.03}_{-0.02}$ & $ 0.24^{+0.06}_{-0.04}$ & $ 0.47^{+0.12}_{-0.07}$ & $ 0.91^{+0.17}_{-0.29}$ & $ 1.72^{+0.10}_{-0.06}$ & $ 1.16^{+0.03}_{-0.02}$ & $ 0.24^{+0.06}_{-0.04}$ & $ 0.57^{+0.14}_{-0.08}$ & $ 1.15^{+0.20}_{-0.32}$ & $ 1.85^{+0.14}_{-0.07}$  \\ 
34 & $ 1.14^{+0.09}_{-0.07}$ & $ 0.24^{+0.09}_{-0.10}$ & $ 0.45^{+0.15}_{-0.20}$ & $ 2.85^{+0.73}_{-1.29}$ & $ 1.88^{+0.14}_{-0.25}$ & $ 1.14^{+0.09}_{-0.07}$ & $ 0.24^{+0.09}_{-0.10}$ & $ 0.54^{+0.17}_{-0.24}$ & $ 3.51^{+0.83}_{-1.40}$ & $ 2.04^{+0.17}_{-0.31}$  \\ 
35 & $ 1.04^{+0.01}_{-0.03}$ & $ 0.13^{+0.03}_{-0.03}$ & $ 0.25^{+0.06}_{-0.06}$ & $ 0.37^{+0.32}_{-0.04}$ & $ 1.32^{+0.07}_{-0.05}$ & $ 1.04^{+0.01}_{-0.03}$ & $ 0.13^{+0.03}_{-0.03}$ & $ 0.30^{+0.08}_{-0.07}$ & $ 0.73^{+0.32}_{-0.08}$ & $ 1.41^{+0.08}_{-0.06}$  \\ 
36 & $ 0.72^{+0.01}_{-0.01}$ & $ 0.06^{+0.01}_{-0.01}$ & $ 0.08^{+0.02}_{-0.01}$ & $ 1.05^{+0.15}_{-0.08}$ & $ 0.90^{+0.02}_{-0.01}$ & $ 0.72^{+0.01}_{-0.01}$ & $ 0.06^{+0.01}_{-0.01}$ & $ 0.10^{+0.02}_{-0.01}$ & $ 1.32^{+0.18}_{-0.10}$ & $ 0.95^{+0.03}_{-0.01}$  \\ 
37 & $ 1.65^{+0.04}_{-0.03}$ & $ 1.64^{+0.18}_{-0.16}$ & $ 2.57^{+0.28}_{-0.25}$ & $ 1.95^{+0.36}_{-0.19}$ & $ 4.41^{+0.27}_{-0.23}$ & $ 1.65^{+0.04}_{-0.03}$ & $ 1.64^{+0.18}_{-0.16}$ & $ 3.09^{+0.35}_{-0.28}$ & $ 3.11^{+0.55}_{-0.31}$ & $ 5.05^{+0.36}_{-0.27}$  \\ 
38 & $ 0.95^{+0.02}_{-0.04}$ & $ 0.24^{+0.04}_{-0.03}$ & $ 0.43^{+0.09}_{-0.06}$ & $ 1.09^{+0.44}_{-0.14}$ & $ 1.49^{+0.09}_{-0.06}$ & $ 0.95^{+0.02}_{-0.04}$ & $ 0.24^{+0.04}_{-0.03}$ & $ 0.51^{+0.11}_{-0.07}$ & $ 1.42^{+0.45}_{-0.16}$ & $ 1.60^{+0.11}_{-0.07}$  \\ 
39 & $ 0.15^{+0.02}_{-0.03}$ & $ 0.57^{+0.05}_{-0.05}$ & $ 0.52^{+0.05}_{-0.04}$ & $ 0.87^{+0.12}_{-0.14}$ & $ 0.76^{+0.03}_{-0.04}$ & $ 0.15^{+0.02}_{-0.03}$ & $ 0.57^{+0.05}_{-0.05}$ & $ 0.71^{+0.10}_{-0.07}$ & $ 1.31^{+0.16}_{-0.16}$ & $ 0.99^{+0.09}_{-0.07}$  \\ 
40 & $ 0.97^{+0.01}_{-0.00}$ & $ 0.01^{+0.00}_{-0.00}$ & $ 0.03^{+0.01}_{-0.00}$ & $ 0.82^{+0.12}_{-0.06}$ & $ 1.09^{+0.02}_{-0.00}$ & $ 0.97^{+0.01}_{-0.00}$ & $ 0.01^{+0.00}_{-0.00}$ & $ 0.06^{+0.02}_{-0.01}$ & $ 1.00^{+0.15}_{-0.08}$ & $ 1.14^{+0.03}_{-0.01}$  \\ 
41 & $ 0.35^{+0.02}_{-0.02}$ & $ 0.16^{+0.03}_{-0.03}$ & $ 0.31^{+0.07}_{-0.05}$ & $ 1.65^{+0.31}_{-0.25}$ & $ 0.82^{+0.08}_{-0.05}$ & $ 0.35^{+0.02}_{-0.02}$ & $ 0.16^{+0.03}_{-0.03}$ & $ 0.38^{+0.09}_{-0.06}$ & $ 2.02^{+0.38}_{-0.29}$ & $ 0.93^{+0.10}_{-0.07}$  \\ 
42 & $ 2.88^{+0.03}_{-0.05}$ & $ 0.66^{+0.09}_{-0.05}$ & $ 0.62^{+0.08}_{-0.05}$ & $ 0.75^{+0.42}_{-0.40}$ & $ 3.57^{+0.07}_{-0.06}$ & $ 2.88^{+0.03}_{-0.05}$ & $ 1.41^{+0.18}_{-0.13}$ & $ 0.82^{+0.10}_{-0.08}$ & $ 1.83^{+0.91}_{-0.98}$ & $ 3.88^{+0.09}_{-0.09}$  \\ 
\hline	
\end{tabular}
\par
{{\bf Table B2} Infrared and bolometric luminosities from the FR06 model fits.
}
\end{center}
\end{table*}

\begin{table*}
\label{tab:complums2}
\begin{center}
\begin{tabular}{lcccccccccc}
\hline
   & \multicolumn{5}{c}{Infrared Luminosities}   &  \multicolumn{5}{c}{Bolometric Luminosities}  \\         
ID & L$_{Sb}$             & L$_{AGN}^{o}$        & L$_{AGN}^{c}$            & L$_{host}$           & L$_{Tot}^{c}$        & L$_{Sb}$             & L$_{AGN}^{o}$        & L$_{AGN}^{c}$        & L$_{host}$           & L$_{Tot}^{c}$            \\
   & \multicolumn{3}{c}{$10^{12}$L$_{\odot}$}                               & $10^{11}$L$_{\odot}$ & $10^{12}$L$_{\odot}$ & \multicolumn{3}{c}{$10^{12}$L$_{\odot}$}                           & $10^{11}$L$_{\odot}$ & $10^{12}$L$_{\odot}$ \\
\hline	
1 & $ 1.48^{+0.07}_{-0.03}$ & $ 0.05^{+0.02}_{-0.03}$ & $ 0.09^{+0.05}_{-0.04}$ & $ 1.39^{+0.35}_{-0.28}$ & $ 1.71^{+0.10}_{-0.05}$ & $ 1.48^{+0.07}_{-0.03}$ & $ 0.05^{+0.02}_{-0.03}$ & $ 0.17^{+0.20}_{-0.08}$ & $ 1.72^{+0.42}_{-0.32}$ & $ 1.82^{+0.24}_{-0.09}$  \\ 
2 & $ 6.64^{+0.35}_{-0.26}$ & $ 2.27^{+0.79}_{-0.43}$ & $ 14.15^{+6.14}_{-3.30}$ & $ 1.52^{+1.37}_{-0.58}$ & $ 20.95^{+5.96}_{-3.01}$ & $ 6.64^{+0.35}_{-0.26}$ & $ 2.27^{+0.79}_{-0.43}$ & $ 80.78^{+34.20}_{-19.33}$ & $ 2.16^{+1.53}_{-0.66}$ & $ 87.64^{+34.03}_{-19.04}$  \\ 
3 & $ 0.80^{+0.01}_{-0.00}$ & $ 0.55^{+0.03}_{-0.07}$ & $ 0.73^{+0.04}_{-0.09}$ & $ 0.23^{+0.03}_{-0.03}$ & $ 1.55^{+0.04}_{-0.09}$ & $ 0.80^{+0.01}_{-0.00}$ & $ 0.55^{+0.03}_{-0.07}$ & $ 0.77^{+0.04}_{-0.10}$ & $ 0.57^{+0.07}_{-0.08}$ & $ 1.63^{+0.04}_{-0.10}$  \\ 
4 & $ 3.79^{+0.10}_{-0.04}$ & $ 0.05^{+0.00}_{-0.01}$ & $ 0.07^{+0.00}_{-0.01}$ & $ 0.51^{+0.18}_{-0.09}$ & $ 3.91^{+0.11}_{-0.05}$ & $ 3.79^{+0.10}_{-0.04}$ & $ 0.05^{+0.00}_{-0.01}$ & $ 0.07^{+0.01}_{-0.01}$ & $ 0.69^{+0.20}_{-0.10}$ & $ 3.93^{+0.12}_{-0.05}$  \\ 
5 & $ 2.04^{+0.02}_{-0.02}$ & $ 0.12^{+0.02}_{-0.04}$ & $ 0.15^{+0.04}_{-0.05}$ & $ 0.60^{+0.15}_{-0.13}$ & $ 2.25^{+0.06}_{-0.05}$ & $ 2.04^{+0.02}_{-0.02}$ & $ 0.12^{+0.02}_{-0.04}$ & $ 0.16^{+0.10}_{-0.06}$ & $ 0.73^{+0.18}_{-0.14}$ & $ 2.27^{+0.10}_{-0.06}$  \\ 
6 & $ 0.99^{+0.01}_{-0.03}$ & $ 0.20^{+0.04}_{-0.02}$ & $ 0.16^{+0.04}_{-0.01}$ & $ 0.48^{+0.25}_{-0.07}$ & $ 1.20^{+0.03}_{-0.02}$ & $ 0.99^{+0.01}_{-0.03}$ & $ 0.20^{+0.04}_{-0.02}$ & $ 0.20^{+0.05}_{-0.02}$ & $ 0.77^{+0.26}_{-0.08}$ & $ 1.27^{+0.05}_{-0.03}$  \\ 
7 & $ 1.07^{+0.63}_{-0.12}$ & $ 0.22^{+0.19}_{-0.21}$ & $ 0.51^{+0.90}_{-0.49}$ & $ 2.23^{+1.81}_{-2.23}$ & $ 1.81^{+0.87}_{-0.44}$ & $ 1.07^{+0.64}_{-0.12}$ & $ 0.22^{+0.19}_{-0.21}$ & $ 1.13^{+4.63}_{-1.10}$ & $ 2.70^{+2.42}_{-2.70}$ & $ 2.48^{+4.63}_{-1.01}$  \\ 
8 & $ 1.17^{+0.03}_{-0.02}$ & $ 0.11^{+0.02}_{-0.01}$ & $ 0.15^{+0.02}_{-0.02}$ & $ 1.43^{+0.35}_{-0.38}$ & $ 1.46^{+0.03}_{-0.02}$ & $ 1.17^{+0.03}_{-0.02}$ & $ 0.11^{+0.02}_{-0.01}$ & $ 0.16^{+0.03}_{-0.02}$ & $ 2.22^{+0.36}_{-0.54}$ & $ 1.55^{+0.04}_{-0.05}$  \\ 
9 & $ 1.98^{+0.03}_{-0.09}$ & $ 2.30^{+0.41}_{-0.19}$ & $ 2.13^{+0.56}_{-0.20}$ & $ 0.00^{+0.00}_{-0.00}$ & $ 4.11^{+0.54}_{-0.26}$ & $ 1.99^{+0.04}_{-0.09}$ & $ 5.19^{+0.80}_{-0.37}$ & $ 9.08^{+2.34}_{-2.25}$ & $ 0.00^{+0.00}_{-0.00}$ & $ 11.07^{+2.33}_{-2.29}$  \\ 
10 & $ 1.91^{+0.04}_{-0.02}$ & $ 0.47^{+0.08}_{-0.08}$ & $ 0.61^{+0.10}_{-0.10}$ & $ 0.94^{+0.15}_{-0.23}$ & $ 2.61^{+0.11}_{-0.10}$ & $ 1.93^{+0.04}_{-0.02}$ & $ 0.47^{+0.08}_{-0.08}$ & $ 0.65^{+0.11}_{-0.11}$ & $ 2.18^{+0.51}_{-0.62}$ & $ 2.80^{+0.12}_{-0.12}$  \\ 
11 & $ 1.27^{+0.03}_{-0.02}$ & $ 0.15^{+0.04}_{-0.03}$ & $ 0.48^{+0.21}_{-0.12}$ & $ 0.17^{+0.02}_{-0.02}$ & $ 1.77^{+0.20}_{-0.11}$ & $ 1.27^{+0.03}_{-0.02}$ & $ 0.15^{+0.04}_{-0.03}$ & $ 1.24^{+1.02}_{-0.40}$ & $ 0.39^{+0.04}_{-0.04}$ & $ 2.55^{+1.01}_{-0.39}$  \\ 
12 & $ 1.47^{+0.01}_{-0.02}$ & $ 0.10^{+0.02}_{-0.01}$ & $ 0.12^{+0.02}_{-0.01}$ & $ 0.29^{+0.08}_{-0.05}$ & $ 1.62^{+0.02}_{-0.02}$ & $ 1.48^{+0.01}_{-0.02}$ & $ 0.10^{+0.02}_{-0.01}$ & $ 0.13^{+0.02}_{-0.01}$ & $ 0.59^{+0.16}_{-0.10}$ & $ 1.67^{+0.03}_{-0.02}$  \\ 
13 & $ 1.40^{+0.02}_{-0.01}$ & $ 0.14^{+0.02}_{-0.03}$ & $ 0.19^{+0.03}_{-0.04}$ & $ 0.70^{+0.12}_{-0.15}$ & $ 1.66^{+0.04}_{-0.04}$ & $ 1.40^{+0.02}_{-0.01}$ & $ 0.14^{+0.02}_{-0.03}$ & $ 0.21^{+0.03}_{-0.04}$ & $ 1.05^{+0.16}_{-0.20}$ & $ 1.72^{+0.05}_{-0.04}$  \\ 
14 & $ 0.85^{+0.04}_{-0.01}$ & $ 0.08^{+0.01}_{-0.02}$ & $ 0.11^{+0.02}_{-0.02}$ & $ 1.42^{+0.14}_{-0.21}$ & $ 1.11^{+0.03}_{-0.02}$ & $ 0.85^{+0.04}_{-0.01}$ & $ 0.08^{+0.01}_{-0.02}$ & $ 0.12^{+0.02}_{-0.02}$ & $ 1.70^{+0.17}_{-0.24}$ & $ 1.14^{+0.03}_{-0.03}$  \\ 
15 & $ 1.24^{+0.02}_{-0.03}$ & $ 0.15^{+0.01}_{-0.02}$ & $ 0.20^{+0.02}_{-0.02}$ & $ 0.13^{+0.06}_{-0.03}$ & $ 1.45^{+0.02}_{-0.03}$ & $ 1.24^{+0.02}_{-0.03}$ & $ 0.15^{+0.01}_{-0.02}$ & $ 0.21^{+0.02}_{-0.03}$ & $ 0.28^{+0.11}_{-0.06}$ & $ 1.48^{+0.02}_{-0.03}$  \\ 
16 & $ 1.39^{+0.00}_{-0.01}$ & $ 0.53^{+0.07}_{-0.06}$ & $ 0.70^{+0.09}_{-0.08}$ & $ 0.55^{+0.11}_{-0.06}$ & $ 2.14^{+0.09}_{-0.09}$ & $ 1.39^{+0.00}_{-0.01}$ & $ 0.53^{+0.07}_{-0.06}$ & $ 0.75^{+0.10}_{-0.09}$ & $ 1.14^{+0.23}_{-0.11}$ & $ 2.25^{+0.10}_{-0.09}$  \\ 
17 & $ 1.56^{+0.03}_{-0.03}$ & $ 0.06^{+0.01}_{-0.01}$ & $ 0.09^{+0.02}_{-0.01}$ & $ 1.49^{+0.35}_{-0.33}$ & $ 1.80^{+0.03}_{-0.01}$ & $ 1.56^{+0.03}_{-0.03}$ & $ 0.06^{+0.01}_{-0.01}$ & $ 0.09^{+0.02}_{-0.01}$ & $ 1.77^{+0.42}_{-0.37}$ & $ 1.83^{+0.03}_{-0.02}$  \\ 
18 & $ 0.71^{+0.02}_{-0.04}$ & $ 0.64^{+0.21}_{-0.10}$ & $ 0.84^{+0.27}_{-0.13}$ & $ 2.57^{+0.93}_{-0.49}$ & $ 1.81^{+0.32}_{-0.17}$ & $ 0.71^{+0.02}_{-0.04}$ & $ 0.64^{+0.21}_{-0.10}$ & $ 0.90^{+0.29}_{-0.14}$ & $ 3.60^{+1.18}_{-0.68}$ & $ 1.97^{+0.37}_{-0.20}$  \\ 
19 & $ 1.81^{+0.05}_{-0.03}$ & $ 0.08^{+0.01}_{-0.01}$ & $ 0.11^{+0.01}_{-0.01}$ & $ 0.89^{+0.51}_{-0.17}$ & $ 2.01^{+0.04}_{-0.02}$ & $ 1.81^{+0.05}_{-0.03}$ & $ 0.08^{+0.01}_{-0.01}$ & $ 0.12^{+0.01}_{-0.02}$ & $ 1.46^{+0.54}_{-0.21}$ & $ 2.08^{+0.05}_{-0.02}$  \\ 
20 & $ 0.96^{+0.03}_{-0.04}$ & $ 0.06^{+0.01}_{-0.01}$ & $ 0.08^{+0.01}_{-0.01}$ & $ 0.74^{+0.39}_{-0.27}$ & $ 1.12^{+0.01}_{-0.01}$ & $ 0.96^{+0.03}_{-0.04}$ & $ 0.06^{+0.01}_{-0.01}$ & $ 0.09^{+0.01}_{-0.01}$ & $ 0.98^{+0.39}_{-0.23}$ & $ 1.15^{+0.03}_{-0.01}$  \\ 
21 & $ 0.88^{+0.02}_{-0.01}$ & $ 0.06^{+0.01}_{-0.01}$ & $ 0.08^{+0.02}_{-0.01}$ & $ 0.22^{+0.03}_{-0.02}$ & $ 0.99^{+0.02}_{-0.01}$ & $ 0.88^{+0.02}_{-0.01}$ & $ 0.06^{+0.01}_{-0.01}$ & $ 0.09^{+0.02}_{-0.01}$ & $ 0.52^{+0.07}_{-0.05}$ & $ 1.03^{+0.03}_{-0.01}$  \\ 
22 & $ 1.11^{+0.02}_{-0.01}$ & $ 0.15^{+0.03}_{-0.02}$ & $ 0.12^{+0.03}_{-0.01}$ & $ 0.22^{+0.05}_{-0.04}$ & $ 1.25^{+0.03}_{-0.01}$ & $ 1.11^{+0.02}_{-0.01}$ & $ 0.15^{+0.03}_{-0.02}$ & $ 0.13^{+0.04}_{-0.02}$ & $ 0.44^{+0.09}_{-0.07}$ & $ 1.29^{+0.04}_{-0.02}$  \\ 
23 & $ 2.74^{+0.04}_{-0.04}$ & $ 0.10^{+0.03}_{-0.02}$ & $ 0.13^{+0.04}_{-0.03}$ & $ 0.38^{+0.08}_{-0.04}$ & $ 2.91^{+0.04}_{-0.04}$ & $ 2.75^{+0.04}_{-0.04}$ & $ 0.10^{+0.03}_{-0.02}$ & $ 0.14^{+0.05}_{-0.03}$ & $ 0.93^{+0.19}_{-0.11}$ & $ 2.98^{+0.06}_{-0.04}$  \\ 
24 & $ 2.10^{+0.01}_{-0.02}$ & $ 0.04^{+0.01}_{-0.01}$ & $ 0.05^{+0.01}_{-0.02}$ & $ 0.93^{+0.12}_{-0.15}$ & $ 2.25^{+0.01}_{-0.03}$ & $ 2.11^{+0.01}_{-0.02}$ & $ 0.04^{+0.01}_{-0.01}$ & $ 0.06^{+0.01}_{-0.02}$ & $ 1.12^{+0.15}_{-0.16}$ & $ 2.28^{+0.01}_{-0.04}$  \\ 
25 & $ 0.77^{+0.04}_{-0.05}$ & $ 0.17^{+0.04}_{-0.07}$ & $ 0.24^{+0.05}_{-0.12}$ & $ 1.58^{+0.45}_{-0.52}$ & $ 1.17^{+0.05}_{-0.13}$ & $ 0.77^{+0.04}_{-0.05}$ & $ 0.17^{+0.04}_{-0.07}$ & $ 0.28^{+0.06}_{-0.14}$ & $ 2.28^{+0.56}_{-0.73}$ & $ 1.29^{+0.08}_{-0.16}$  \\ 
26 & $ 2.23^{+0.04}_{-0.02}$ & $ 0.12^{+0.02}_{-0.02}$ & $ 0.16^{+0.02}_{-0.03}$ & $ 0.52^{+0.12}_{-0.09}$ & $ 2.44^{+0.06}_{-0.03}$ & $ 2.23^{+0.04}_{-0.02}$ & $ 0.12^{+0.02}_{-0.02}$ & $ 0.17^{+0.03}_{-0.03}$ & $ 1.20^{+0.39}_{-0.23}$ & $ 2.52^{+0.08}_{-0.04}$  \\ 
27 & $ 1.64^{+0.09}_{-0.14}$ & $ 0.02^{+0.01}_{-0.00}$ & $ 0.02^{+0.01}_{-0.00}$ & $ 2.01^{+0.69}_{-0.87}$ & $ 1.86^{+0.15}_{-0.19}$ & $ 1.64^{+0.09}_{-0.14}$ & $ 0.02^{+0.01}_{-0.00}$ & $ 0.02^{+0.02}_{-0.00}$ & $ 2.40^{+0.82}_{-1.04}$ & $ 1.91^{+0.16}_{-0.21}$  \\ 
28 & $ 2.97^{+0.02}_{-0.01}$ & $ 0.30^{+0.03}_{-0.03}$ & $ 0.66^{+0.08}_{-0.06}$ & $ 1.37^{+0.13}_{-0.30}$ & $ 3.77^{+0.08}_{-0.07}$ & $ 2.97^{+0.02}_{-0.01}$ & $ 0.30^{+0.03}_{-0.03}$ & $ 1.20^{+0.17}_{-0.11}$ & $ 1.88^{+0.17}_{-0.37}$ & $ 4.35^{+0.17}_{-0.12}$  \\ 
29 & $ 1.45^{+0.01}_{-0.00}$ & $ 0.06^{+0.01}_{-0.01}$ & $ 0.07^{+0.02}_{-0.01}$ & $ 0.30^{+0.06}_{-0.12}$ & $ 1.55^{+0.02}_{-0.02}$ & $ 1.45^{+0.01}_{-0.00}$ & $ 0.06^{+0.01}_{-0.01}$ & $ 0.08^{+0.02}_{-0.01}$ & $ 0.50^{+0.07}_{-0.14}$ & $ 1.58^{+0.02}_{-0.02}$  \\ 
30 & $ 0.66^{+0.01}_{-0.02}$ & $ 0.01^{+0.00}_{-0.00}$ & $ 0.02^{+0.00}_{-0.00}$ & $ 0.53^{+0.12}_{-0.09}$ & $ 0.74^{+0.01}_{-0.02}$ & $ 0.66^{+0.01}_{-0.02}$ & $ 0.01^{+0.00}_{-0.00}$ & $ 0.04^{+0.01}_{-0.01}$ & $ 0.97^{+0.16}_{-0.13}$ & $ 0.80^{+0.02}_{-0.03}$  \\ 
31 & $ 0.87^{+0.00}_{-0.00}$ & $ 0.08^{+0.01}_{-0.01}$ & $ 0.11^{+0.01}_{-0.01}$ & $ 0.36^{+0.04}_{-0.03}$ & $ 1.01^{+0.02}_{-0.01}$ & $ 0.87^{+0.00}_{-0.00}$ & $ 0.08^{+0.01}_{-0.01}$ & $ 0.11^{+0.02}_{-0.01}$ & $ 0.85^{+0.10}_{-0.08}$ & $ 1.07^{+0.02}_{-0.02}$  \\ 
32 & $ 0.71^{+0.01}_{-0.02}$ & $ 0.11^{+0.02}_{-0.01}$ & $ 0.14^{+0.03}_{-0.02}$ & $ 0.99^{+0.25}_{-0.22}$ & $ 0.95^{+0.03}_{-0.04}$ & $ 0.71^{+0.01}_{-0.02}$ & $ 0.11^{+0.02}_{-0.01}$ & $ 0.15^{+0.03}_{-0.02}$ & $ 1.44^{+0.26}_{-0.20}$ & $ 1.01^{+0.04}_{-0.04}$  \\ 
33 & $ 1.13^{+0.04}_{-0.04}$ & $ 0.09^{+0.02}_{-0.03}$ & $ 0.12^{+0.02}_{-0.03}$ & $ 0.13^{+0.06}_{-0.04}$ & $ 1.26^{+0.03}_{-0.05}$ & $ 1.13^{+0.04}_{-0.04}$ & $ 0.09^{+0.02}_{-0.03}$ & $ 0.12^{+0.02}_{-0.04}$ & $ 0.28^{+0.09}_{-0.08}$ & $ 1.28^{+0.03}_{-0.05}$  \\ 
34 & $ 1.43^{+0.05}_{-0.08}$ & $ 0.10^{+0.02}_{-0.03}$ & $ 0.14^{+0.03}_{-0.05}$ & $ 2.28^{+0.44}_{-0.83}$ & $ 1.80^{+0.05}_{-0.11}$ & $ 1.43^{+0.05}_{-0.08}$ & $ 0.10^{+0.02}_{-0.03}$ & $ 0.15^{+0.04}_{-0.05}$ & $ 3.01^{+0.40}_{-1.20}$ & $ 1.88^{+0.04}_{-0.15}$  \\ 
35 & $ 1.19^{+0.01}_{-0.01}$ & $ 0.06^{+0.01}_{-0.01}$ & $ 0.08^{+0.01}_{-0.01}$ & $ 0.41^{+0.24}_{-0.05}$ & $ 1.31^{+0.02}_{-0.01}$ & $ 1.19^{+0.01}_{-0.01}$ & $ 0.06^{+0.01}_{-0.01}$ & $ 0.08^{+0.01}_{-0.01}$ & $ 0.77^{+0.26}_{-0.05}$ & $ 1.35^{+0.02}_{-0.01}$  \\ 
36 & $ 0.71^{+0.02}_{-0.08}$ & $ 0.03^{+0.03}_{-0.00}$ & $ 0.03^{+0.16}_{-0.00}$ & $ 0.99^{+0.45}_{-0.25}$ & $ 0.84^{+0.07}_{-0.02}$ & $ 0.71^{+0.02}_{-0.09}$ & $ 0.03^{+0.03}_{-0.00}$ & $ 0.04^{+0.68}_{-0.01}$ & $ 1.27^{+0.51}_{-0.31}$ & $ 0.88^{+0.57}_{-0.03}$  \\ 
37 & $ 2.43^{+0.03}_{-0.12}$ & $ 0.72^{+0.13}_{-0.10}$ & $ 0.68^{+0.12}_{-0.08}$ & $ 1.01^{+0.12}_{-0.15}$ & $ 3.20^{+0.11}_{-0.11}$ & $ 2.43^{+0.03}_{-0.12}$ & $ 0.72^{+0.13}_{-0.10}$ & $ 0.79^{+0.14}_{-0.11}$ & $ 2.20^{+0.30}_{-0.34}$ & $ 3.45^{+0.13}_{-0.15}$  \\ 
38 & $ 1.27^{+0.02}_{-0.06}$ & $ 0.09^{+0.04}_{-0.01}$ & $ 0.12^{+0.13}_{-0.01}$ & $ 0.39^{+0.16}_{-0.06}$ & $ 1.43^{+0.08}_{-0.03}$ & $ 1.27^{+0.02}_{-0.06}$ & $ 0.09^{+0.04}_{-0.01}$ & $ 0.13^{+0.28}_{-0.02}$ & $ 0.76^{+0.16}_{-0.09}$ & $ 1.48^{+0.24}_{-0.03}$  \\ 
39 & $ 0.21^{+0.01}_{-0.02}$ & $ 0.56^{+0.07}_{-0.07}$ & $ 0.70^{+0.09}_{-0.09}$ & $ 0.43^{+0.15}_{-0.07}$ & $ 0.95^{+0.09}_{-0.08}$ & $ 0.21^{+0.01}_{-0.02}$ & $ 0.56^{+0.07}_{-0.07}$ & $ 1.60^{+0.19}_{-0.23}$ & $ 0.90^{+0.30}_{-0.20}$ & $ 1.90^{+0.19}_{-0.21}$  \\ 
40 & $ 1.08^{+0.01}_{-0.00}$ & $ 0.01^{+0.00}_{-0.00}$ & $ 0.03^{+0.01}_{-0.00}$ & $ 0.21^{+0.03}_{-0.01}$ & $ 1.13^{+0.01}_{-0.00}$ & $ 1.08^{+0.01}_{-0.00}$ & $ 0.01^{+0.00}_{-0.00}$ & $ 0.04^{+0.03}_{-0.00}$ & $ 0.42^{+0.06}_{-0.03}$ & $ 1.17^{+0.03}_{-0.00}$  \\ 
41 & $ 0.43^{+0.01}_{-0.01}$ & $ 0.02^{+0.00}_{-0.00}$ & $ 0.02^{+0.00}_{-0.00}$ & $ 0.96^{+0.14}_{-0.13}$ & $ 0.55^{+0.01}_{-0.01}$ & $ 0.43^{+0.01}_{-0.01}$ & $ 0.02^{+0.00}_{-0.00}$ & $ 0.02^{+0.00}_{-0.00}$ & $ 1.30^{+0.15}_{-0.16}$ & $ 0.59^{+0.01}_{-0.01}$  \\ 
42 & $ 2.75^{+0.05}_{-0.03}$ & $ 0.71^{+0.08}_{-0.11}$ & $ 0.62^{+0.07}_{-0.09}$ & $ 1.47^{+0.36}_{-0.62}$ & $ 3.52^{+0.07}_{-0.12}$ & $ 2.75^{+0.05}_{-0.03}$ & $ 1.69^{+0.21}_{-0.22}$ & $ 2.16^{+0.52}_{-0.26}$ & $ 2.19^{+0.31}_{-0.79}$ & $ 5.13^{+0.52}_{-0.27}$  \\ 
\hline	
\end{tabular}
\par
{{\bf Table B3} Infrared and bolometric luminosities from the SKIRTOR model fits.
}
\end{center}
\end{table*}

\begin{table*}
\label{tab:complums3}
\begin{center}
\begin{tabular}{lcccccccccc}
\hline
   & \multicolumn{5}{c}{Infrared Luminosities}   &  \multicolumn{5}{c}{Bolometric Luminosities}  \\         
ID & L$_{Sb}$             & L$_{AGN}^{o}$        & L$_{AGN}^{c}$            & L$_{host}$           & L$_{Tot}^{c}$        & L$_{Sb}$             & L$_{AGN}^{o}$        & L$_{AGN}^{c}$        & L$_{host}$           & L$_{Tot}^{c}$            \\
   & \multicolumn{3}{c}{$10^{12}$L$_{\odot}$}                               & $10^{11}$L$_{\odot}$ & $10^{12}$L$_{\odot}$ & \multicolumn{3}{c}{$10^{12}$L$_{\odot}$}                           & $10^{11}$L$_{\odot}$ & $10^{12}$L$_{\odot}$ \\
\hline	
1 & $ 1.72^{+0.07}_{-0.06}$ & $ 0.27^{+0.10}_{-0.09}$ & $ 0.21^{+0.07}_{-0.06}$ & $ 0.74^{+0.33}_{-0.29}$ & $ 2.00^{+0.10}_{-0.07}$ & $ 1.72^{+0.07}_{-0.06}$ & $ 0.27^{+0.10}_{-0.09}$ & $ 0.21^{+0.07}_{-0.06}$ & $ 1.04^{+0.39}_{-0.31}$ & $ 2.03^{+0.11}_{-0.07}$  \\ 
2 & $ 10.01^{+0.22}_{-0.13}$ & $ 0.62^{+0.07}_{-0.13}$ & $ 0.53^{+0.06}_{-0.12}$ & $ 0.55^{+0.66}_{-0.21}$ & $ 10.60^{+0.16}_{-0.14}$ & $ 10.01^{+0.22}_{-0.13}$ & $ 0.65^{+0.07}_{-0.12}$ & $ 0.81^{+0.08}_{-0.19}$ & $ 0.99^{+0.64}_{-0.36}$ & $ 10.92^{+0.14}_{-0.17}$  \\ 
3 & $ 1.07^{+0.06}_{-0.04}$ & $ 0.71^{+0.09}_{-0.09}$ & $ 0.91^{+0.12}_{-0.13}$ & $ 0.03^{+0.03}_{-0.03}$ & $ 1.99^{+0.09}_{-0.11}$ & $ 1.07^{+0.06}_{-0.04}$ & $ 0.76^{+0.10}_{-0.09}$ & $ 1.12^{+0.15}_{-0.14}$ & $ 0.08^{+0.08}_{-0.07}$ & $ 2.20^{+0.11}_{-0.13}$  \\ 
4 & $ 2.72^{+0.19}_{-0.11}$ & $ 1.35^{+0.13}_{-0.25}$ & $ 1.29^{+0.12}_{-0.29}$ & $ 0.00^{+0.00}_{-0.00}$ & $ 4.01^{+0.06}_{-0.17}$ & $ 2.73^{+0.18}_{-0.11}$ & $ 1.35^{+0.13}_{-0.25}$ & $ 1.29^{+0.12}_{-0.29}$ & $ 0.00^{+0.00}_{-0.00}$ & $ 4.02^{+0.06}_{-0.18}$  \\ 
5 & $ 2.02^{+0.02}_{-0.03}$ & $ 0.40^{+0.05}_{-0.04}$ & $ 0.50^{+0.06}_{-0.05}$ & $ 0.15^{+0.02}_{-0.02}$ & $ 2.53^{+0.04}_{-0.04}$ & $ 2.02^{+0.02}_{-0.03}$ & $ 0.40^{+0.05}_{-0.04}$ & $ 0.51^{+0.06}_{-0.05}$ & $ 0.26^{+0.02}_{-0.03}$ & $ 2.56^{+0.04}_{-0.04}$  \\ 
6 & $ 0.80^{+0.17}_{-0.04}$ & $ 0.61^{+0.06}_{-0.20}$ & $ 0.43^{+0.05}_{-0.14}$ & $ 0.31^{+0.05}_{-0.04}$ & $ 1.26^{+0.08}_{-0.03}$ & $ 0.80^{+0.17}_{-0.04}$ & $ 0.61^{+0.06}_{-0.20}$ & $ 0.43^{+0.05}_{-0.14}$ & $ 0.60^{+0.07}_{-0.09}$ & $ 1.29^{+0.08}_{-0.03}$  \\ 
7 & $ 1.46^{+0.04}_{-0.03}$ & $ 0.05^{+0.01}_{-0.01}$ & $ 0.04^{+0.00}_{-0.01}$ & $ 0.16^{+0.13}_{-0.13}$ & $ 1.52^{+0.02}_{-0.01}$ & $ 1.46^{+0.04}_{-0.03}$ & $ 0.11^{+0.07}_{-0.03}$ & $ 0.13^{+0.04}_{-0.03}$ & $ 0.22^{+0.18}_{-0.18}$ & $ 1.61^{+0.06}_{-0.03}$  \\ 
8 & $ 1.05^{+0.03}_{-0.02}$ & $ 0.22^{+0.03}_{-0.03}$ & $ 0.21^{+0.03}_{-0.04}$ & $ 2.02^{+0.23}_{-0.30}$ & $ 1.46^{+0.03}_{-0.04}$ & $ 1.05^{+0.03}_{-0.02}$ & $ 0.22^{+0.03}_{-0.03}$ & $ 0.21^{+0.03}_{-0.04}$ & $ 2.68^{+0.32}_{-0.44}$ & $ 1.53^{+0.04}_{-0.06}$  \\ 
9 & $ 0.40^{+0.42}_{-0.16}$ & $ 3.14^{+0.77}_{-0.37}$ & $ 2.71^{+0.66}_{-0.33}$ & $ 9.10^{+1.64}_{-4.71}$ & $ 4.03^{+0.64}_{-0.36}$ & $ 0.40^{+0.42}_{-0.16}$ & $ 3.31^{+0.82}_{-0.39}$ & $ 4.14^{+1.01}_{-0.52}$ & $ 12.92^{+2.25}_{-6.79}$ & $ 5.83^{+0.92}_{-0.72}$  \\ 
10 & $ 2.24^{+0.08}_{-0.08}$ & $ 0.53^{+0.03}_{-0.14}$ & $ 0.75^{+0.07}_{-0.28}$ & $ 0.84^{+0.29}_{-0.24}$ & $ 3.08^{+0.06}_{-0.24}$ & $ 2.26^{+0.08}_{-0.09}$ & $ 0.54^{+0.03}_{-0.13}$ & $ 1.00^{+0.14}_{-0.40}$ & $ 1.80^{+0.72}_{-0.61}$ & $ 3.44^{+0.13}_{-0.37}$  \\ 
11 & $ 1.23^{+0.07}_{-0.07}$ & $ 0.30^{+0.09}_{-0.07}$ & $ 0.25^{+0.07}_{-0.07}$ & $ 0.18^{+0.01}_{-0.02}$ & $ 1.50^{+0.02}_{-0.02}$ & $ 1.23^{+0.07}_{-0.07}$ & $ 0.30^{+0.09}_{-0.07}$ & $ 0.25^{+0.07}_{-0.07}$ & $ 0.41^{+0.03}_{-0.03}$ & $ 1.53^{+0.02}_{-0.02}$  \\ 
12 & $ 1.55^{+0.08}_{-0.01}$ & $ 0.06^{+0.01}_{-0.04}$ & $ 0.04^{+0.00}_{-0.02}$ & $ 0.03^{+0.05}_{-0.03}$ & $ 1.59^{+0.07}_{-0.01}$ & $ 1.56^{+0.08}_{-0.01}$ & $ 0.07^{+0.01}_{-0.04}$ & $ 0.04^{+0.01}_{-0.02}$ & $ 0.05^{+0.09}_{-0.05}$ & $ 1.60^{+0.07}_{-0.01}$  \\ 
13 & $ 1.22^{+0.08}_{-0.05}$ & $ 0.49^{+0.07}_{-0.11}$ & $ 0.44^{+0.07}_{-0.11}$ & $ 0.62^{+0.50}_{-0.15}$ & $ 1.72^{+0.05}_{-0.06}$ & $ 1.22^{+0.08}_{-0.05}$ & $ 0.49^{+0.07}_{-0.11}$ & $ 0.44^{+0.07}_{-0.11}$ & $ 0.97^{+0.55}_{-0.20}$ & $ 1.76^{+0.06}_{-0.06}$  \\ 
14 & $ 0.75^{+0.05}_{-0.02}$ & $ 0.11^{+0.05}_{-0.02}$ & $ 0.40^{+0.31}_{-0.07}$ & $ 0.74^{+0.31}_{-0.25}$ & $ 1.23^{+0.34}_{-0.09}$ & $ 0.76^{+0.05}_{-0.02}$ & $ 0.11^{+0.05}_{-0.02}$ & $ 0.72^{+0.50}_{-0.12}$ & $ 0.99^{+0.38}_{-0.29}$ & $ 1.58^{+0.54}_{-0.14}$  \\ 
15 & $ 1.20^{+0.12}_{-0.02}$ & $ 0.39^{+0.05}_{-0.39}$ & $ 0.48^{+0.07}_{-0.48}$ & $ 0.16^{+0.05}_{-0.02}$ & $ 1.70^{+0.07}_{-0.37}$ & $ 1.20^{+0.12}_{-0.02}$ & $ 0.39^{+0.05}_{-0.39}$ & $ 0.49^{+0.07}_{-0.48}$ & $ 0.35^{+0.10}_{-0.05}$ & $ 1.72^{+0.07}_{-0.37}$  \\ 
16 & $ 1.10^{+0.10}_{-0.11}$ & $ 1.10^{+0.14}_{-0.17}$ & $ 0.88^{+0.11}_{-0.13}$ & $ 0.54^{+0.09}_{-0.07}$ & $ 2.03^{+0.02}_{-0.07}$ & $ 1.10^{+0.10}_{-0.11}$ & $ 1.10^{+0.14}_{-0.17}$ & $ 0.88^{+0.11}_{-0.13}$ & $ 1.02^{+0.17}_{-0.14}$ & $ 2.08^{+0.05}_{-0.07}$  \\ 
17 & $ 1.50^{+0.03}_{-0.03}$ & $ 0.16^{+0.04}_{-0.03}$ & $ 0.14^{+0.03}_{-0.02}$ & $ 1.38^{+0.23}_{-0.24}$ & $ 1.78^{+0.04}_{-0.02}$ & $ 1.50^{+0.04}_{-0.03}$ & $ 0.16^{+0.04}_{-0.03}$ & $ 0.14^{+0.03}_{-0.02}$ & $ 1.78^{+0.30}_{-0.28}$ & $ 1.82^{+0.04}_{-0.03}$  \\ 
18 & $ 0.62^{+0.09}_{-0.12}$ & $ 1.20^{+0.38}_{-0.18}$ & $ 0.98^{+0.33}_{-0.13}$ & $ 2.47^{+0.92}_{-0.35}$ & $ 1.84^{+0.35}_{-0.08}$ & $ 0.62^{+0.09}_{-0.12}$ & $ 1.20^{+0.38}_{-0.18}$ & $ 1.07^{+0.38}_{-0.14}$ & $ 3.40^{+1.23}_{-0.47}$ & $ 2.03^{+0.43}_{-0.11}$  \\ 
19 & $ 1.59^{+0.04}_{-0.02}$ & $ 0.23^{+0.04}_{-0.04}$ & $ 0.26^{+0.05}_{-0.05}$ & $ 1.57^{+0.23}_{-0.34}$ & $ 2.01^{+0.05}_{-0.05}$ & $ 1.60^{+0.04}_{-0.02}$ & $ 0.23^{+0.04}_{-0.04}$ & $ 0.26^{+0.05}_{-0.05}$ & $ 2.14^{+0.29}_{-0.40}$ & $ 2.07^{+0.06}_{-0.05}$  \\ 
20 & $ 0.92^{+0.02}_{-0.01}$ & $ 0.17^{+0.03}_{-0.03}$ & $ 0.40^{+0.10}_{-0.08}$ & $ 0.54^{+0.10}_{-0.18}$ & $ 1.38^{+0.08}_{-0.07}$ & $ 0.92^{+0.02}_{-0.01}$ & $ 0.17^{+0.03}_{-0.03}$ & $ 0.45^{+0.12}_{-0.09}$ & $ 0.81^{+0.16}_{-0.23}$ & $ 1.45^{+0.10}_{-0.08}$  \\ 
21 & $ 0.80^{+0.03}_{-0.03}$ & $ 0.32^{+0.04}_{-0.04}$ & $ 0.41^{+0.04}_{-0.05}$ & $ 0.47^{+0.07}_{-0.15}$ & $ 1.26^{+0.03}_{-0.04}$ & $ 0.80^{+0.03}_{-0.03}$ & $ 0.32^{+0.04}_{-0.04}$ & $ 0.41^{+0.04}_{-0.05}$ & $ 0.78^{+0.08}_{-0.15}$ & $ 1.29^{+0.03}_{-0.04}$  \\ 
22 & $ 1.03^{+0.05}_{-0.05}$ & $ 0.45^{+0.06}_{-0.04}$ & $ 0.39^{+0.05}_{-0.04}$ & $ 0.23^{+0.05}_{-0.04}$ & $ 1.44^{+0.07}_{-0.05}$ & $ 1.03^{+0.05}_{-0.05}$ & $ 0.47^{+0.06}_{-0.04}$ & $ 0.43^{+0.06}_{-0.05}$ & $ 0.51^{+0.13}_{-0.09}$ & $ 1.51^{+0.08}_{-0.06}$  \\ 
23 & $ 2.86^{+0.05}_{-0.04}$ & $ 0.06^{+0.02}_{-0.02}$ & $ 0.07^{+0.02}_{-0.02}$ & $ 0.00^{+0.00}_{-0.00}$ & $ 2.93^{+0.05}_{-0.04}$ & $ 2.86^{+0.05}_{-0.04}$ & $ 0.14^{+0.02}_{-0.03}$ & $ 0.18^{+0.02}_{-0.04}$ & $ 0.00^{+0.00}_{-0.00}$ & $ 3.04^{+0.04}_{-0.05}$  \\ 
24 & $ 2.25^{+0.05}_{-0.02}$ & $ 0.08^{+0.02}_{-0.03}$ & $ 0.06^{+0.01}_{-0.02}$ & $ 0.45^{+0.13}_{-0.13}$ & $ 2.36^{+0.04}_{-0.01}$ & $ 2.25^{+0.05}_{-0.02}$ & $ 0.08^{+0.02}_{-0.03}$ & $ 0.06^{+0.01}_{-0.02}$ & $ 0.61^{+0.16}_{-0.14}$ & $ 2.38^{+0.04}_{-0.01}$  \\ 
25 & $ 0.80^{+0.02}_{-0.05}$ & $ 0.23^{+0.06}_{-0.03}$ & $ 0.35^{+0.12}_{-0.05}$ & $ 1.72^{+0.25}_{-0.26}$ & $ 1.32^{+0.10}_{-0.05}$ & $ 0.80^{+0.02}_{-0.05}$ & $ 0.23^{+0.06}_{-0.03}$ & $ 0.38^{+0.19}_{-0.05}$ & $ 2.49^{+0.39}_{-0.42}$ & $ 1.43^{+0.17}_{-0.08}$  \\ 
26 & $ 2.05^{+0.12}_{-0.08}$ & $ 0.27^{+0.07}_{-0.07}$ & $ 0.23^{+0.13}_{-0.05}$ & $ 1.65^{+0.84}_{-0.71}$ & $ 2.45^{+0.13}_{-0.04}$ & $ 2.05^{+0.12}_{-0.08}$ & $ 0.27^{+0.07}_{-0.07}$ & $ 0.23^{+0.13}_{-0.05}$ & $ 2.30^{+0.93}_{-0.73}$ & $ 2.52^{+0.14}_{-0.03}$  \\ 
27 & $ 2.21^{+0.03}_{-0.06}$ & $ 0.02^{+0.03}_{-0.01}$ & $ 0.01^{+0.02}_{-0.01}$ & $ 1.32^{+0.38}_{-1.08}$ & $ 2.36^{+0.06}_{-0.16}$ & $ 2.22^{+0.03}_{-0.07}$ & $ 0.02^{+0.02}_{-0.01}$ & $ 0.02^{+0.03}_{-0.01}$ & $ 1.70^{+0.48}_{-1.18}$ & $ 2.41^{+0.07}_{-0.16}$  \\ 
28 & $ 2.89^{+0.07}_{-0.05}$ & $ 0.47^{+0.09}_{-0.10}$ & $ 0.60^{+0.11}_{-0.13}$ & $ 0.42^{+0.07}_{-0.05}$ & $ 3.53^{+0.11}_{-0.08}$ & $ 2.89^{+0.07}_{-0.05}$ & $ 0.47^{+0.09}_{-0.10}$ & $ 0.63^{+0.11}_{-0.13}$ & $ 0.91^{+0.15}_{-0.11}$ & $ 3.61^{+0.11}_{-0.09}$  \\ 
29 & $ 1.45^{+0.00}_{-0.01}$ & $ 0.07^{+0.02}_{-0.01}$ & $ 0.08^{+0.02}_{-0.01}$ & $ 0.41^{+0.13}_{-0.04}$ & $ 1.57^{+0.02}_{-0.01}$ & $ 1.45^{+0.00}_{-0.01}$ & $ 0.07^{+0.02}_{-0.01}$ & $ 0.08^{+0.02}_{-0.01}$ & $ 0.61^{+0.13}_{-0.05}$ & $ 1.59^{+0.02}_{-0.01}$  \\ 
30 & $ 0.68^{+0.03}_{-0.03}$ & $ 0.29^{+0.05}_{-0.05}$ & $ 0.27^{+0.05}_{-0.04}$ & $ 0.86^{+0.16}_{-0.13}$ & $ 1.04^{+0.04}_{-0.03}$ & $ 0.68^{+0.03}_{-0.03}$ & $ 0.29^{+0.05}_{-0.05}$ & $ 0.27^{+0.05}_{-0.04}$ & $ 1.36^{+0.19}_{-0.17}$ & $ 1.09^{+0.05}_{-0.04}$  \\ 
31 & $ 0.32^{+0.07}_{-0.04}$ & $ 0.96^{+0.03}_{-0.05}$ & $ 1.24^{+0.15}_{-0.05}$ & $ 0.76^{+0.30}_{-0.22}$ & $ 1.64^{+0.15}_{-0.02}$ & $ 0.33^{+0.07}_{-0.04}$ & $ 0.96^{+0.03}_{-0.05}$ & $ 1.24^{+0.15}_{-0.05}$ & $ 1.22^{+0.34}_{-0.24}$ & $ 1.69^{+0.16}_{-0.02}$  \\ 
32 & $ 0.56^{+0.05}_{-0.02}$ & $ 0.19^{+0.04}_{-0.04}$ & $ 0.18^{+0.04}_{-0.05}$ & $ 0.90^{+0.16}_{-0.34}$ & $ 0.83^{+0.03}_{-0.03}$ & $ 0.56^{+0.05}_{-0.02}$ & $ 0.19^{+0.04}_{-0.04}$ & $ 0.18^{+0.04}_{-0.05}$ & $ 1.43^{+0.14}_{-0.30}$ & $ 0.89^{+0.03}_{-0.03}$  \\ 
33 & $ 1.29^{+0.02}_{-0.02}$ & $ 0.29^{+0.03}_{-0.06}$ & $ 0.36^{+0.03}_{-0.08}$ & $ 0.27^{+0.05}_{-0.05}$ & $ 1.68^{+0.02}_{-0.07}$ & $ 1.29^{+0.02}_{-0.02}$ & $ 0.29^{+0.03}_{-0.06}$ & $ 0.36^{+0.03}_{-0.08}$ & $ 0.57^{+0.11}_{-0.10}$ & $ 1.71^{+0.02}_{-0.07}$  \\ 
34 & $ 1.21^{+0.12}_{-0.08}$ & $ 0.56^{+0.08}_{-0.14}$ & $ 1.38^{+0.20}_{-0.31}$ & $ 3.13^{+0.60}_{-0.82}$ & $ 2.90^{+0.18}_{-0.25}$ & $ 1.21^{+0.12}_{-0.08}$ & $ 0.56^{+0.08}_{-0.14}$ & $ 1.50^{+0.25}_{-0.35}$ & $ 4.11^{+0.71}_{-1.01}$ & $ 3.12^{+0.23}_{-0.30}$  \\ 
35 & $ 1.05^{+0.05}_{-0.04}$ & $ 0.45^{+0.07}_{-0.06}$ & $ 0.97^{+0.16}_{-0.20}$ & $ 0.38^{+0.07}_{-0.05}$ & $ 2.06^{+0.13}_{-0.17}$ & $ 1.06^{+0.05}_{-0.04}$ & $ 0.45^{+0.07}_{-0.06}$ & $ 1.03^{+0.18}_{-0.22}$ & $ 0.74^{+0.11}_{-0.08}$ & $ 2.16^{+0.15}_{-0.19}$  \\ 
36 & $ 0.67^{+0.01}_{-0.01}$ & $ 0.00^{+0.00}_{-0.00}$ & $ 0.00^{+0.00}_{-0.00}$ & $ 0.75^{+0.09}_{-0.12}$ & $ 0.74^{+0.01}_{-0.02}$ & $ 0.67^{+0.01}_{-0.01}$ & $ 0.00^{+0.00}_{-0.00}$ & $ 0.01^{+0.00}_{-0.00}$ & $ 1.00^{+0.12}_{-0.16}$ & $ 0.78^{+0.01}_{-0.02}$  \\ 
37 & $ 2.17^{+0.08}_{-0.09}$ & $ 1.46^{+0.16}_{-0.21}$ & $ 2.17^{+0.20}_{-0.35}$ & $ 1.03^{+0.20}_{-0.10}$ & $ 4.45^{+0.18}_{-0.28}$ & $ 2.17^{+0.08}_{-0.09}$ & $ 1.49^{+0.15}_{-0.22}$ & $ 3.31^{+0.23}_{-0.55}$ & $ 2.42^{+0.59}_{-0.19}$ & $ 5.73^{+0.26}_{-0.49}$  \\ 
38 & $ 1.01^{+0.03}_{-0.03}$ & $ 0.17^{+0.09}_{-0.03}$ & $ 0.38^{+0.20}_{-0.16}$ & $ 0.34^{+0.12}_{-0.03}$ & $ 1.42^{+0.19}_{-0.14}$ & $ 1.01^{+0.03}_{-0.03}$ & $ 0.17^{+0.09}_{-0.03}$ & $ 0.42^{+0.31}_{-0.19}$ & $ 0.69^{+0.15}_{-0.06}$ & $ 1.50^{+0.31}_{-0.17}$  \\ 
39 & $ 0.00^{+0.01}_{-0.00}$ & $ 0.76^{+0.04}_{-0.04}$ & $ 1.09^{+0.05}_{-0.05}$ & $ 0.73^{+0.18}_{-0.19}$ & $ 1.16^{+0.03}_{-0.04}$ & $ 0.00^{+0.01}_{-0.00}$ & $ 0.77^{+0.05}_{-0.04}$ & $ 1.88^{+0.08}_{-0.09}$ & $ 1.15^{+0.26}_{-0.23}$ & $ 1.99^{+0.08}_{-0.08}$  \\ 
40 & $ 0.59^{+0.11}_{-0.01}$ & $ 0.45^{+0.02}_{-0.13}$ & $ 0.53^{+0.02}_{-0.12}$ & $ 0.60^{+0.07}_{-0.10}$ & $ 1.18^{+0.04}_{-0.04}$ & $ 0.59^{+0.11}_{-0.01}$ & $ 0.45^{+0.02}_{-0.13}$ & $ 0.53^{+0.02}_{-0.12}$ & $ 0.80^{+0.10}_{-0.11}$ & $ 1.20^{+0.04}_{-0.04}$  \\ 
41 & $ 0.44^{+0.04}_{-0.03}$ & $ 0.12^{+0.03}_{-0.02}$ & $ 0.16^{+0.05}_{-0.03}$ & $ 1.07^{+0.37}_{-0.49}$ & $ 0.70^{+0.05}_{-0.03}$ & $ 0.44^{+0.04}_{-0.03}$ & $ 0.12^{+0.03}_{-0.02}$ & $ 0.16^{+0.05}_{-0.03}$ & $ 1.45^{+0.41}_{-0.52}$ & $ 0.74^{+0.05}_{-0.04}$  \\ 
42 & $ 2.42^{+0.08}_{-0.22}$ & $ 1.67^{+0.17}_{-0.20}$ & $ 1.35^{+0.14}_{-0.15}$ & $ 2.78^{+0.88}_{-0.43}$ & $ 4.05^{+0.15}_{-0.23}$ & $ 2.42^{+0.08}_{-0.22}$ & $ 2.03^{+0.19}_{-0.28}$ & $ 1.36^{+0.15}_{-0.17}$ & $ 6.96^{+1.43}_{-1.16}$ & $ 4.48^{+0.20}_{-0.29}$  \\ 
\hline	
\end{tabular}
\par
{{\bf Table B4} Infrared and bolometric luminosities from the Siebenmorgen15 model fits.}

\end{center}
\end{table*}

\label{lastpage}

\end{document}